%% file: Paper_arxiv.tex
\documentclass[12pt, draftclsnofoot, onecolumn]{IEEEtran}
\IEEEoverridecommandlockouts

\makeatletter
\def\markboth#1#2{\def\leftmark{\@IEEEcompsoconly{\sffamily}\MakeUppercase{\protect#1}}%
\def\rightmark{\@IEEEcompsoconly{\sffamily}\MakeUppercase{\protect#2}}}
\makeatother

\PassOptionsToPackage{bookmarks={false}}{hyperref}
\usepackage[utf8x]{inputenc}
\usepackage[english]{babel}
\usepackage{ucs}
\usepackage{amsmath}
\usepackage{amsfonts}
\usepackage{amssymb}
\usepackage{amsthm}
\usepackage{array}
\usepackage{verbatim}
\usepackage{listings}
\usepackage{psfrag}
\usepackage{stfloats}

\usepackage{algorithm}
\usepackage{url}
\usepackage{enumerate}
\usepackage{cite}
\usepackage[usenames,dvipsnames,svgnames,table]{xcolor}

\ifCLASSINFOpdf
  \usepackage[pdftex]{graphicx}
  \usepackage[caption=false]{subfig}
  \usepackage[outdir=./]{epstopdf}
  \graphicspath{{./Figures/}}
   \DeclareGraphicsExtensions{.eps,.ps,.png,.pdf}
\else
  \usepackage[dvipdf]{graphicx}
  \usepackage{subfigure}
  \graphicspath{{./Figures/}}
   \DeclareGraphicsExtensions{.eps,.ps,.pdf}
\fi

\usepackage[acronyms,nonumberlist,nopostdot,nomain,nogroupskip]{glossaries}
\input{acronyms.tex}
\usepackage{tablefootnote}
\usepackage{booktabs}

\usepackage{tabularx}

\usepackage{tikz}
\usepackage{pgfplots}
\pgfplotsset{compat=newest}
\pgfplotsset{plot coordinates/math parser=false}
\usetikzlibrary{plotmarks,patterns,decorations.pathreplacing,backgrounds,calc,arrows,arrows.meta,spy,matrix}
\usepgfplotslibrary{patchplots,groupplots}
\usepackage{tikzscale}
\usepackage{hyperref}

\usepackage{algpseudocode, algorithmicx}
\newcommand\bigO[1]{$\mathcal{O}(#1)$}
\newtheorem{lem}{Lemma}
\usepackage[binary-units=true]{siunitx}

\input{myTikz.tex}

\usepackage{soul} 

\begin{document}

\title{Resource Management for 5G NR\\Integrated Access and Backhaul:\\a Semi-centralized Approach}

\author{\IEEEauthorblockN{
Matteo~Pagin,
Tommaso~Zugno,~\IEEEmembership{Student Member,~IEEE,} 
Michele~Polese,~\IEEEmembership{Member,~IEEE,} 
Michele~Zorzi,~\IEEEmembership{Fellow,~IEEE}%
}
\thanks{Matteo Pagin, Tommaso Zugno and Michele Zorzi are with the Department of Information Engineering, University of Padova, Padova, Italy.~Email: $\{$paginmatte, zugnotom, zorzi$\} \,$@dei.unipd.it}%
\thanks{Michele Polese is with the Institute for the Wireless Internet of Things, Northeastern University, Boston, MA USA.~Email: m.polese@northeastern.edu}
}

\makeatletter
\patchcmd{\@maketitle}
  {\addvspace{0.5\baselineskip}\egroup}
  {\addvspace{-2\baselineskip}\egroup}
  {}
  {}
\makeatother

\maketitle

\begin{tikzpicture}[remember picture,overlay]
\node[anchor=north,yshift=0pt] at (current page.north) {\fbox{\parbox{\dimexpr\textwidth-\fboxsep-\fboxrule\relax}{
\centering\footnotesize This paper has been submitted to IEEE for publication. Copyright may change without notice.}}};
\end{tikzpicture}

\begin{abstract}
The next generations of mobile networks will be deployed as ultra-dense networks, to match the demand for increased capacity and the challenges that communications in the higher portion of the spectrum (i.e., the mmWave band) introduce. Ultra-dense networks, however, require pervasive, high-capacity backhaul solutions, and deploying fiber optic to all base stations is generally considered to be too expensive for network operators. The \gls{3gpp} has thus introduced \gls{iab}, a wireless backhaul solution in which the access and backhaul links share the same hardware, protocol stack, and also spectrum. The multiplexing of different links in the same frequency bands, however, introduces interference and capacity sharing issues, thus calling for the introduction of advanced scheduling and coordination schemes. This paper proposes a semi-centralized resource allocation scheme for \gls{iab} networks, designed to be flexible, with low complexity, and compliant with the \gls{3gpp} \gls{iab} specifications. We develop a version of the \gls{mwm} problem that can be applied on a spanning tree that represents the \gls{iab} network and whose complexity is linear in the number of \gls{iab}-nodes. The proposed solution is compared with state-of-the-art distributed approaches through end-to-end, full-stack system-level simulations with a \gls{3gpp}-compliant channel model, protocol stack, and a diverse set of user applications. The results show how that our scheme can increase the throughput of cell-edge users up to 5 times, while decreasing the overall network congestion with an end-to-end delay reduction of up to 25 times.
\end{abstract}


\glsresetall

\glsunset{nr}

\section{Introduction}

Future generations of cellular networks will provide groundbreaking network capacity, in conjunction with a significantly lower delay and ubiquitous coverage~\cite{Ericss_rep}. \gls{5g} and beyond deployments will support new mobile broadband use cases, e.g., Augmented (AR) and \gls{vr}, and expand into new vertical markets, enabling an unprecedented degree of automation in industrial scenarios, \gls{v2x} communications, remote medical surgery and smart electrical grids.

To this end, the \gls{3gpp} has introduced various technological advancements with the specifications of the \gls{5g} \gls{ran} and \gls{cn}, namely \gls{nr} and \gls{5gc}~\cite{3gpp_38_300}. In particular, \gls{nr} features a user and control plane split, a flexible \gls{ofdm} frame structure, and the support for \gls{mmwave} communications, while the \gls{cn} introduces virtualization and slicing~\cite{yousaf2017nfv}. 

Specifically, the \gls{mmwave} band, i.e., the portion of the spectrum between 30 GHz and 300 GHz, represents the major technological enabler toward the Gbit/s capacity target. These frequencies are characterized by the availability of vast chunks of contiguous and currently unused spectrum, in stark contrast with the crowded sub-6 GHz frequencies. However, \glspl{mmwave} exhibit unfavorable propagation characteristics such as high isotropic losses and a marked susceptibility to blockages and signal attenuation~\cite{khan2011mmwave, rangan2014millimeter}.
These issues can be partially mitigated using beamforming through large antenna arrays, thanks to the small wavelengths and advances in low-power \gls{cmos} RF circuits~\cite{hemadeh2017millimeter}; nevertheless, their introduction alone is not enough for meeting the high service availability requirement. 
Therefore, \glspl{mmwave} networks also need densification, to decrease the average distance between mobile terminals and base stations and improve the average \gls{sinr}. The theoretical effectiveness of this technique is well understood~\cite{gomez2017capacity}; however, achieving dense \gls{5g} deployments is extremely challenging from a practical point of view. Specifically, providing a fiber backhaul among base stations and toward the \gls{cn} is deemed economically impractical, even more so in the initial \gls{5g} deployments~\cite{polese2020integrated}.

Recently, wireless backhaul solutions for \gls{5g} networks have emerged as a viable strategy toward cost effective, dense \gls{mmwave} deployments. Notably, the \gls{3gpp} has promoted \gls{iab}~\cite{3gpp_38_874}, i.e., a wireless backhaul architecture which dynamically splits the overall system bandwidth for backhaul and access purposes. \gls{iab} has been integrated in the latest release of the \gls{3gpp} \gls{nr} specifications. Prior research has highlighted that \gls{iab} represents a cost-performance trade-off~\cite{stoch_geom2, polese2020integrated}, as base stations need to multiplex access and backhaul resources, and as the wireless backhaul at \glspl{mmwave} is less reliable than a fiber connection. In particular, \gls{iab} networks may suffer from excessive buffering (and, consequently, high latency and low throughput) when a suboptimal partition of access and backhaul resources is selected, thus hampering the benefits that the high bandwidth \gls{mmwave} links introduce~\cite{polese2020integrated,polese2018end}. Therefore, it is fundamental to solve these non-trivial challenges to enable a smooth integration of \gls{iab} in \gls{5g} and beyond deployments.

\subsection{Contributions}

This article tackles the access and backhaul partitioning problem by proposing an optimal, semi-centralized resource allocation scheme for \gls{3gpp} \gls{iab} networks, based on the \gls{mwm} problem on graphs.  It receives periodic \gls{kpi} reports from the nodes of the \gls{iab} deployment, constructs a spanning tree that represents the deployment, and uses a simplified, low-complexity version of the \gls{mwm} to partition the links between access and backhaul. After a feedback step, each node can then schedule the resources at a subframe-level among the connected devices. To the best of our knowledge, this is the first \gls{mwm}-based resource allocation framework for \gls{3gpp} \gls{iab} networks at \glspl{mmwave}, designed with three goals, i.e., it is flexible, integrated with the \gls{3gpp} specifications, and has low complexity. 

The flexibility makes it possible to easily adapt the resource allocation strategy to different requirements, use cases, and classes of traffic for \gls{5g} networks. We achieve this by developing a generic optimization algorithm, which identifies with a configurable periodicity the access and backhaul partition that optimizes a certain utility function. The selection of the utility function prioritizes the optimization of different metrics, e.g., throughput or latency, which in turn can be mapped to different classes of traffic. 
To achieve the second goal, i.e., the compliance with the \gls{3gpp} \gls{iab} specifications, the resource allocation framework relies only on information that can be actually exchanged and reported in a \gls{3gpp} deployment. In this regard, we also review the latest updates related to the \gls{3gpp} \gls{iab} standardization activities.

Finally, the algorithm operates with a low complexity, i.e., we propose a version of the \gls{mwm} algorithm that can be applied on spanning trees with linear complexity in the number of nodes in the network infrastructure, and demonstrate its equivalence to the generic (and more complex) \gls{mwm}. Additionally, the proposed framework also relies on a feedback exchange that is linear in the number of base stations, and is thus decoupled from the number of users.

Furthermore, we evaluate the performance of the proposed scheme with an end-to-end, full-stack system-level simulation, using the ns-3 mmWave module~\cite{mezzavilla2018end} and its \gls{iab} extension~\cite{polese2018end}. This represents the first evaluation of an optimized resource allocation scheme for \gls{iab} with a simulator that is based on a \gls{3gpp}-compliant protocol stack, uses \gls{3gpp} channel models, and integrates realistic applications and transport protocols. The extended performance evaluation highlights how the proposed scheme improves the throughput of a diverse set of applications, with a 5-fold increase for the worst case users, with different packet sizes and transport protocols, while decreasing the latency and buffering at intermediate nodes by up to 25 times for the smallest packet sizes.

\subsection{State of the Art}

This section reviews relevant research on resources allocation in a multi-hop wireless network, deployed through either \gls{iab} or other wireless mesh solutions~\cite{gambiroza2004end}.

The literature adopts different approaches to model and solve the resource allocation problem. The first, discussed in~\cite{qualcomm1, qualcomm2, kulkarni2018max, lei2020deep, rasekh2015interference, bilal2014time, cruz2003optimal} is based on conventional optimization techniques.
Specifically, the authors of~\cite{qualcomm1} present a simple and thus tractable system model and find the minimal number of \glspl{gnb} featuring a wired backhaul that are needed to sustain a given traffic load. Their work is further extended in~\cite{qualcomm2}, which provides an analysis of the performance benefits introduced by additional, fiber-less \glspl{gnb}. 
In~\cite{kulkarni2018max}, the mobile network is modeled as a noise-limited, $k$-ring deployment. Such model is then used to obtain closed-form expressions for the max-min rates achieved by \glspl{ue} in the network. Moreover,~\cite{lei2020deep} proposes a system model which leads to an NP-hard optimization problem, even though it considers single-hop backhaul networks only, and uses deep \gls{rl} to reduce its computation complexity. In~\cite{rasekh2015interference}, the joint routing and resource allocation problem is tackled via a \gls{lp} technique. Notably, this work assumes that data can be transmitted (received) toward (from) multiple nodes at the same time.
Similarly, the authors of~\cite{bilal2014time} formulate a \gls{tdd}, multi-hop resource allocation optimization problem which leverages the directionality of \gls{mmwave} antennas, albeit in the context of \gls{wpans}. Since such problem is also NP-hard, a sub-optimum solution is found. Finally,~\cite{cruz2003optimal} focuses on joint link scheduling, routing and power allocation in multi-hop wireless networks. As in previous cases the obtained optimization problem is not tractable: in this instance such obstacle is overcome by studying the dual problem via an iterative approach.


The second approach relies on stochastic geometry to model \gls{iab} networks~\cite{stoch_geom1, stoch_geom2}. Specifically,~\cite{stoch_geom1} determines the rate coverage probability of \gls{iab} networks and compares different access/backhaul resource partitioning strategies. Similarly,~\cite{stoch_geom2} provides a comparison of orthogonal and integrated resource allocation policies, although limited to single-hop wireless networks.  

Another significant body of literature leverages \glspl{mc} to study \gls{iab} networks; some of these works can be interpreted as a direct application of such theory~\cite{singh2018throughput, ji2012throughput}, while others~\cite{vu2018path,garcia2015analysis, gomez2016optimal, gomez2019optimal} exploit a more complex framework.  
The papers which belong to the former class are based on the  pioneering work of~\cite{tassiulas1990stability}, which inspects the stability of generic multi-hop wireless networks and formulates a throughput-maximizing algorithm known as \textit{back-pressure}. In particular,~\cite{singh2018throughput} focuses on the optimization of the timely-throughput, i.e., takes into account that packets usually have an arrival deadline. Such problem is then addressed by formulating a \gls{mdp}, leading to a distributed resource allocation algorithm. Similarly,~\cite{ji2012throughput} proposes an algorithm that also targets throughput optimality but, contrary to the back-pressure algorithm, manages to avoid the need for per-flow information.
On the other hand, the body of literature which belongs to the latter class uses the \gls{mc}-derived \gls{num} framework first introduced in~\cite{kelly1997charging} and ~\cite{kelly1998rate}. Specifically, the authors of~\cite{vu2018path} focus on satisfying the \gls{urllc} \gls{qos} requirements by jointly optimizing routing and resource allocation. Then, 
the problem is solved using both convex optimization and \gls{rl} techniques. In~\cite{garcia2015analysis}, an in-depth analysis of a \gls{mmwave}, multi-hop wireless system is presented, proposing and comparing three different interference frameworks, under the assumption of a dynamic \gls{tdd} system. This work is extended in~\cite{gomez2016optimal} and~\cite{gomez2019optimal}, which consider respectively a \gls{sdma} and a \gls{mu}-\gls{mimo} capable system.

Finally, only a small portion of the literature~\cite{polese2018end, polese2018iab, polese2020integrated} analyzes the end-to-end performance of \gls{iab} networks. Specifically, the authors of~\cite{polese2018end} extend the ns-3 \gls{mmwave} module, introducing realistic \gls{iab} functionalities which are then used to characterize the benefit of deploying wireless relays in \gls{mmwave} networks. Their work is extended in~\cite{polese2018iab}, where path selection policies are formulated and their impact on the system performance is inspected. A further end-to-end analysis of \gls{iab} networks is carried out in~\cite{polese2020integrated}, providing insights into the potentials of this technology and the related open research challenges.

Concluding, the literature exhibits the presence of algorithms relying on a varying degree of assumptions on the network topology and the knowledge of system. Furthermore, most of the aforementioned studies lack an end-to-end, full-stack system-level analysis of the proposed solution.
Conversely, this paper proposes a semi-centralized resource allocation scheme, which also has a low complexity, both computationally and in terms of required feedback. Moreover, we provide considerations on how our proposed solution can be implemented and deployed in standard-compliant \gls{3gpp} \gls{iab} networks, and compare such solution to the state of the art with an end-to-end, realistic performance analysis.

\subsection{Paper structure}
The remainder of this paper is organized as follows. Sec.~\ref{Sec:Sys-model} describes our assumptions and the system model. Then, Sec.~\ref{Sec:scheme_main} presents a novel scheme for resource partitioning in \gls{mmwave} \gls{iab} networks, along with considerations on how it can be implemented in \gls{3gpp} NR. Finally, Sec.~\ref{Sec:perf_eval} describes the performance evaluation results and Sec.~\ref{Sec:conc} concludes this paper.

\section{IAB networks}
\label{Sec:Sys-model}

The following paragraphs identify the characteristics and constraints of \gls{mmwave} \gls{iab}, according to the \gls{3gpp} design guidelines presented in~\cite{3gpp_38_874} and the specifications of~\cite{3gpp_38_174}.

\subsection{Network topology}
\begin{figure}[tbp]
	\centering
  	\subfloat[\gls{st} and \gls{dag} topologies.\label{Fig:IAB_topology}]{
  	\includegraphics[width=0.46\linewidth]{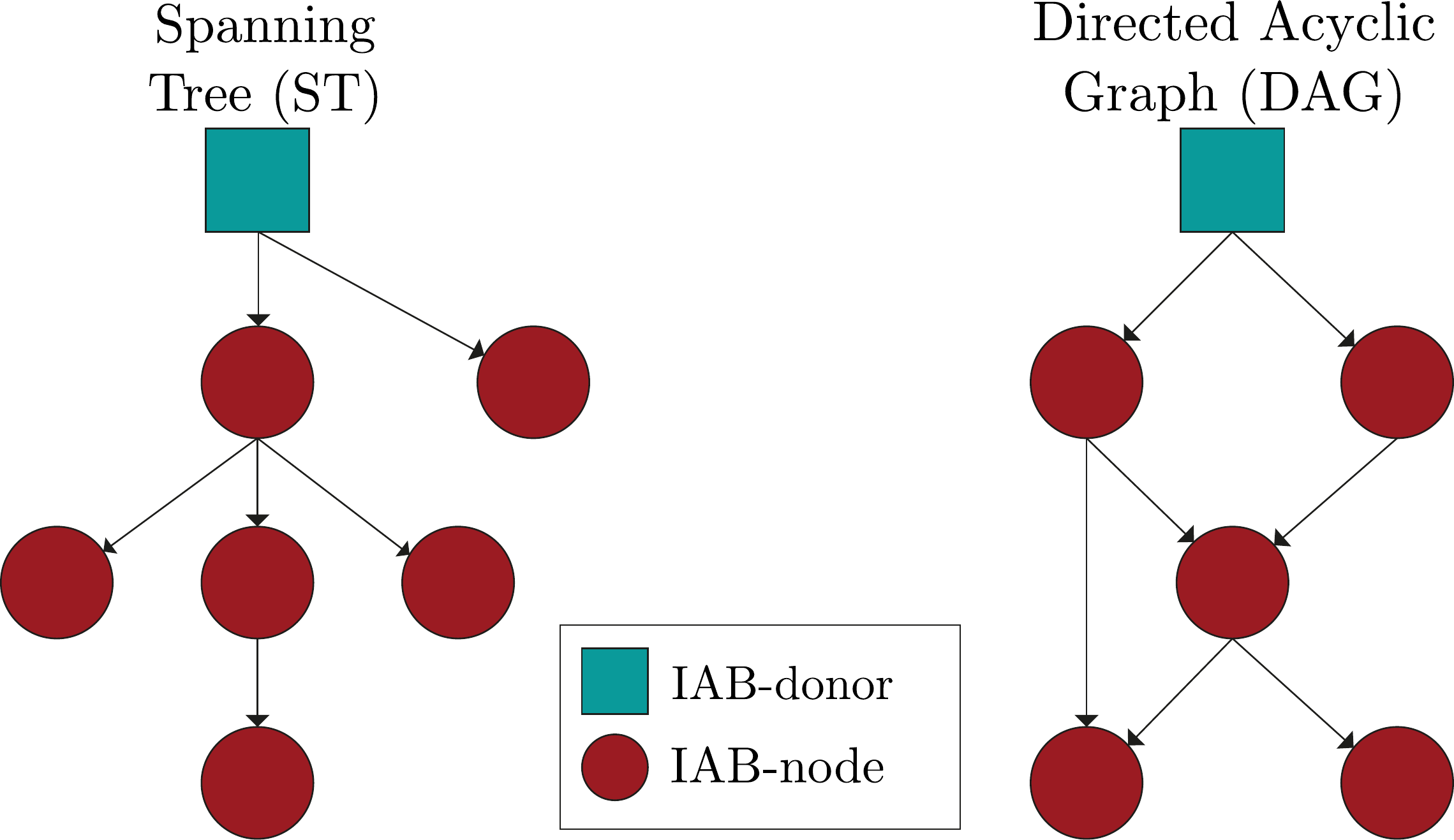}}
  	\hfill
  	\subfloat[System model notation.\label{Fig:IAB_notation}]{
  	\includegraphics[width=0.46\linewidth]{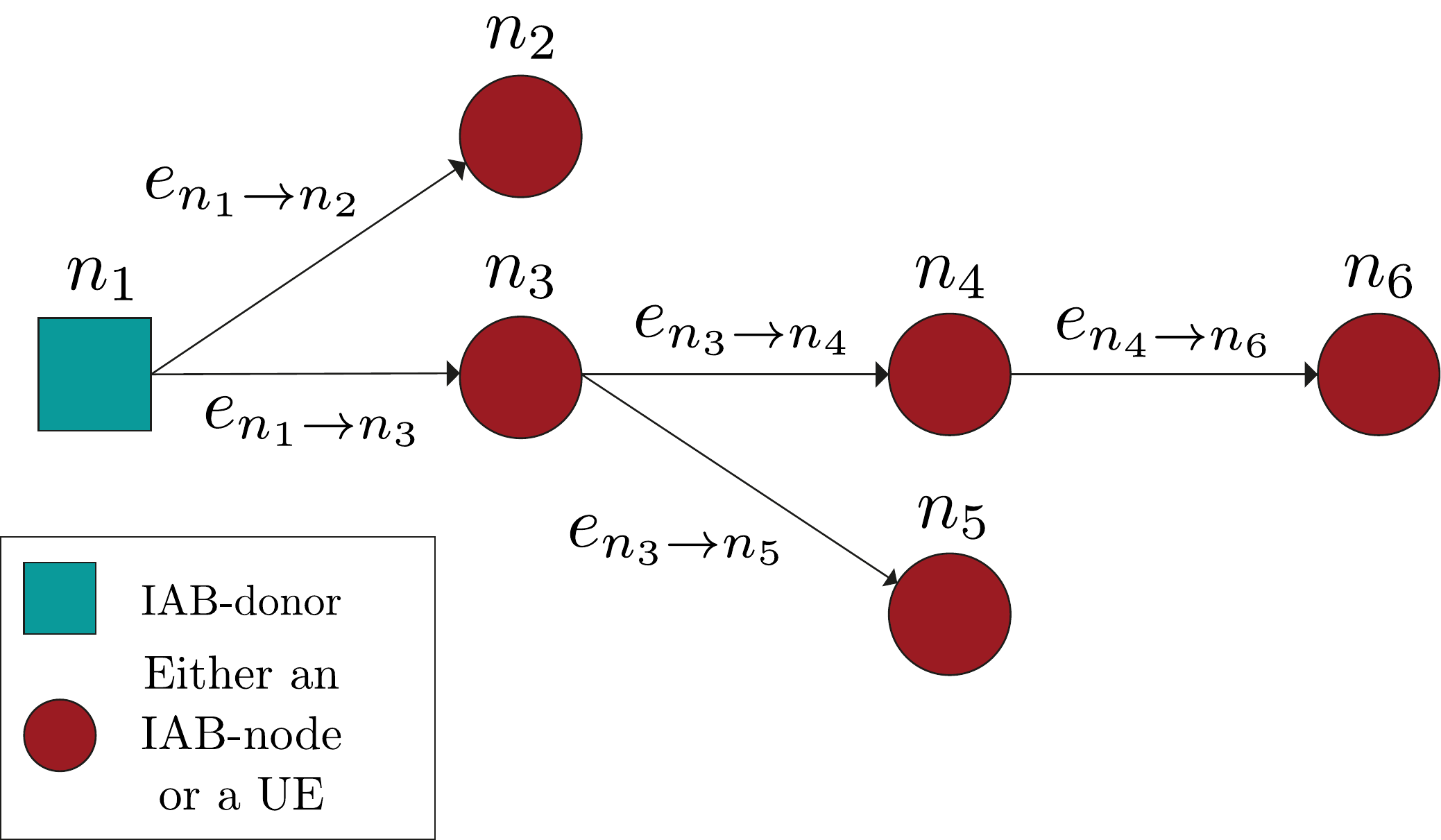}}
    \caption{Comparison of the \gls{iab} network topologies analyzed in~\cite{3gpp_38_874} and related notation.}
    \label{Fig:IAB_top_not}
    \vspace{-.6cm}  
\end{figure}

In general, an \gls{iab} network is a deployment where a percentage of \glspl{gnb} (i.e., the \gls{iab}-nodes) use wireless backhaul connections to connect to a few \glspl{gnb} (i.e., the \gls{iab}-donors) which feature a wired connection to the core network, as can be seen in Fig.~\ref{Fig:IAB_top_not}.
Moreover, these deployments exhibit a \textit{multi-hop} topology where a strict parent-child relation is present. The former can be represented by the \gls{iab}-donor itself or an \gls{iab}-node; the latter by either \gls{ue}s or downstream \gls{iab}-nodes. In~\cite{3gpp_38_874}, no a priori limit on the number of backhaul hops is introduced. As a consequence, \gls{3gpp} argues that \gls{iab} protocols should provide sufficient flexibility with respect to the number of backhaul hops. 
Moreover, the \gls{si} on \gls{iab}~\cite{3gpp_38_874} highlights the support for both the topologies depicted in Fig.~\ref{Fig:IAB_topology}, i.e., \gls{st} and \gls{dag} \gls{iab}. Clearly, the former exhibits less complexity but, at the same time, poses some limits in terms of network performance: the possible presence of obstacles may result in a service interruption, due to the unique backhaul route established by the \glspl{ue}. 
On the other hand, a \gls{dag} topology offers routing redundancy, which can be used not only to decrease the probability of experiencing a ``topological blockage," but also for load balancing purposes.



\subsection{Multiple access schemes and scheduling}
An in-band, dynamic partitioning of the access and backhaul spectrum resources is currently preferred by \gls{3gpp}~\cite{3gpp_38_874, 3gpp_38_174}, together with half-duplex operations of the \gls{iab}-nodes. Moreover, most of the literature suggests that \gls{5g} \gls{mmwave} systems will operate in a \gls{tdd} fashion~\cite{khan2011mmwave, dutta2017frame}. This choice is mainly driven by the stringent latency requirements which the next generation of mobile networks will be required to support, and by the usage of analog or hybrid beamforming. The usage of \gls{fdd}, in conjunction with the presence of large chunks of bandwidth, would lead to severe resource under-utilization and make channel estimation more difficult.

Based on these considerations, the system model exhibits a \gls{tdd}, \gls{tdma}-based scheduling where the access/backhaul interfaces are multiplexed in a half-duplex manner. It follows that at any given time instant, each node of the \gls{iab} network cannot be simultaneously involved in more than one transmission or reception. In particular, \gls{iab}-nodes cannot schedule time and frequency resources which are already allocated by their parent for backhaul communications which involve them.
Finally, the introduction of resource coordination mechanisms and related signaling is explicitly supported in the \gls{iab} specification drafts~\cite{3gpp_38_874, 3gpp_38_174}. Nevertheless, these solutions must reuse as much as possible the available NR specifications and
require at most minimal changes to the Rel.15 \gls{5gc} and \gls{nr}.


\subsection{System model}
\label{Subsec:sys_model}

According to these assumptions and referring to Fig.~\ref{Fig:IAB_notation}., a generic \gls{iab} network can be modeled as a directed graph $\mathcal{G} = \{\mathcal{N},  \mathcal{E} \}$, where the set of nodes $\mathcal{N} \equiv \{ n_1, \, n_2, \, \ldots \, n_{\vert \mathcal{N} \vert }  \}$ comprises the \gls{iab}-donor, the various \gls{iab}-nodes and the \glspl{ue}. 
Accordingly, the set of directed edges $\mathcal{E} \equiv \{ e_{1} , \, e_{2}, \, \ldots \, \allowbreak e_{\vert \mathcal{E} \vert} \} \equiv \{ e_{n_j \to n_k} \}_{j, k} $, where the edge $e_{n_j \to n_k}$ originates at the parent node $n_j$ and terminates at the children $n_k$, comprises in all the active cell attachments, either of mobile terminals to a \gls{gnb} or from \gls{iab}-nodes towards their parent node.
Since the goal of this paper is to study backhaul/access resource partitioning policies, this generic model can be actually simplified: in fact, all the \glspl{ue} connected to a given \gls{gnb} can be represented by a single node in $\mathcal{G}$ without any loss of generality. Similarly, the same holds true for their links toward the serving \gls{gnb}, which can then be represented by a single edge. 
Furthermore, this work focuses on \gls{st} topologies only. Nevertheless, the proposed framework can be easily extended to the case of a \gls{dag} \gls{iab} network: it suffices to introduce a preliminary routing step where an \gls{st} $\mathcal{G^{'}}$ is computed from the actual \gls{dag} network $\mathcal{G}$, for example by using the strategies presented in~\cite{polese2018iab}. Such process can be possibly repeated at each allocation instance, effectively removing any constraint on the network topology. 

We define as \textit{feasible schedule} any set of links $\mathcal{E'} \subseteq \mathcal{E}$ such that none of them share a common vertex, i.e., $ \forall \, e_{n_j \to n_k} \neq e_{n_l \to n_m} \in \mathcal{E'}$ it holds that $n_{j} \neq n_{m}$ and $n_{l} \neq n_{k}$. Let then $f_u$ be a utility \textit{additive map}, namely, a function such that the overall utility experienced by the system when scheduling edges $e_1$ and $e_2$ satisfies $f_u (e_1, e_2) = f_u (e_1) + f_u(e_2)$. Let also $\mathcal{W} \equiv \{ w_1, \, w_2, \, \ldots \, w_{\vert \mathcal{E} \vert } \}$ be the set of positive weights whose generic entry $w_j$ represents the utility which is obtained when scheduling the $j$-th edge, namely, $w_j \equiv f_u (e_j)$. Then, the overall utility of the system is $\mathcal{U} \equiv \sum_{e_k \, \in \, \mathcal{E'}} f_u(e_k) = \sum_{e_k \, \in \, \mathcal{E'}} w_k $.
The goal is to find the feasible set $\mathcal{E'}^{*}$ which maximizes the overall utility, i.e., $\underset{\mathcal{E'}}{\mathrm{argmax}} \,\, \mathcal{U} $. In computer science, this task is typically referred to as the \textit{Maximum Weighted Matching} problem~\cite{Korte2002}.

Finding the \gls{mwm} of a given graph, in the general case, is not trivial from a computational point of view. 
In fact, the fastest known \gls{mwm} algorithm for generic graphs has a complexity of \bigO{\vert V \vert \vert E \vert + \vert V \vert^2 \log{\vert V \vert}}~\cite{1990Gabow}, posing serious limitations to the suitability of such algorithm to \gls{5g} and beyond networks, which target a connection density of 1 million devices per km\textsuperscript{2}. However, we argue that under the aforementioned assumptions on the system model, which restrict the network to an \gls{st} topology, it is possible to design an \gls{mwm}-based centralized resource partitioning framework which exhibits linear complexity with respect to the network size and which, as a result, is able to satisfy the scalability requirements highlighted by \gls{3gpp} in~\cite{3gpp_38_874}.

\section{Semi-centralized resource allocation scheme for IAB networks}
\label{Sec:scheme_main}

This section presents an \gls{mwm} algorithm for \gls{st} topologies (Sec.~\ref{Sec:T-MWM}), an efficient and \gls{mwm}-based centralized resource partitioning framework for \gls{iab} networks (Sec.~\ref{Sec:Cent-scheme}) and some considerations about its implementation (Sec.~\ref{Sec:ns3-impl}).
\subsection{\gls{mwm} for \gls{st} graphs}
\label{Sec:T-MWM}

As the first of our contributions, we present an algorithm, hereby called \texttt{T-MWM}, which computes the \gls{mwm} of an \gls{st} in linear time. In particular, \texttt{T-MWM} is a bottom-up algorithm which, upon receiving as input a weighted \gls{st} $\mathcal{G}$ described by its edge list $\mathbf{E}$ and the corresponding weight list $\mathbf{W}$, produces as output a set of active edges $\mathbf{E^*}$ which are an \gls{mwm} of $\mathcal{G}$. 
Furthermore, $\mathbf{E}$ is from now on assumed to exhibit the following invariant: each \gls{iab} parent precedes its children in the list, hence avoiding the need for a recursion. 
This is automatically obtained as each \gls{iab} child connects after its parent, and is thus added to the list in a subsequent position.
Nevertheless, this assumption can be easily relaxed, albeit at the cost of losing as a side-effect the bottom-up design.

As can be seen in Alg.~\ref{Alg:tmwm}, the \texttt{T-MWM} algorithm basically performs two traversals. During the first one, the utility yielded by the various nodes and their children is computed. Then, during the second traversal, this knowledge is used for computing an \gls{mwm} of the network. 

The correctness of this procedure can be easily proved. Consider the sub-tree of $\mathcal{G}$ whose root is represented by the generic node $n_k$. Let also $\mathbf{F}(n_k)$ be the utility yielded by a \gls{mwm} of such sub-tree which activates a link originating from $n_k$, and $\mathbf{G}(n_k)$, conversely, the utility provided when the \gls{mwm} contains no links which originate from such node. Then, the  correctness of the first phase of the algorithm, namely the computation of the $\mathbf{F}$ and $\mathbf{G}$ vectors, follows directly from the following Lemmas.

\begin{algorithm}[tbp!]
\small
    \caption{Tree-Maximum Weighted Matching}
    \label{Alg:tmwm}
    \hspace*{\algorithmicindent} \textbf{Input:} A weighted \gls{st} $\mathcal{G}$ encoded by a list $\mathbf{E}$, which associates each node in $\mathcal{G}$ to its edges, and the corresponding weights list $\mathbf{W}$. \\ 
    \hspace*{\algorithmicindent} \textbf{Output:} An \gls{mwm} $\mathbf{E^*}$ of $\mathcal{G}$. 
    \begin{algorithmic}[1] 
        \Procedure{\texttt{T-MWM}}{$\mathbf{E}, \mathbf{W}$} 
            \State $\mathbf{F} \gets \mathbf{0}$; $\mathbf{G} \gets \mathbf{0}$ \Comment{Initialize the utility vectors to zero vectors}
            \State $\mathbf{E^*} \gets \{ \}$ \Comment{Initialize the set of active edges as empty}
            \For{each node $n_k \in \mathbf{E}$} \Comment{\parbox[t]{.43\linewidth}{Iterate over the various nodes, in ascending order w.r.t. to their depth in $\mathcal{G}$}}
            \State $maxUtil \gets 0$;  $\mathbf{maxEdge}(n_k) \gets \{ \}$
            \For{each edge $e_{k,j} \equiv (n_k, n_j)$} \Comment{Iterate over its edges}
                \State $\mathbf{G}(n_k) \gets \mathbf{G}(n_k) + \mathbf{F}(n_j)$
                \State $currUtil \gets \mathbf{W}(e_{k, j}) + \mathbf{G}(n_j) - \mathbf{F}(n_j)$
            \If{$currUtil > maxUtil$}
            	\State $maxUtil \gets currUtil$;  $\mathbf{maxEdge}(n_k) \gets e_{k, j}$ 
           		\State $\mathbf{F}(n_k) \gets \mathbf{G}(n_k) + maxUtil$ 
            \EndIf
            \EndFor\label{edgesFor}
            \EndFor\label{nodesFor}
			
			\For{each node $n_k \in \mathbf{E}$} \Comment{\parbox[t]{.43\linewidth}{Iterate over the various nodes, in ascending order wrt to their depth in $\mathcal{G}$}}			
			
				\If { $\mathbf{F}(n_k) \geq \mathbf{G}(n_k)$}
					\State $\mathbf{E^*} \gets \mathbf{E^*} \cup e_{k, \mathbf{maxEdge}(n_k)}$
					\State $\mathbf{F}(\mathbf{maxEdge}(n_k)) \gets -1$ \Comment{\parbox[t]{.3\linewidth}{Ensure child does not get activated multiple times}}
				\EndIf\label{activeIf}
			\EndFor\label{edgesFor2}            
            
            \State \textbf{return} $\mathbf{E^*}$
        \EndProcedure
    \end{algorithmic}
\end{algorithm}

\begin{lem}
\label{lemma_ch_par}
Let $n_k$ be an arbitrary internal node of $\mathcal{G}$ and $\{ n_j \}_k$ be the set of its children. Then, any \gls{mwm} of $\mathcal{G}$ must contain an edge which has as one of its vertices either $n_k$ or an element of  $\{ n_j \}_k$.
\end{lem}

\begin{IEEEproof}
Suppose there exists an \gls{mwm} $\mathbf{E^*}$ of $\mathcal{G}$ which does not contain any such edge. Then the set $\hat{\mathbf{E}}^* \equiv \mathbf{E^*} \cup \{ e_{n_k \to n_m} \}$, where $ e_{n_k \to n_m} $ is the edge from $n_j$ to its (arbitrary) child $n_m$ is still a feasible activation set, since no edge in $\mathbf{E^*}$ shares such vertices. Furthermore, since the weights are positive we have that $f_u (\hat{\mathbf{E}}^*) \equiv f_u (\mathbf{E^*}) + \mathbf{W} (e_{n_k \to n_m}) > f_u (\mathbf{E^*}) $, which is clearly a contradiction.
\end{IEEEproof}

\begin{lem}
\label{lemma_utils}
For any internal node $n_k$:
\[ 
\begin{cases}
\mathbf{G}(n_k) = \sum\limits_{ \{ n_j \}_k} \mathbf{F}(n_j) \\
                  
\mathbf{F}(n_k) = \sum\limits_{ \{ n_j \}_k } \mathbf{F}(n_j) + \underset{\{ n_j \}_k }{\max} \{ \mathbf{W} ( e_{n_k \to n_j} ) + \mathbf{G}(n_k) - \mathbf{F}(n_k)\}
\end{cases}
\]
where the set $\{ n_j \}_k$ comprises all the children of $n_k$. 
Conversely, for leaf nodes $\mathbf{F}(n_l) \equiv \mathbf{G}(n_l) \equiv 0 $.
\end{lem}

\begin{IEEEproof}
This lemma can be proved by induction over the height $h_k$ of the sub-tree corresponding to node $n_k$. The base case is $h_k = 0$, i.e., when $n_k$ is a leaf node; in this case, trivially, both $\mathbf{F}(n_k)$ and $\mathbf{G}(n_k)$ are zero since no links exhibit $n_k$ as parent node and the sub-tree of $\mathcal{G}$ which originates in $n_k$ consists of $n_k$ only, respectively.

Consider then the node $n_k$ having a sub-tree of height $h_k > 0$. From Lemma~\ref{lemma_ch_par} we know that either $n_k$ or (at least) one of its children must be included in any \gls{mwm}. If on the one hand we do not activate any edge which originates from $n_k$, then no constraints on the children's activation are introduced. Therefore, in this case the maximum utility is obtained when \textit{all} the children are active, hence $\mathbf{G}(n_k) = \sum_{ \{ n_j \}_k} \mathbf{F}(n_j)$. On the other hand, if we activate an edge from $n_k$ to one of its children $n_m$ then no additional edges which originate from the latter can be added to $\mathbf{E^*}$. It follows that the utility obtained in this instance reads:
\[ \sum_{ \{ n_j \neq n_m\}_k } \mathbf{F}(n_j) +  \mathbf{W} ( e_{n_k \to n_m} ) + \mathbf{G}(n_m) \]
and can be rewritten as:
\[ \sum_{ \{ n_j \}_k } \mathbf{F}(n_j) + \mathbf{W} ( e_{n_k \to n_m} ) + \mathbf{G}(n_m) - \mathbf{F}(n_m) \]
Such utility is maximized when $n_m$ is chosen as $ \underset{ \{ n_j \}_k }{\mathrm{argmax}} \, \{ \mathbf{W} ( e_{n_k \to n_j} ) + \mathbf{G}(n_j) - \mathbf{F}(n_j) \}$, yielding:
\[ \mathbf{F}(n_k) = \sum\limits_{ \{ n_j \}_k } \mathbf{F}(n_j) + \underset{\{ n_j \}_k }{\max} \, \{ \mathbf{W} ( e_{n_k \to n_j} ) + \mathbf{G}(n_k) - \mathbf{F}(n_k)\} \qedhere \] 
\end{IEEEproof}
Finally, the validity of the last phase of \texttt{T-MWM} follows from Lemma~\ref{lemma_active_edges}.
\begin{lem}
\label{lemma_active_edges}
Given an \gls{st} $\mathcal{G}$, let $\mathcal{G}_k$ be its sub-tree of root $n_k$. Then, an \gls{mwm} of $\mathcal{G}$ can be computed by performing, in a recursive fashion and starting from the root, the following procedure:
\begin{enumerate}
\item If $ \, \mathbf{F}(n_k) \geq \mathbf{G}(n_k) $, add to $\mathbf{E^*}$ the edge from $n_k$ to $n_m$, where the latter is defined as $n_m \equiv \underset{ \{ n_j \}_k }{\mathrm{argmax}} \, \{ \mathbf{W} ( e_{n_k \to n_j} ) + \mathbf{G}(n_j) - \mathbf{F}(n_j) \}$. Then, repeat recursively on all the sub-trees corresponding to $n_k$'s children $\{ n_j\}_k \, \mathrm{s.t.} \, n_j \neq n_m$ and on the children of $n_m$ itself.
\item If $ \, \mathbf{F}(n_k) < \mathbf{G}(n_k) $, repeat recursively on  all the sub-trees which exhibit the children of $n_k$ as their root.
\end{enumerate}
\end{lem}

\begin{IEEEproof}
This Lemma follows directly from the definitions of $\mathbf{F}$ and $\mathbf{G}$ and the previous Lemmas. Specifically, the above procedure always yields a feasible activation, i.e., a matching of $\mathcal{G}$.  In fact, in either options we never recurse on a node which has already been activated, hence no pair of edges $\in \mathbf{E}$ can share any vertices. Furthermore, due to the properties of $\mathbf{F}$ and $\mathbf{G}$ and their validity for each sub-tree in $\mathcal{G}$, the edges of $\mathbf{E^*}$ comprise a \textit{maximal} matching, i.e., they yield the maximum possible utility among all the feasible schedules. 
\end{IEEEproof}

Regarding the computational complexity of the proposed algorithm, it can be observed that during the first phase the main loop effectively scans each edge of $\mathcal{G}$, hence exhibiting a complexity \bigO{\vert E \vert}. Moreover,
the second phase of \texttt{T-MWM} has complexity \bigO{\vert V \vert}, since it loops through all the network nodes.
Therefore, we can conclude that the overall asymptotic complexity of the algorithm is \bigO{\vert V \vert + \vert E \vert}, or, equivalently, \bigO{\vert V \vert} since in an \gls{st} the number of edges equals $\vert V \vert - 1$.

\subsection{Semi-centralized resource partitioning scheme}
\label{Sec:Cent-scheme}
Based on the system model introduced in Sec.~\ref{Sec:Sys-model}, and the \texttt{T-MWM} algorithm, we present a generic optimization framework which partially centralizes the backhaul/access resource partitioning process, in compliance with the guidelines of~\cite{3gpp_38_874}. 
The goal of this framework is to aid the distributed schedulers, adapting the number of \gls{ofdm} symbols allocated to the backhaul and access interfaces to the phenomena which exhibit a sufficiently slow evolution over time, i.e., large scale fading and local congestion. 
This optimization is undertaken with respect to a generic additive utility function $f_u$. An \gls{iab} network of arbitrary size is considered, composed of a single \gls{iab}-donor, multiple \gls{iab}-nodes and a (possibly time-varying) number of \glspl{ue} which connect to both types of \glspl{gnb}. 
Furthermore, let the topology of the \gls{iab} network be pre-computed, for instance by using the policies of~\cite{polese2018iab}, and assume that a central controller is installed on the \gls{iab}-donor. 

The proposed framework can be subdivided into the following phases, which are periodically repeated every $T_{alloc}$ subframes:
\begin{enumerate}
\item \label{Enum_frameword:item_one} \textbf{Initial setup}. This step consists in the computation of the simplified \gls{iab} network graph $\mathcal{G} \equiv \{ \mathcal{V}, \mathcal{E} \}$. Specifically, $\mathcal{V}$ is composed of the donor, the various \gls{iab}-nodes and, possibly, of additional nodes which represent the set of \glspl{ue} that are connected to a given \gls{gnb}. Accordingly, $\mathcal{E}$ contains the active cell associations of the aforementioned nodes. This process, depicted in Fig.~\ref{Fig:Phase0}, must be repeated every time the \gls{iab} topology changes, i.e., whenever a new \gls{ue} performs its \gls{ia} procedure or an \gls{iab}-node connects to a different parent due to a \gls{rlf}.

\begin{figure}[t]
\centering
\includegraphics[width=0.65\linewidth]{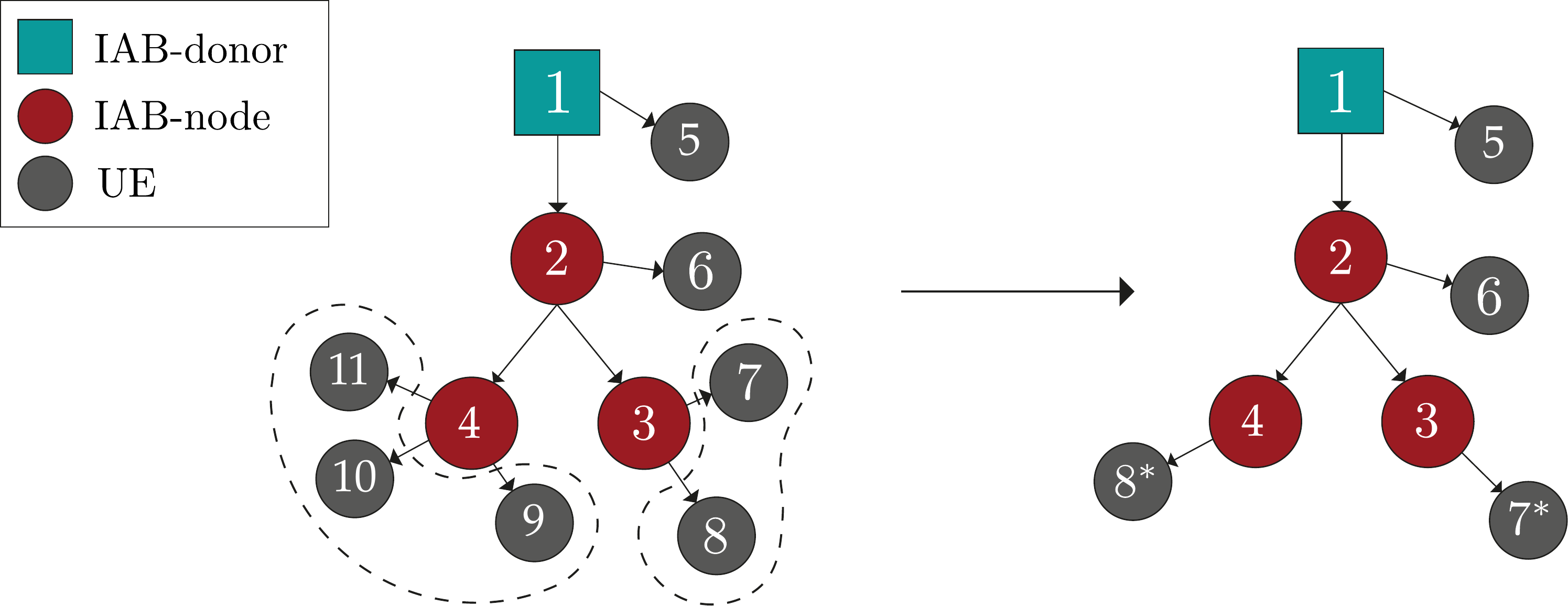}
\caption{Creation of the \gls{iab} network graph. The original topology, exhibiting the actual cell attachments, is depicted on the left. Conversely, the reduced topology is shown on the right.}
\label{Fig:Phase0}
\vspace{-.6cm}
\end{figure}

\item \label{Enum_frameword:item_two} \textbf{Information collection}. During this phase, the various \gls{iab}-nodes send to the central controller a pre-established set of information for each of their children in $\mathcal{G}$. For instance, this feedback may consist in their congestion status and/or  information regarding their channel quality. To such end, the implementation of this paper uses modified versions of pre-existing \gls{nr} Release 16 \glspl{ce}, as strongly recommended in the \gls{iab} SI~\cite{3gpp_38_874}. However, the scheme does not actually impose any limitations in such regard.
\item \label{Enum_frameword:item_three} \textbf{Centralized scheduling indication}. Upon reception of the feedback information, the central controller calculates the set of weights $\mathcal{W}$ accordingly. Then, an \gls{mwm} of $\mathcal{G}$ is computed using the \texttt{T-MWM} algorithm and producing as output the activation set $\mathbf{E^*}$, which maximizes the overall utility of the system with respect to the chosen utility function. Subsequently, $\mathbf{E^*}$ is used in order to create a set of \textit{favored} downstream nodes, i.e., of children which will be served with the highest priority by their parent, as depicted in Fig.~\ref{Fig:Phase2}. Finally, these scheduling indications are forwarded to the various \gls{iab}-nodes which act as parents in the edges of $\mathbf{E^*}$.

\begin{figure}[t]
\centering
\includegraphics[width=0.65\linewidth]{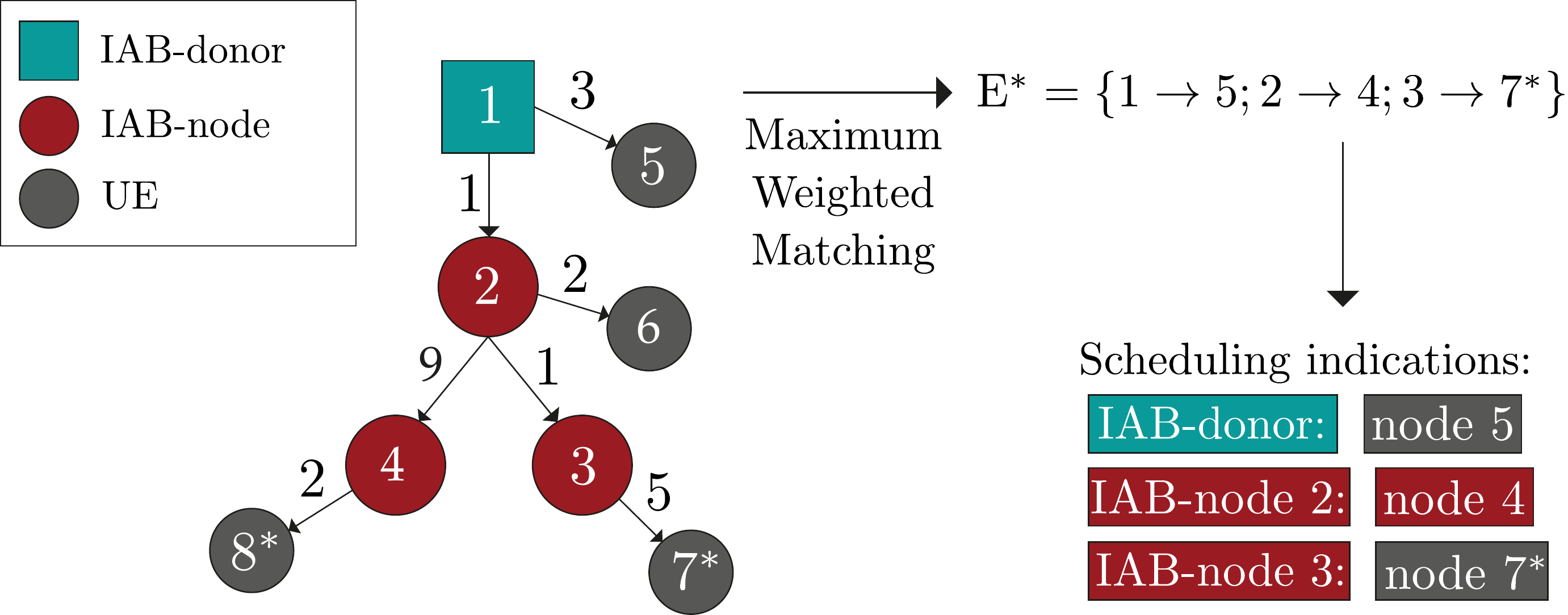}
\caption{Computation of the \gls{mwm} and of the corresponding scheduling indications.}
\label{Fig:Phase2}
\vspace{-.6cm}
\end{figure}

\item \textbf{Distributed scheduling allocation}. During this phase, the various \gls{iab}-nodes make use of the indications received by the central controller, if available, in order to perform the actual scheduling (which is, therefore, predominantly distributed). Specifically, the favored nodes are served with the highest priority, while the remaining downstream nodes are scheduled if and only if the resource allocation of the former does not exhaust the available \gls{ofdm} symbols.
\end{enumerate}
It is important to note that since $\mathcal{G}$ contains only the \gls{iab}-nodes, the donor and at most one ``representative" \gls{ue} per \gls{gnb}, the proposed scheme effectively performs only the backhaul/access resource partitioning in a centralized manner. On the other hand, the actual \gls{mac}-level scheduling is still undertaken in a distributed fashion, albeit leveraging the indications produced by the central controller. The major advantages which this two-tier design exhibits, compared to a completely centralized solution, are the presence of a relatively light signaling overhead and the ability to promptly react to fast channel variations, for instance caused by small scale fading.  

\subsection{Implementation of centralized allocation schemes in mmWave \gls{iab} networks}
\label{Sec:ns3-impl}

The remainder of this section discusses how the proposed scheme can be implemented in \gls{iab} deployments, with references to how the \gls{3gpp} specifications can support it.

Basically, the resource allocation framework requires (i) a central controller, which is installed on the \gls{iab}-donor, or could be deployed in a \gls{ric} following the O-RAN architecture~\cite{bonati2020open}; and (ii) a scheduler which exchanges resource coordination information with the former and computes the weights for the resource allocation.
In particular, referring to the aforementioned phases of the proposed scheme, the following implementation considerations can be made.


\subsubsection{Initial setup}
The setup of the various centralized mechanisms is subdivided into two sub-phases: an initial configuration, where the relevant entities are initialized, and a periodic update of the topology information.
The former takes place when the \gls{iab}-nodes are first connected to the network. During this phase, the controller acquires preliminary topology information, by leveraging the configuration messages which are already exchanged during the usual Rel.16 \gls{ia} procedure, generating a map which associates the depth in the \gls{iab} network to a pair of child-parent global identifiers (which from now on will be referred to as ``IDs"). Since this phase takes place when no \gls{ue} has performed its \gls{ia} procedure yet, the exchanged topology information concerns the donor and \gls{iab}-nodes only.

Moreover, the central controller is in charge of periodically updating the topology information. In order to minimize the signaling overhead, this process does not require any additional control information: in fact, the status information which is already collected in a periodic manner can be leveraged in such regard. Specifically, the periodic info received from the various \gls{iab}-nodes, which will carry a list of ID-value pairs, is analyzed. The child-parent associations are then compared with the ones known by the controller, updating the latter whenever the two topology information happens to differ.

\subsubsection{Information collection}
The generation of the feedback information is performed in a distributed manner by both the \gls{iab}-nodes and the \gls{iab}-donor. To such end, the current implementation features the forwarding of information on the channel quality and buffer status, in the form of \glspl{cqi} and \glspl{bsr} respectively. This choice is driven by both the will of maximizing the re-utilization of the \gls{nr} Rel.16 specifications and the goal of making use of \gls{mac}-level \glspl{ce} only, hence avoiding the introduction of any constraint regarding the supported \gls{iab}-relaying architecture.



In particular, the \gls{cqi} and \gls{bsr} information is generated by analyzing the corresponding \glspl{ce} which are received by the host \gls{gnb}, and checking whether the source \gls{rnti} belongs to an \gls{iab}-node or to a \gls{ue}. In the first case, the corresponding ID
is retrieved and an entry carrying such identifier along with its \gls{cqi}/\gls{bsr} value is generated. 
The feedback information concerning the \glspl{ue}, instead, is averaged in the case of the \glspl{cqi} and added up for the \glspl{bsr}, to obtain a single value for each \glspl{gnb}.
It can be noted that both \glspl{cqi} and \glspl{bsr} are available to the scheduler, since 
the UL buffer statuses are already periodically reported by the downstream nodes via their \gls{bsr}s and
the DL statuses can be easily retrieved by the former, since the \gls{rlc} buffers reside on the same node as the scheduler itself, i.e., the \gls{gnb}.

Referring to the \gls{3gpp} specifications of~\cite{3gpp_38_874, 3gpp_38_321, 3gpp_38_331}, the buffers occupancy can then be forwarded to the \gls{iab}-donor by introducing a periodic-only \gls{bsr} whose period is controlled by an ad hoc \gls{rrc} timer. Similarly, the channel qualities can be reported by the various \gls{iab}-nodes via additional periodic \glspl{cqi} which would carry only the \gls{cqi} index, hence neglecting the \gls{ri} and \gls{pmi}, since this information would generate unnecessary signaling overhead. These \glspl{ce} would preferably leverage pre-existing \gls{nr} measurements: the main novelty would be the introduction of their periodic reporting to the \gls{iab}-donor. To such end, the \gls{5g} \gls{cqi} and \gls{bsr} data-structures require an additional field which carries the ID, if the chosen \gls{iab}-relaying architecture does not feature an Adaptation Layer~\cite{3gpp_38_874}. Conversely, relaying solutions which support the latter can reuse the \gls{nr} \glspl{ce} and let such layer introduce an additional header.

\subsubsection{Centralized scheduling indication}
\label{Sec:cent_indic}
Periodically, the controller located at the donor makes use of the feedback received by the \gls{iab}-nodes to first compute the weights of the various network links and then generate the centralized scheduling indications. We propose the following policies to compute the weights for the \gls{mwm} problem:

\begin{enumerate}
\item \textbf{\gls{msr}}. 
This policy maximizes the overall \gls{phy}-layer throughput, i.e., the utility function is 
\[ f_u^{\mathrm{MSR}} \equiv \sum_{e_{i \to k} \, \in \, \mathbf{E^*}} c_{i, \, k}, \]
 and the weight assigned to the edge from node $i$ to node $k$ reads $w_{i, \, k} \equiv c_{i, \, k}$, where $c_{i, \, k}$ is the capacity of the link $e_{i \to k}$.
\item \textbf{\gls{ba}}.
This resource partitioning strategy aims at avoiding congestion. Therefore, the system utility is:
\[ f_u^{\mathrm{BA}} \equiv \sum_{e_{i \to k} \, \in \, \mathbf{E^*}} q_{i, \, k}, \]
where the weight $w_{i, \, k}$ reads $q_{i, \, k}$, namely, the amount of buffered data which would reach its next hop in the \gls{iab} network by crossing the link $e_{i \to k}$.
\item \textbf{\gls{mrba}}. 
This represents the most balanced option among the three, since it exploits favorable channel conditions while also preventing network congestion and favoring network fairness. The weight assigned to link $e_{i \to k}$ is:
\[ w_{i, \, k} \equiv c_{i, \, k} + \eta \cdot q_{i, \, k} \cdot \left( \frac{\mu}{\mu_{thr}} \right)^{k}, \]
where $\eta$, $\, \mu_{thr}$ and $k$ are arbitrary parameters and $\mu$ represents the number of subframes which have elapsed since the last time edge $e_{i \to k}$ has been marked as favored.
\end{enumerate}

Once the weights are computed, the controller obtains an \gls{mwm} of the network via an implementation of the aforementioned \texttt{T-MWM}. This function outputs the activation set $\mathbf{E^*}$, i.e., a map associating the ID of the parent \gls{gnb} to the one of its favored downstream node. Notably, this set does not necessarily contain scheduling indications for \textit{each} \gls{iab}-node in the network: an entry corresponding to a given \gls{gnb} is present if and only if such node is indeed active in the \gls{mwm}.
First, this map is used by the controller in order to keep track of which link has not been favored and for how long; this information may then be used to introduce a weight prediction mechanism, improving the robustness of the scheme with respect to the information collection period.
Finally, these scheduling indications are forwarded to the corresponding \gls{iab}-nodes.

\subsubsection{Distributed scheduling allocation}
The last phase of the resource allocation procedure consists in the distributed \gls{mac}-level scheduling. Before assigning the available resources, the various schedulers check whether any indication has been received from the controller. Based on this condition, the buffer occupancy information is then split into two groups.
The first contains the \gls{bsr}s related to the favored \gls{rnti} (if any), with the caveat that if the latter indicates the cumulative access link, then this set contains the \gls{bsr}s of all the \gls{ue}s attached to the host \gls{gnb}, while the other comprises the remaining control information.
The resource allocation process is then undertaken twice: first considering the set of favored \gls{bsr}s only, then the remainder of these \glspl{ce}. 

Thanks to this design, the favored link(s) is (are) scheduled with the highest priority, while the rest of the network only gets the remaining resources. In such a way, the information received by the controller is actually used as an \textbf{indication} and not as the eventual \textbf{resource allocation}. For instance, the \glspl{gnb} are free to override these indications whenever the buffer of the favored child is actually empty, due to discrepancies between its actual status and the related information available to the controller. In such a way, the unavoidable delay between the information collection and the reception of the scheduling does not lead to any resource underutilization. Moreover, this is achieved with minimal changes to the state of the art schedulers, making the proposed scheme relatively easy to implement and deploy in real-world networks.
 
\section{Performance evaluation}
\label{Sec:perf_eval}

We implemented the proposed resource allocation scheme in the popular open source simulator ns-3, exploiting the \gls{mmwave} module~\cite{mezzavilla2018end} and its \gls{iab} extension~\cite{polese2018end}, to characterize the system-level performance of the proposed solution with realistic protocol stacks, scenarios, and user applications.

The ns-3 \gls{mmwave} module is based on~\cite{baldo2011open} and introduces \gls{mmwave} channel models, including the \gls{3gpp} channel model for 5G evaluations~\cite{zugno2020implementation}, and highly customizable \gls{phy} and \gls{mac} layer implementation, with an NR-like flexible \gls{ofdm} numerology and frame structure.
Additionally, the \gls{iab} module~\cite{polese2018end} models wireless relaying functionalities which mimic the specifications presented in~\cite{3gpp_38_874}. Specifically, this module supports both single and multi-hop deployment scenarios, auto-configuration (within the network) of the \gls{iab}-nodes and a detailed \gls{3gpp} protocol stack, allowing wireless researchers to perform system-level analyses of \gls{iab} systems in ns-3.

It is of particular relevance to understand how the scheduling operations are implemented in the \gls{iab} module, since they offer not only the baseline for the proposed scheme, but also valid guidelines for real-world deployments. 
The current ns-3 \gls{iab} schedulers exhibit a \gls{tdma}-based multiplexing between the access and backhaul interfaces. Moreover, scheduling decisions are undertaken in a distributed manner across the \gls{iab} network, i.e., each \gls{gnb} allocates the resources which its access interface offers (to both \gls{ue}s and  \gls{iab}-nodes) independently of the other \gls{gnb}s in the network. 
In fact, in an \gls{iab} network these scheduling decisions are \textit{almost} independent of one another: if a parent node schedules the backhaul interface of a downstream node, clearly the latter will be constrained in its own scheduling decisions, as it will not be allowed to allocate the time resources which have already been scheduled for backhaul transmissions by its parent. Therefore, in a tree-based, multi-hop wireless network the various \glspl{gnb} need to know in advance the scheduling decisions performed by their upstream nodes: to solve this problem, the authors of the \gls{iab} module for ns-3 introduced a ``\textit{look-ahead backhaul-aware scheduling mechanism}"~\cite{polese2018end}. 
Such mechanism features an exchange of \gls{dci} between the access and backhaul interfaces: in such a way, any time resources already scheduled by the parent for backhaul communications can be marked as such by the corresponding downstream node, preventing any overlap with other transmissions.
Furthermore, the \textit{look-ahead} mechanism requires the schedulers of the various \gls{gnb}s to commit to their resource allocation for a given time $T$ at a time $T - k$, where $k - 1$ is the maximum distance (in terms of wireless hops) of any node from the donor. In such a way, the \gls{dci}s will have time to propagate across the \gls{iab} network and reach the farthest node at time $T - 1$, thus allowing its scheduler to perform the resource allocation process at least one radio subframe in advance.



\subsection{Simulation scenario and parameters}
\begin{figure}[tbp]
    \centering
    \subfloat[\label{Fig:Sim_scen}]{
	    \setlength\fwidth{0.38\columnwidth}
	    \setlength\fheight{0.25\columnwidth}
       \raisebox{-.5\height}{\input{Figures/Sim_scenario.tex}}
    }
    \subfloat[\label{Tab:Sim_params}]{
    	\footnotesize
		\begin{tabular}{cc}
		\multicolumn{2}{c}{\textsc{Simulation parameters}}\\
		\hline
		\textsc{Parameter} & \textsc{Value} \\
		\hline
		Number of runs $N_{runs}$ & 25  \\
		\rowcolor{gray!15} Simulation time $T_{sim}$ & 3 s \\
		\gls{mwm} period $T_{alloc}$ & $\{ 1, 2, 4\}$ subframes \\
		\rowcolor{gray!15} Layer 4 protocol & $\{ $UDP, TCP$ \}$ \\
		UDP packet size $s_{UDP}$ & $\{50, 100, 200, 500 \}$ B \\
		\rowcolor{gray!15} Weight policy $f_u$ & $\{$\gls{msr}, \gls{ba}, \gls{mrba}$\}$ \\
		\hline
		\end{tabular}
    }
    \label{Float:Sim_scen_and_params}
    \caption{On the left, a realization of the simulation scenario is depicted; the dotted lines represent the cell-attachments of the \gls{iab}-nodes. On the right, a brief list of simulation parameters is provided.}
    \vspace{-.6cm}
\end{figure}
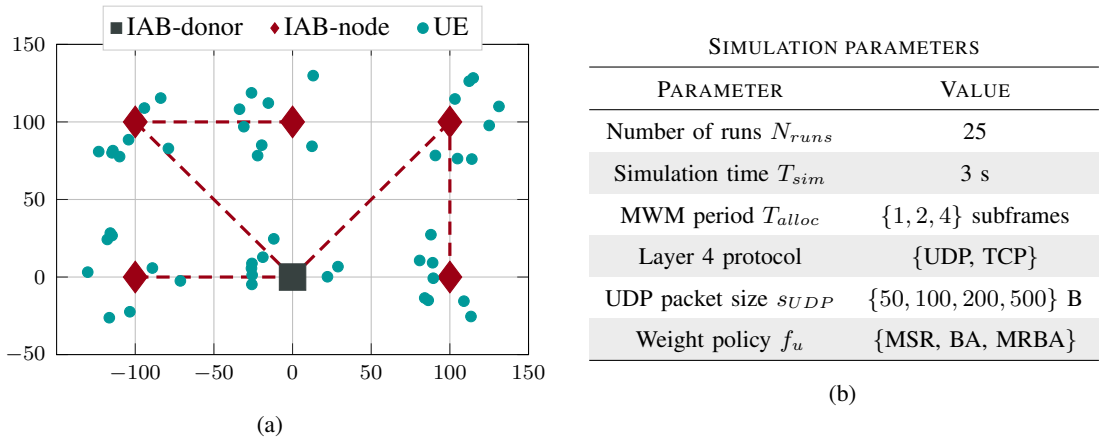

The purpose of these simulations is to understand the performance of the proposed resource partitioning framework in the context of its target deployment, i.e., a multi-hop \gls{iab} network. As a consequence, the reference scenario consists of a dense urban deployment with a single \gls{iab}-donor and multiple \gls{iab}-nodes, as depicted in Fig.~\ref{Fig:Sim_scen}. In particular, the various \glspl{gnb} are distributed along an urban grid where the donor is located at the origin while the \gls{iab}-nodes are deployed along the street intersections, with a minimum inter-site distance of 100 m. The \gls{iab}-nodes attachments are computed using the so-called \textit{HQF} policy presented in~\cite{polese2018iab}; however, this choice does not introduce any loss of generality since such parameter is fixed for all the runs. 
A given number of 
\glspl{ue} are deployed within the surroundings of these base stations, with an initial
position which is randomly sampled from circles of radius $\rho$ and whose centers are the various \glspl{gnb}. 

Both the \gls{iab}-donor and the \gls{iab}-nodes are equipped with a phased array featuring 64 antenna elements, and transmit with a power of 33 dBm; conversely \glspl{ue} are equipped with 16 antenna elements and their transmission power is restricted to 23 dBm. Notably, the presence of additional antenna elements at the \glspl{gnb} is a key (but reasonable) assumption, as it allows base stations to achieve a high beamforming gain. In turn, it is possible to achieve a high capacity, which is fundamental to avoid performance bottlenecks, given the absence of a fiber backhaul.
The \glspl{ue} download data which originates from sources that are installed on a remote host; both the \gls{udp} and the \gls{tcp} are used. For the \gls{udp} simulations, the rate of the sources is varied from 4 to 40 Mbps to introduce different degrees of saturation in the network. Therefore, in these simulations only DL traffic is considered.
Finally, the performance of the proposed policies is hereby compared with the baseline of~\cite{polese2018end}, indicated as ``Dist.'' by examining end-to-end throughput, latency, and a network congestion metric.

\subsection{Throughput}

The first metric which is inspected in this analysis is the end-to-end throughput at the application layer. As a consequence, only the packets which are correctly received at the uppermost layer of the destination node in the network are taken into account. In particular, for each \gls{ue} and each simulation run, the long-term average throughput is computed as follows:
\[ S^{\mathrm{APP}}_{k, n} \equiv \frac{B(T_{sim}, k, n)}{T_{sim}} \]
where $B(t, j)$ is the cumulative number of bits received up to time $t$ by the $k$-th \gls{ue}, during the $n$-th simulation run. Then, the distribution of 
$\mathbf{S}^{\mathrm{APP}}$, namely, the vector containing the collection of the $S^{\mathrm{APP}}_{k, n}$ values across the different runs and \glspl{ue}, is analyzed.

\begin{figure}[tbp]
	\centering
  	\subfloat[$s_{UDP}$ = 100 B, i.e., \gls{udp} rate of 8 Mbps.\label{Fig:Thr_ECDF_100}]{
  	\setlength\fwidth{0.4\columnwidth}
    \setlength\fheight{0.2\columnwidth}
    \input{Figures/Sim_results/E2E_throughput_ECDF_packet_size_100.tex}
  	}
  	\hfill
  	\subfloat[$s_{UDP}$ = 500 B, i.e., \gls{udp} rate of 40 Mbps.\label{Fig:Thr_ECDF_500}]{
    \setlength\fwidth{0.4\columnwidth}
    \setlength\fheight{0.2\columnwidth}
    \input{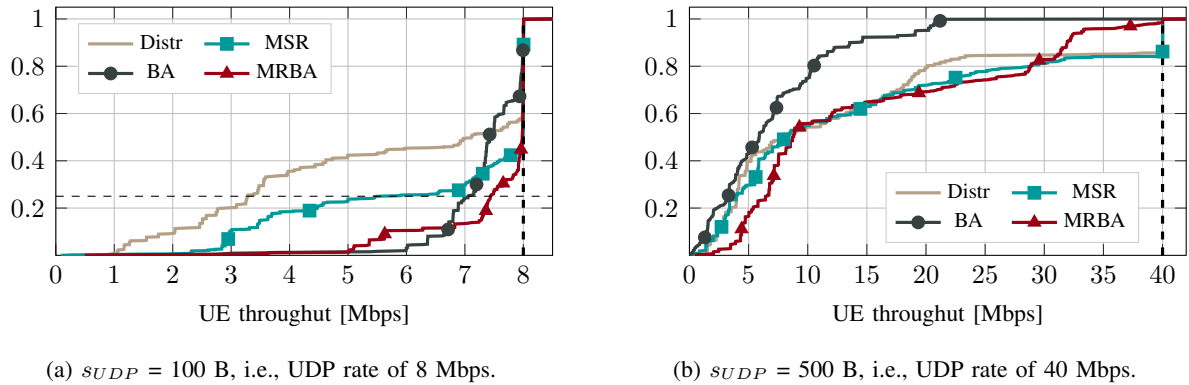}
  	}
    \caption{Per-\gls{ue} end-to-end throughput \glspl{ecdf}. The dashed line represents the rate of the \gls{udp} sources.}
    \label{Fig:thr_ECDF}   
    \vspace{-.6cm} 
\end{figure}

Figs.~\ref{Fig:Thr_ECDF_100} and~\ref{Fig:Thr_ECDF_500} report the \gls{ecdf} of $\mathbf{S}^{\mathrm{APP}}$, for a \gls{udp} packet size of 100 and 500 bytes, respectively, and the policies introduced in Sec.~\ref{Sec:ns3-impl}. In the former, we can notice that the introduction of the centralized framework increases by up to 15\% the percentage of \glspl{ue} whose throughput matches the rate of the \gls{udp} sources.
Moreover, by focusing on the leftmost portion of Fig.~\ref{Fig:Thr_ECDF_100} we can observe another interesting result, concerning the throughput experienced by the \glspl{ue} which do not fulfill their \gls{qos} requirements. In fact, with respect to the first quartile the distributed scheduler achieves the worse performance, with 25\% of the \glspl{ue} obtaining a throughput smaller than 3.3~Mbps. The centralized framework significantly improves these results, even though the extent of such improvements varies quite dramatically across the different policies.
Compared with the distributed case, the \gls{msr} policy achieves a higher throughput with respect to all the percentiles, albeit exhibiting the same high variance of the former. Instead, the \gls{ba} and \gls{mrba} policies have a dramatic impact on the system performance, introducing a 5-fold increase of the worst case throughput coupled with a significantly lower variance.

These results can be explained as follows: since a \gls{udp} packet size of 100 bytes does not saturate the capacity of the access links, the main performance bottleneck of this configuration is represented by the buffering of the aggregated traffic on the intermediate backhaul links. Therefore, the \gls{msr} policy provides only minimal improvements compared to the performance of the distributed scheduler, since it simply favors the links which exhibit a higher \gls{sinr}. Conversely, the prioritization of the most congested links which is introduced by the other two strategies successfully tackles the former problem. 
In particular, the \gls{ba} policy exhibits the highest worst case throughput, albeit at the cost of satisfying the \gls{qos} requirements of approximately 20\% of the \glspl{ue}. On the other hand, the bias towards high \gls{sinr} channels introduced by the \gls{mrba} strategy has the opposite effect, improving mostly the higher percentiles but also outperforming \gls{msr} and the baseline in the lower percentiles. 

By increasing the \gls{udp} packet size to 500 bytes, the network becomes noticeably saturated, as depicted by Fig.~\ref{Fig:Thr_ECDF_500}; in fact, in this instance only a minority of the \glspl{ue} achieves a throughput which is comparable to the source rate. With this configuration, the \gls{ba} strategy achieves the worst performance, providing a significantly lower throughput across all the percentiles. On the other hand, the differences among the behavior of the remaining strategies are less evident. In particular, the \gls{msr} policy exhibits only a slight improvement over the distributed solution, albeit noticeable across the whole \gls{ecdf}. The \gls{mrba}, conversely, introduces performance benefits which mostly affect the bottom percentiles only. However, with this strategy only a limited portion of the \glspl{ue} achieves the target throughput of 40 Mbps. 
As a consequence, we can conclude that with the configuration depicted in Fig.~\ref{Fig:Thr_ECDF_500} the network is approaching the capacity of the \gls{mmwave} channels. Therefore, buffering phenomena are likely occurring at each intermediate \gls{iab}-node. Moreover, we can say that in a saturated network the congestion is so severe that prioritizing the bottleneck links is not enough: we also need to take into account the channel conditions and prioritize the links which not only are congested, but also have the ``biggest chance" of getting rid of the buffered data due to the temporary better channel quality.

\begin{figure}[tbp]
  \subfloat[First quartile.\label{Fig:Throughput_first_quartile}]{
    \setlength\fwidth{0.37\columnwidth}
    \setlength\fheight{0.2\columnwidth}
    \input{Figures/Sim_results/E2E_first_quartile_throughput.tex}
	}
  \hfill
\subfloat[Third quartile.\label{Fig:Throughput_third_quartile}]{
    \setlength\fwidth{0.37\columnwidth}
    \setlength\fheight{0.2\columnwidth}
    \input{Figures/Sim_results/E2E_third_quartile_throughput.tex}
	}
  \vspace*{-3mm}
   \caption{End-to-end throughput quartiles, for $s_{UDP}$ $\in$ $\{50, 100, 200, 500 \}$ B.}
  \label{Fig:Throughput_quartiles}
  \vspace{-.6cm} 
\end{figure}
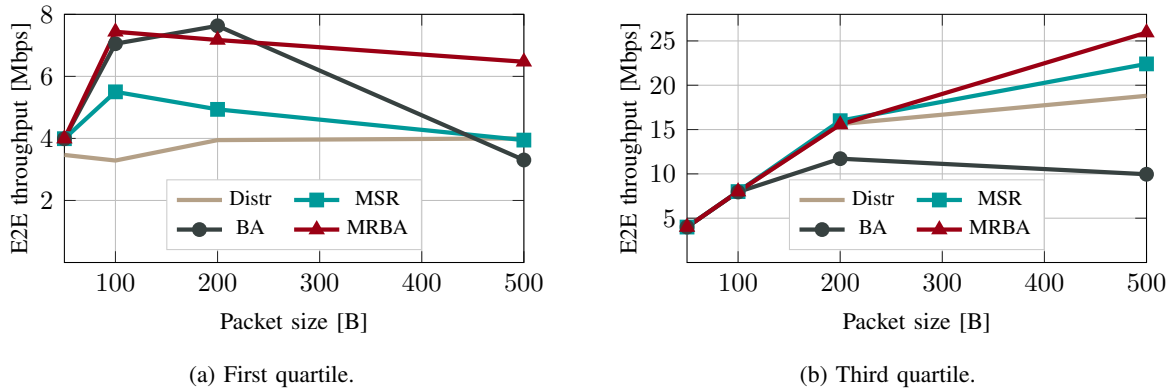

Finally, Fig.~\ref{Fig:Throughput_quartiles} presents the first and third quartiles of $\mathbf{S}^{\mathrm{APP}}$ as a function of the \gls{udp} packet size $s_{UDP}$. It can be noted that, with respect to the first quartile, the \gls{mrba} outperforms all the other policies by delivering a throughput which is up to 90\% higher than the one obtained by the distributed scheduler. In fact, Fig.~\ref{Fig:Throughput_third_quartile} shows how \gls{mrba} achieves also the best third quartile, albeit the improvement over the distributed solution is less dramatic. 
Furthermore, we can observe how the positive impact of the \gls{ba} strategy is inversely proportional to the saturation in the network. We can then conclude that the bias it introduces loses its effectiveness as the buffering phenomena start to affect the majority of the \gls{iab}-nodes.

\subsection{Latency}

Just like the aforementioned metric, the latency is measured end-to-end at the application layer. Thanks to this choice, the resulting delay accurately represents the system-level performance, as it includes the latency which is introduced at each hop in the \gls{iab} network.

\begin{figure}[tbp]
	\centering
    \subfloat[ECDF, for $s_{UDP}$ = 100 B.\label{Fig:Del_ECDF_100}]{
    \setlength\fwidth{0.48\columnwidth}
    \setlength\fheight{0.2\columnwidth}
    \input{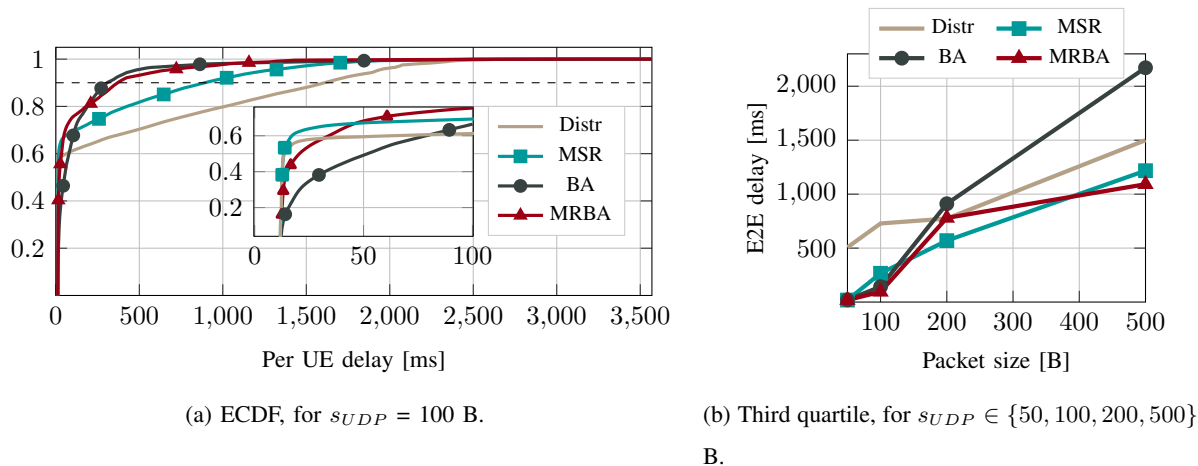}
    }
    \hfill
  \subfloat[Third quartile, for $s_{UDP}$ $\in$ $\{50, 100, 200, 500 \}$ B.\label{Fig:Delay_third_quartile}]{
    \setlength\fwidth{0.24\columnwidth}
    \setlength\fheight{0.2\columnwidth}
     \input{Figures/Sim_results/E2E_third_quartile_delay.tex}
    }
  \vspace*{-3mm}
   \caption{Per-\gls{ue} end-to-end delay statistics.}
  \label{Fig:Delay_stats}
    \vspace{-.6cm} 
\end{figure}

In particular, for each packet correctly received at the uppermost layer of its final destination, the following quantity is traced:
\[ D^{\mathrm{APP}}_{i} \equiv \sum_{l_k \, \in \, \mathcal{E}_{i}} D^{l_k}_i \] 
where $\mathcal{E}_{i}$  comprises the links in the \gls{iab} network that are crossed by the $i$-th packet, while the term $D^{l_k}_i$ indicates its point-to-point latency over the path link $l_i$. 
Finally, these values are collected for each of the various runs into the vector $\mathbf{D}^{\mathrm{APP}}$ and its statistical properties are inspected.

Fig.~\ref{Fig:Del_ECDF_100} shows the empirical \gls{ecdf} of $\mathbf{D}^{\mathrm{APP}}$ for a packet size of 100 bytes. It can be noticed that, in this case, the 90th percentile achieved by the \gls{ba} and the \gls{mrba} policies are approximately 20 \% smaller than the one obtained by the distributed scheduler. Moreover, these strategies manage to dramatically reduce the number of packets received with extremely high delay, i.e., in the order of seconds, showing the dramatic impact of buffering in the baseline configuration. Conversely, the \gls{msr} policy provides the best performance with respect to the best case delay only, although it still outperforms quite  significantly the distributed strategy. 

These trends are exacerbated by Fig.~\ref{Fig:Delay_third_quartile}, which shows the third quartile of $\mathbf{D}^{\mathrm{APP}}$ as a function of the \gls{udp} packet size $s_{UDP}$. In fact, we can notice that the effectiveness of the \gls{ba} policy is inversely proportional to the network saturation; the opposite holds true with respect to the \gls{msr} strategy. It follows that, for \gls{udp} rates in the order of 5 to 10 Mbps, the network is mainly plagued by local congestion which causes the insurgence of buffering in some of the nodes. Conversely, as the rate of the \gls{udp} sources increases the system shifts to a capacity-limited regime, a phenomenon which explains the dominance of the \gls{msr} and \gls{mrba} policies.

\subsection{Network congestion}

The network congestion is measured by collecting, every $T_{alloc}$ subframes, the \gls{rlc} buffers status of the various nodes into the vector $\mathbf{B}^{\mathrm{RLC}}$. It must be noted that, since \gls{rlc} \gls{am} is used, these values will indicate data which is related to both new packets and possible retransmissions.

\begin{figure}[tbp]
\centering
  \subfloat[Medians, toward \glspl{ue}.\label{Fig:BSR_median_ue}]{
    \setlength\fwidth{0.19\columnwidth}
    \setlength\fheight{0.2\columnwidth}
    \input{Figures/Sim_results/Median_BSR_UE.tex}
    }
  \hfill
  \subfloat[Medians, toward \gls{iab}-nodes.\label{Fig:BSR_median_node}]{
    \setlength\fwidth{0.19\columnwidth}
    \setlength\fheight{0.2\columnwidth}
    \input{Figures/Sim_results/Median_BSR_node.tex}
    }
  \hfill
  \subfloat[Third quartile vs. depth in the \gls{iab} network, for $s_{UDP}$ = 200 B.\label{Fig:BSR_third_quartile_depth}]{
    \setlength\fwidth{0.23\columnwidth}
    \setlength\fheight{0.2\columnwidth}
    \input{Figures/Sim_results/Third_quartile_BSR_vs_depth.tex}
    }  
  \vspace*{-3mm}
   \caption{Buffer occupancy statistics, for $s_{UDP}$ $\in$ $\{50, 100, 200, 500 \}$~B.}
  \label{Fig:BSR_median}
    \vspace{-.6cm} 
\end{figure}
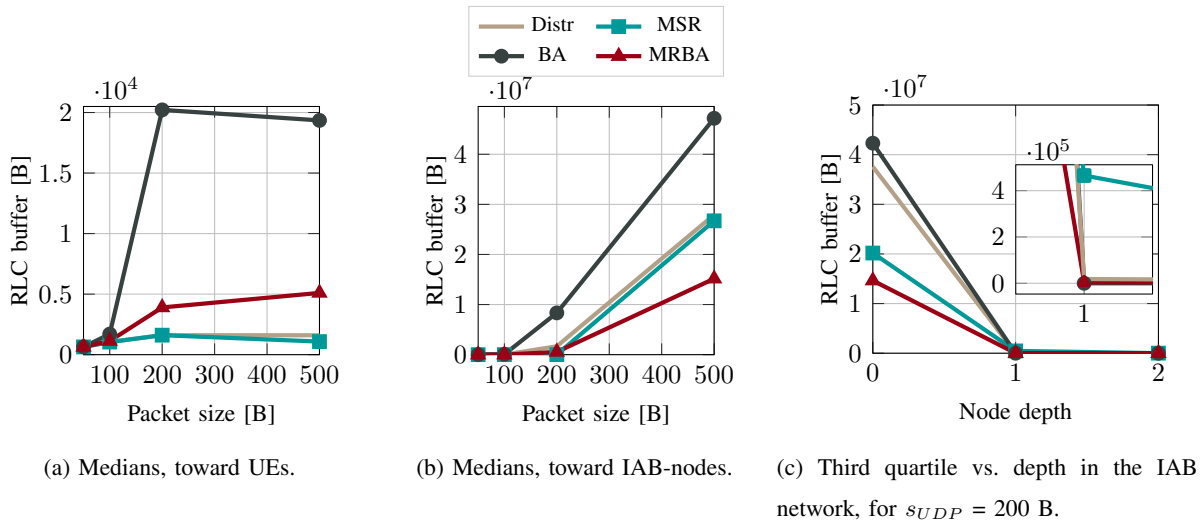

Figs.~\ref{Fig:BSR_median_ue} and~\ref{Fig:BSR_median_node} show the median of $\mathbf{B}^{\mathrm{RLC}}$, for traffic flows whose next hop in the network is represented by either \glspl{ue} or \gls{iab}-nodes respectively. The \gls{ba} strategy achieves the worst performance in this metric, leading to unstable systems in the cases of $s_{UDP}$ = $\{$200, 500$\}$ B. 
A reason for this behavior can be found in the ``locality" of the \gls{ba} policy criteria and the lack of influence of the past allocations on the weights. These characteristics may lead to favoring the same link in a repeated manner, hence offering little remedy to the end-to-end congestion.
On the other hand, the buffer occupancy achieved by the \gls{msr} strategy depicts a system behavior which, in accordance with previous observations, is extremely similar to that of the distributed case. Interestingly, with these configurations the network congestion occurs primarily at the donor and, in general, on the backhaul links towards \gls{iab}-nodes. This phenomenon can be explained as follows: even though, on average, the channel qualities of the backhaul links experience a better \gls{sinr}, the maximum number of such links which can be concurrently activated is limited, due to the \gls{tdd} configuration. Therefore, the \gls{msr} policy may introduce a bias towards the access links instead, since their activation yields the highest sum capacity, despite their lower channel quality.
Finally, the \gls{mrba} policy achieves the lowest amount of \gls{rlc} buffering. Specifically, Fig.~\ref{Fig:BSR_median_node} shows that, compared to the \gls{msr} and distributed strategies, the median buffer occupancy among backhaul links is up to 60\% smaller, albeit at the cost of slightly more congested \gls{ue} buffers. 

Finally, Fig.~\ref{Fig:BSR_third_quartile_depth} depicts the third quartiles of $\mathbf{B}^{\mathrm{RLC}}$ as a function of the depth of the corresponding \gls{gnb} in the \gls{iab} network. It is possible to notice that, regardless of the policy in use, the amount of buffering at the various \glspl{gnb} generally decreases as their distance to the donor increases. This follows from the fact that nodes which have a lower depth exhibit, on average, a bigger subtending tree; therefore the amount of traffic which makes use of their backhaul links is significantly higher. 

\subsection{Performance with \gls{tcp} traffic}


This subsection extends the aforementioned analysis by inspecting the performance of the proposed scheme in the case of \gls{tcp} traffic. Specifically, a \gls{tcp} full-buffer source model is used, and the various centralized resource allocation policies are compared against the distributed scheduler.

\begin{figure}[tbp]
\centering
  \subfloat[Delay \gls{ecdf}.\label{Fig:TCP_delay_ECDF_256}]{
    \setlength\fwidth{0.36\columnwidth}
    \setlength\fheight{0.2\columnwidth}
    \input{Figures/Sim_results/TCP_E2E_delay_ECDF_packet_size_256.tex}
    }
  \hfill
  \subfloat[Throughput \gls{ecdf}.\label{Fig:TCP_throughput_ECDF_256}]{
    \setlength\fwidth{0.36\columnwidth}
    \setlength\fheight{0.2\columnwidth}
    \input{Figures/Sim_results/TCP_E2E_throughput_ECDF_packet_size_256.tex}
    }
   \caption{End-to-end delay and throughput statistics, for \gls{tcp} layer 4 protocol.}
  \label{Fig:TCP_ECDFs}
    \vspace{-.6cm} 

\end{figure}
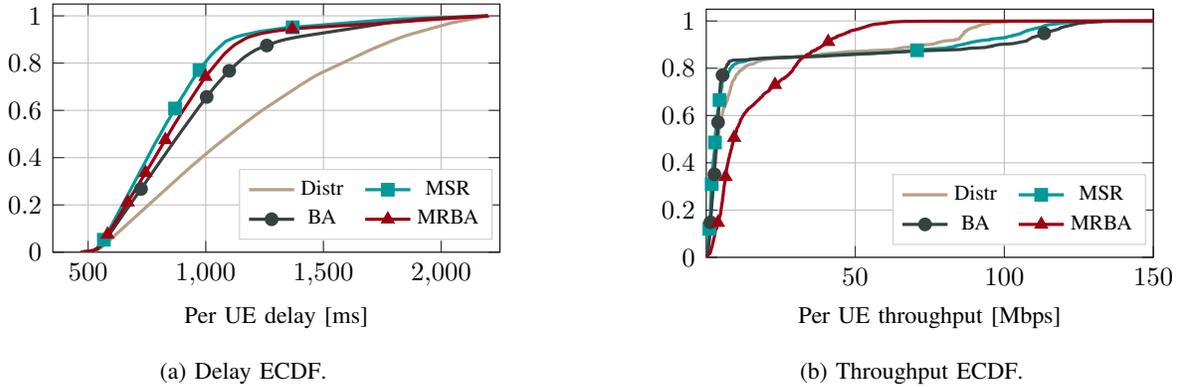

Fig.~\ref{Fig:TCP_delay_ECDF_256} shows the \gls{ecdf} of the end-to-end delay experienced by the successfully received packets. Similarly to the \gls{udp} case, the distributed scheduler exhibits the worse performance in this regard. However, the behavior of the centralized policies is remarkably different. In particular, with this configuration the \gls{msr} policy provides the best results, followed quite closely by the \gls{mrba} and \gls{ba} strategies. Fig.~\ref{Fig:TCP_throughput_ECDF_256}, which depicts the statistics of the end-to-end throughput achieved by the various \glspl{ue}, helps explain these results. The net effect of the \gls{ba} and \gls{msr} policies is, approximately, a 20\% increase of the peak throughput. Conversely, the \gls{mrba} strategy causes a redistribution of the achieved data rate, massively improving the lower quartiles (up to the 80-th), albeit at the expense of the maximum throughput. 

Therefore, we can conclude that regardless of the specific policies used, the proposed scheme improves the system performance by limiting the insurgence of local buffering, aiding the end-to-end congestion control mechanism offered by \gls{tcp}. Furthermore, it can be noted that both a prioritization of the most congested links and of the channels featuring a higher quality results in performance benefits in the average case, although it also causes a decrease of the network fairness. On the other hand, the \gls{mrba} policy manages to optimize the backhaul/access resource partitioning, while introducing an increase in the throughput fairness at the same time.


\subsection{Further considerations}

\begin{figure}[tbp]
  \subfloat[Combined per \gls{ue} end-to-end throughput first quartile and end-to-end delay third quartile, as a function of the centralized allocation period $T_{alloc}$.\label{Fig:Alloc_period_impact}]{
    \setlength\fwidth{0.45\columnwidth}
    \setlength\fheight{0.2\columnwidth}
    \input{Figures/Sim_results/Perf_vs_alloc_period.tex}
    }
  \hfill
  \subfloat[\gls{mwm} runtime as a function of the number of \gls{iab}-nodes in the network.\label{Fig:MWM_runtime}]{
    \setlength\fwidth{0.3\columnwidth}
    \setlength\fheight{0.2\columnwidth}
     \input{Figures/Sim_results/MWM_runtime_vs_nodes.tex}
    }
  
   \caption{Considerations on the formulated assumptions.}
  \label{Fig:Assump}
      \vspace{-.6cm} 

\end{figure}
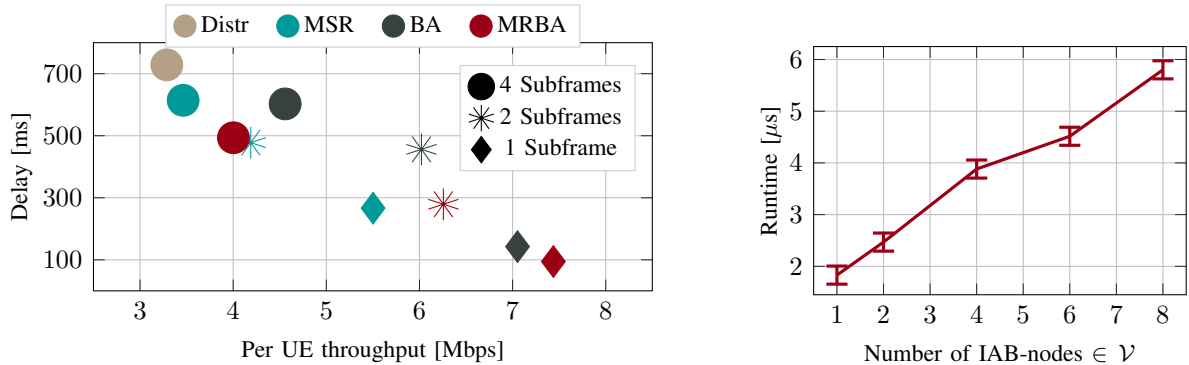

It is of particular relevance to analyze the performance of the centralized policies when relaxing the most restrictive hypothesis, i.e., the capability of reliably exchanging feedback information in a timely manner, and to understand how restrictive such assumption actually is. 
To such end, Fig.~\ref{Fig:Alloc_period_impact} shows the performance of the proposed framework as a function of the centralized allocation period $T_{alloc}$. In particular, each of the depicted points represents the joint end-to-end throughput and delay achieved with the different configurations.
As expected, in general the effectiveness of the various centralized policies progressively deteriorates as the frequency of the scheduling indications decreases. Interestingly, the \gls{ba} policy exhibits the lowest performance degradation with respect to an increase of the allocation period, which suggests that this phenomenon has a slower evolution over time compared to the one exhibited by the channels quality. Nevertheless, the key takeaway is that all of the proposed allocation strategies outperform the distributed solution, across both metrics. However, the trend depicted by Fig.~\ref{Fig:Alloc_period_impact} also suggests that there exists a threshold value of $T_{alloc}$ after which the performance of the proposed frameworks brings only marginal performance benefits.

Additionally, the running time of the \gls{mwm} algorithm presented in Sec.~\ref{Sec:T-MWM} was analyzed, in order to understand whether it may partially invalidate the timely feedback assumption. Specifically, Fig.~\ref{Fig:MWM_runtime} presents the statistics of the various \gls{mwm} execution times, obtained on a machine equipped with an i7-6700 4-core processor clocked at 3.4~{GHz}. 
The first observation which can be made is that this empirical analysis confirms the previously estimated asymptotic complexity, depicting a running time which exhibits a linear dependence on the number of \glspl{gnb} in the network. Furthermore, it can be noted that the runtime of the \gls{mwm} algorithm does not exceed 6~$\mu$s, even for a significant number of \gls{iab}-nodes connected to the same \gls{iab}-donor. As a consequence, we can conclude that the execution times of the centralized allocation process do not pose any threat to the timely feedback assumption, since they are reasonably smaller than the duration of the minimum centralized allocation period.

\section{Conclusions}
\label{Sec:conc}

In this paper we proposed a centralized resource partitioning scheme for \gls{5g} and beyond \gls{iab} networks, coupled with a set of allocation policies. We showed that the introduction of this light resource allocation cooperation dramatically improves the end-to-end throughput and delay achieved by the system already, preventing (or at the very least limiting) the insurgence of network congestion on the backhaul links. Specifically, the \gls{mrba} policy exhibits the most promising results, offering up to a 5-fold increase in the worst-case throughput and approximately a 50\% smaller worst-case latency, compared to the distributed scheduler. On the other hand, the effectiveness of the \gls{ba} and \gls{msr} policies varies quite significantly across the specific system configuration and inspected metric. 

We provided considerations on the implementation of a centralized resource allocation controller in real world deployments. In particular, we acknowledged that the proposed scheme relies on the assumption of \gls{iab}-nodes being capable of exchanging timely feedback information with the \gls{iab}-donor. Even though the amount of signaling data which the proposed solution requires is quite low, and its performance is quite robust with respect to an increase of the central allocation period, we argue that this remains a significant constraint. Moreover, such drawback is exacerbated by the unfavorable \glspl{mmwave} propagation characteristics.
As a consequence, we deem that centralized solutions, which rely on timely exchange of control information with the \gls{iab}-donor, are likely to require dedicated control channels, possibly at sub-6~{GHz}, in order to grant the utmost priority and reliability to the feedback information. Therefore, we can conclude that the aforementioned framework can bring dramatic performance benefits to \gls{iab} networks, although its introduction in \gls{5g} and beyond deployments requires additional research efforts.

For this reason, as part of our future work we plan to design machine-learning algorithms which predict the network evolution at the \gls{iab}-donor. This improvement will allow us to relax the timely feedback assumption, by increasing the minimum centralized allocation period which leads to performance benefits over distributed strategies. 
Moreover, we foresee to implement mechanisms which adapt the parameters of the \gls{mrba} policy to the system load and configuration, and additional resource partitioning strategies. 
Finally, the generalization of the proposed framework to \gls{sdma} systems will be studied. 
The use of such multiple access scheme should significantly improve the performance of \gls{mmwave} wireless backhauling by introducing the possibility of concurrently serving multiple terminals, provided that they exhibit a sufficient distance among them.

\glsresetall
\glsunset{nr}

\bibliographystyle{IEEEtran}
\bibliography{bibl.bib}

\end{document}

%% file: acronyms.tex
\newacronym{3gpp}{3GPP}{3rd Generation Partnership Project}
\newacronym{adc}{ADC}{Analog to Digital Converter}
\newacronym{afbw}{AFBW}{Average Fading Bandwidth}
\newacronym{5g}{5G}{5th generation}
\newacronym{4g}{4G}{4th generation}
\newacronym{aimd}{AIMD}{Additive Increase Multiplicative Decrease}
\newacronym{am}{AM}{Acknowledged Mode}
\newacronym{amf}{AMF}{Access and Mobility Management Function}
\newacronym{an}{AN}{Access Network}
\newacronym{amc}{AMC}{Adaptive Modulation and Coding}
\newacronym{aqm}{AQM}{Active Queue Management}
\newacronym{awgn}{AGWN}{Additive White Gaussian Noise}
\newacronym{balia}{BALIA}{Balanced Link Adaptation}
\newacronym{bsr}{BSR}{Buffer Status Report}
\newacronym{msr}{MSR}{Max Sum-Rate}
\newacronym{ba}{BA}{Backlog Avoidance}
\newacronym{mrba}{MRBA}{Max-Rate Backlog Avoidance}
\newacronym{bdp}{BDP}{Bandwidth-Delay Product}
\newacronym{bf}{BF}{Beamforming}
\newacronym{v2x}{V2X}{Vehicle-to-Everything}
\newacronym{vr}{VR}{Virtual Reality}
\newacronym{mdp}{MDP}{Markov Decision Process}
\newacronym{mwm}{MWM}{Maximum Weighted Matching}
\newacronym{tdm}{TDM}{Time Division Multiplexing}
\newacronym{fdm}{FDM}{Frequency Division Multiplexing}
\newacronym{sdm}{SDM}{Space Division Multiplexing}
\newacronym{st}{ST}{Spanning Tree}
\newacronym{rl}{RL}{Reinforcement Learning}
\newacronym{lp}{LP}{Linear Programming}
\newacronym[firstplural=Deep Neural Networks (DNNs)]{dnn}{DNN}{Deep Neural Network}
\newacronym{dag}{DAG}{Directed Acyclic Graph}
\newacronym{gtp}{GTP}{GPRS Tunneling Protocol}
\newacronym{nas}{NAS}{Non-Access Stratum}
\newacronym{isp}{ISP}{Internet Service Provider}
\newacronym{ngnm}{NGNM}{Next Generation Mobile Networks Alliance}
\newacronym{cc}{CC}{Component Carrier}
\newacronym{drb}{DRB}{Data Radio Bearer}
\newacronym{qos}{QoS}{Quality of Service}
\newacronym{ca}{CA}{Carrier Aggregation}
\newacronym{sdap}{SDAP}{Service Data Adaptation Protocol}
\newacronym{lc}{LC}{Logical Channel}
\newacronym{rnti}{RNTI}{Radio Network Temporary Identifier}
\newacronym{qci}{QCI}{Quality Class Identifier}
\newacronym{cdf}{CDF}{Cumulative Distribution Function}
\newacronym{ecdf}{ECDF}{Empirical Cumulative Distribution Function}
\newacronym{cmos}{CMOS}{Complementary Metal-Oxide Semiconductor}
\newacronym{cn}{CN}{Core Network}
\newacronym{cqi}{CQI}{Channel Quality Information}
\newacronym{cir}{CIR}{Channel Impulse Response}
\newacronym{cp}{CP}{Control Plane}
\newacronym{cu}{CU}{Central Unit}
\newacronym{du}{DU}{Distributed Unit}
\newacronym{csirs}{CSI-RS}{Channel State Information - Reference Signal}
\newacronym{dc}{DC}{Dual Connectivity}
\newacronym{imsi}{IMSI}{International Mobile Subscriber Identity}
\newacronym{dce}{DCE}{Direct Code Execution}
\newacronym{dci}{DCI}{Downlink Control Information}
\newacronym{uci}{UCI}{Uplink Control Information}
\newacronym{dl}{DL}{Downlink}
\newacronym{dmr}{DMR}{Deadline Miss Ratio}
\newacronym{dmrs}{DMRS}{DeModulation Reference Signal}
\newacronym{e2e}{E2E}{End-to-End}
\newacronym{ecn}{ECN}{Explicit Congestion Notification}
\newacronym{edf}{EDF}{Earliest Deadline First}
\newacronym{enb}{eNB}{evolved Node Base}
\newacronym{embb}{eMBB}{enhanced Mobile Broadband}
\newacronym{epc}{EPC}{Evolved Packet Core}
\newacronym{es}{ES}{Edge Server}
\newacronym{fdma}{FDMA}{Frequency Division Multiple Access}
\newacronym{fdd}{FDD}{Frequency Division Duplexing}
\newacronym[firstplural=Radio Access Technologies (RATs)]{rat}{RAT}{Radio Access Technology}
\newacronym[firstplural=Markov Chains (MCs)]{mc}{MC}{Markov Chain}
\newacronym{milp}{MILP}{Mixed-Integer Linear Programming}
\newacronym{fs}{FS}{Fast Switching}
\newacronym{ip}{IP}{Internet Protocol}
\newacronym{fr}{FR}{Frequency Range}
\newacronym{ftp}{FTP}{File Transfer Protocol}
\newacronym{gnb}{gNB}{Next Generation Node Base}
\newacronym{arq}{ARQ}{Automatic Repeat reQuest}
\newacronym{harq}{HARQ}{Hybrid Automatic Repeat reQuest}
\newacronym{hetnet}{HetNet}{Heterogeneous Network}
\newacronym{hh}{HH}{Hard Handover}
\newacronym{hol}{HOL}{Head-of-Line}
\newacronym{ia}{IA}{Initial Access}
\newacronym{imt}{IMT}{International Mobile Telecommunication}
\newacronym{iot}{IoT}{Internet of Things}
\newacronym{lcr}{LCR}{Level Crossing Rate}
\newacronym{lcf}{LCF}{Level Crossing Frequency}
\newacronym{los}{LoS}{Line-of-Sight}
\newacronym{lte}{LTE}{Long Term Evolution}
\newacronym{m2m}{M2M}{Machine to Machine}
\newacronym{mac}{MAC}{Medium Access Control}
\newacronym{num}{NUM}{Network Utility Maximization}
\newacronym{ri}{RI}{Rank Index}
\newacronym{pmi}{PMI}{Precoding Matrix Index}
\newacronym{mcs}{MCS}{Modulation and Coding Scheme}
\newacronym{mec}{MEC}{Mobile Edge Cloud}
\newacronym{mi}{MI}{Mutual Information}
\newacronym{mu}{MU}{Multi-User}
\newacronym{mimo}{MIMO}{Multiple Input, Multiple Output}
\newacronym{mmwave}{mmWave}{millimeter wave}
\newacronym{mptcp}{MPTCP}{Multipath TCP}
\newacronym{mr}{MR}{Maximum Rate}
\newacronym{mt}{MT}{Mobile Termination}
\newacronym{mss}{MSS}{Maximum Segment Size}
\newacronym{mtd}{MTD}{Machine-Type Device}
\newacronym{mtu}{MTU}{Maximum Transmission Unit}
\newacronym{nfv}{NFV}{Network Function Virtualization}
\newacronym{nf}{NF}{Network Function}
\newacronym{nlos}{NLoS}{Non-Line-of-Sight}
\newacronym{nr}{NR}{New Radio}
\newacronym{csi}{CSI}{Channel State Information}
\newacronym{o2i}{O2I}{Outdoor-to-Indoor}
\newacronym{ofdm}{OFDM}{Orthogonal Frequency Division Multiplexing}
\newacronym{pdcch}{PDCCH}{Physical Downlink Control Channel}
\newacronym{pdcp}{PDCP}{Packet-Data Convergence Protocol}
\newacronym{pdsch}{PDSCH}{Physical Downlink Shared Channel}
\newacronym{pdu}{PDU}{Packet Data Unit}
\newacronym{sdu}{SDU}{Service Data Unit}
\newacronym{pf}{PF}{Proportional Fair}
\newacronym{ce}{CE}{Control Element}
\newacronym{pgw}{PGW}{Packet Gateway}
\newacronym{phy}{PHY}{Physical}
\newacronym{pbch}{PBCH}{Physical Broadcast Channel}
\newacronym[plural=\gls{mme}s,firstplural=Mobility Management Entities (MMEs)]{mme}{MME}{Mobility Management Entity}
\newacronym{prb}{PRB}{Physical Resource Block}
\newacronym{pss}{PSS}{Primary Synchronization Signal}
\newacronym{pucch}{PUCCH}{Physical Uplink Control Channel}
\newacronym{pusch}{PUSCH}{Physical Uplink Shared Channel}
\newacronym{rach}{RACH}{Random Access Channel}
\newacronym{ran}{RAN}{Radio Access Network}
\newacronym{ngran}{NG-RAN}{Next Generation RAN}
\newacronym{red}{RED}{Random Early Detection}
\newacronym{rf}{RF}{Radio Frequency}
\newacronym{rlc}{RLC}{Radio Link Control}
\newacronym{rlf}{RLF}{Radio Link Failure}
\newacronym{rrc}{RRC}{Radio Resource Control}
\newacronym{rrm}{RRM}{Radio Resource Management}
\newacronym{rr}{RR}{Round Robin}
\newacronym{rs}{RS}{Remote Server}
\newacronym{rsrp}{RSRP}{Reference Signal Received Power}
\newacronym{rss}{RSS}{Received Signal Strength}
\newacronym{rtt}{RTT}{Round Trip Time}
\newacronym{rw}{RW}{Receive Window}
\newacronym{rx}{RX}{Receiver}
\newacronym{sa}{SA}{standalone}
\newacronym{sack}{SACK}{Selective Acknowledgment}
\newacronym{sap}{SAP}{Service Access Point}
\newacronym{sch}{SCH}{Secondary Cell Handover}
\newacronym{scoot}{SCOOT}{Split Cycle Offset Optimization Technique}
\newacronym{sdma}{SDMA}{Spatial Division Multiple Access}
\newacronym{sdn}{SDN}{Software Defined Networking}
\newacronym{sinr}{SINR}{Signal-to-Interference-plus-Noise Ratio}
\newacronym{sir}{SIR}{Signal-to-Interference Ratio}
\newacronym{sm}{SM}{Saturation Mode}
\newacronym{snr}{SNR}{Signal-to-Noise Ratio}
\newacronym{son}{SON}{Self-Organizing Network}
\newacronym{ss}{SS}{Synchronization Signal}
\newacronym{srs}{SRS}{Sounding Reference Signal}
\newacronym{sss}{SSS}{Secondary Synchronization Signal}
\newacronym{tb}{TB}{Transport Block}
\newacronym{tcp}{TCP}{Transmission Control Protocol}
\newacronym{tdd}{TDD}{Time Division Duplexing}
\newacronym{tdma}{TDMA}{Time Division Multiple Access}
\newacronym{tfl}{TfL}{Transport for London}
\newacronym{tm}{TM}{Transparent Mode}
\newacronym{trp}{TRP}{Transmitter Receiver Pair}
\newacronym{tti}{TTI}{Transmission Time Interval}
\newacronym{ttt}{TTT}{Time-to-Trigger}
\newacronym{tx}{TX}{Transmitter}
\newacronym{qam}{QAM}{Quadrature Amplitude Modulation}
\newacronym{ue}{UE}{User Equipment}
\newacronym{ul}{UL}{Uplink}
\newacronym{uml}{UML}{Unified Modeling Language}
\newacronym{um}{UM}{Unacknowledged Mode}
\newacronym{uma}{UMa}{Urban Macro}
\newacronym{utc}{UTC}{Urban Traffic Control}
\newacronym{vm}{VM}{Virtual Machine}
\newacronym{rsrq}{RSRQ}{Reference Signal Received Quality}
\newacronym{rssi}{RSSI}{Received Signal Strength Indicator}
\newacronym{crs}{CRS}{Cell Reference Signal}
\newacronym{nsa}{NSA}{Non Stand Alone}
\newacronym{mrdc}{MR-DC}{Multi \gls{rat} \gls{dc}}
\newacronym{eutra}{E-UTRA}{Evolved Universal Terrestrial Radio Access}
\newacronym{endc}{EN-DC}{E-UTRAN-\gls{nr} \gls{dc}}
\newacronym{5gc}{5GC}{5G Core}
\newacronym{si}{SI}{Study Item}
\newacronym{iab}{IAB}{Integrated Access and Backhaul}
\newacronym{wf}{WF}{Wired-first}
\newacronym{hqf}{HQF}{Highest-quality-first}
\newacronym{pa}{PA}{Position-aware}
\newacronym{mlr}{MLR}{Maximum-local-rate}
\newacronym{wbf}{WBF}{Wired Bias Function}
\newacronym{mib}{MIB}{Master Information Block}
\newacronym{sib}{SIB}{Secondary Information Block}
\newacronym{kpi}{KPI}{Key Performance Indicator}
\newacronym{ppp}{PPP}{Poisson Point Process}
\newacronym{mpc}{MPC}{Multi Path Component}
\newacronym{rt}{RT}{Ray Tracer}
\newacronym{aoa}{AoA}{Angle of Arrival}
\newacronym{aod}{AoD}{Angle of Departure}
\newacronym{scm}{SCM}{Spatial Channel Model}
\newacronym{inr}{INR}{Interference to Noise Ratio}
\newacronym{qd}{QD}{Quasi Deterministic}
\newacronym{wlan}{WLAN}{Wireless Local Area Network}
\newacronym{cad}{CAD}{Computer-aided Design}
\newacronym{ap}{AP}{Access Point}
\newacronym{sta}{STA}{Station}
\newacronym{urllc}{URLLC}{Ultra-Reliable Low-Latency Communication}
\newacronym{udp}{UDP}{User Datagram Protocol}
\newacronym{upf}{UPF}{User Plane Function}
\newacronym{umi}{UMi}{Urban Microcell}
\newacronym{uma2}{UMa}{Urban Macrocell}
\newacronym{rma}{RMa}{Rural Macrocell}
\newacronym{in}{In}{Indoor Office}
\newacronym{wpans}{WPANs}{Wireless Personal Area Networks}
\newacronym{ric}{RIC}{RAN Intelligent Controller}

%% file: myTikz.tex
\usepackage{tikz}
\usepackage{pgfplots}
\pgfplotsset{compat=newest} 
\pgfplotsset{plot coordinates/math parser=false} 
\newlength\fheight
\newlength\fwidth
\usetikzlibrary{plotmarks,patterns,decorations.pathreplacing,backgrounds,calc,arrows,arrows.meta,spy,matrix}
\usepgfplotslibrary{patchplots,groupplots}
\usepackage{tikzscale}

\tikzstyle{startstop} = [rectangle, rounded corners, minimum width=2cm, minimum height=0.5cm,text centered, draw=black]
\tikzstyle{io} = [trapezium, trapezium left angle=70, trapezium right angle=110, minimum width=3cm, minimum height=1cm, text centered, draw=black]
\tikzstyle{process} = [rectangle, minimum width=2cm, minimum height=0.5cm, text centered, draw=black, align=center]
\tikzstyle{decision} = [ellipse, minimum width=2cm, minimum height=1cm, text centered, draw=black]
\tikzstyle{arrow} = [thick,<->,>=stealth]
\tikzstyle{line} = [thick,>=stealth]
\tikzstyle{darrow} = [thick,<->,>=stealth,dashed]
\tikzstyle{sarrow} = [thick,->,>=stealth]
\tikzstyle{larrow} = [line width=0.1mm,dashdotted,<->,>=stealth]

\pgfplotsset{every tick label/.append style={font=\scriptsize}, 
             every axis/.append style={
             width=\fwidth, height=\fheight, at={(0\fwidth,0\fheight)}, 
             xlabel style={font=\footnotesize\color{white!15!black}},
             xmajorgrids,
             ylabel style={yshift=-0.15cm, font=\footnotesize\color{white!15!black}},
             ymajorgrids,
             legend style={font=\footnotesize\color{white!15!black}},
             /pgfplots/ybar legend/.style={/pgfplots/legend image code/.code={\draw[##1,/tikz/.cd,yshift=-0.25em](0cm,0cm) rectangle (10pt,1em);},},
             }}

\definecolor{SchoolColor}{RGB}{0.71, 0, 0.106}
\definecolor{chaptergrey}{rgb}{0.61, 0, 0.09} 
\definecolor{midgrey}{rgb}{0.4, 0.4, 0.4}
\definecolor{chaptergreen}{rgb}{0.09, 0.612, 0}
\definecolor{chapterpurple}{rgb}{0.522, 0, 0.612}
\definecolor{chapterlightgreen}{rgb}{0, 0.612, 0.522}

\makeatletter
\def\grd@save@target#1{%
  \def\grd@target{#1}}
\def\grd@save@start#1{%
  \def\grd@start{#1}}
\tikzset{
  grid with coordinates/.style={
    to path={%
      \pgfextra{%
        \edef\grd@@target{(\tikztotarget)}%
        \tikz@scan@one@point\grd@save@target\grd@@target\relax
        \edef\grd@@start{(\tikztostart)}%
        \tikz@scan@one@point\grd@save@start\grd@@start\relax
        \draw[minor help lines] (\tikztostart) grid (\tikztotarget);
        \draw[major help lines] (\tikztostart) grid (\tikztotarget);
        \grd@start
        \pgfmathsetmacro{\grd@xa}{\the\pgf@x/1cm}
        \pgfmathsetmacro{\grd@ya}{\the\pgf@y/1cm}
        \grd@target
        \pgfmathsetmacro{\grd@xb}{\the\pgf@x/1cm}
        \pgfmathsetmacro{\grd@yb}{\the\pgf@y/1cm}
        \pgfmathsetmacro{\grd@xc}{\grd@xa + \pgfkeysvalueof{/tikz/grid with coordinates/major step x}}
        \pgfmathsetmacro{\grd@yc}{\grd@ya + \pgfkeysvalueof{/tikz/grid with coordinates/major step y}}
        \foreach \x in {\grd@xa,\grd@xc,...,\grd@xb}
        \node[anchor=north] at (\x,\grd@ya) {\pgfmathprintnumber{\x}};
        \foreach \y in {\grd@ya,\grd@yc,...,\grd@yb}
        \node[anchor=east] at (\grd@xa,\y) {\pgfmathprintnumber{\y}};
      }
    }
  },
  minor help lines/.style={
    help lines,
    gray,
    line cap =round,
    xstep=\pgfkeysvalueof{/tikz/grid with coordinates/minor step x},
    ystep=\pgfkeysvalueof{/tikz/grid with coordinates/minor step y}
  },
  major help lines/.style={
    help lines,
    line cap =round,
    line width=\pgfkeysvalueof{/tikz/grid with coordinates/major line width},
    xstep=\pgfkeysvalueof{/tikz/grid with coordinates/major step x},
    ystep=\pgfkeysvalueof{/tikz/grid with coordinates/major step y}
  },
  grid with coordinates/.cd,
  minor step x/.initial=.5,
  minor step y/.initial=.2,
  major step x/.initial=1,
  major step y/.initial=1,
  major line width/.initial=1pt,
}
\makeatother

%% file: Figures/Sim_scenario.tex
\begin{tikzpicture}

    \definecolor{UNIPDRED}{RGB}{155,0,20}
	\definecolor{COMPLEMENTARY}{RGB}{0,153,153}
	\definecolor{DARKGREY}{RGB}{55,65,64}
	\definecolor{BLACK}{RGB}{20, 20, 20}
    
    \begin{axis}[
    width=\fwidth,
    height=\fheight,
    at={(0\fwidth,0\fheight)},
    scale only axis,
    legend style={
    	/tikz/every even column/.append style={column sep=0.3cm},
    	at={(0.5,0.98)},
    	anchor=south, 
    	draw=white!80!black, 
    	font=\small
    	},
    legend columns=3,
    xlabel style={font=\footnotesize},
    xlabel={},
    xtick={-100, -50, 0, 50, 100, 150},
    xmajorgrids,
    xmin=-150, xmax=150,
    xtick style={color=white!15!black},
    ylabel style={font=\footnotesize},
    ylabel={},
    ymajorgrids,
    ymin=-50, ymax=150,
    ytick style={color=white!15!black}
	]
	
	\addplot[very thick, UNIPDRED, forget plot, dash pattern=on 5 pt off 3 pt] 
	table {%
	-100 0 
	0 0 
	};
	\addplot[very thick, UNIPDRED, forget plot, dash pattern=on 5 pt off 3 pt] 
	table {%
	100 100 
	0 0 
	};
	\addplot[very thick, UNIPDRED, forget plot, dash pattern=on 5 pt off 3 pt] 
	table {%
	-100 100 
	0 0 
	};
	\addplot[very thick, UNIPDRED, forget plot, dash pattern=on 5 pt off 3 pt] 
	table {%
	100 0 
	100 100 
	};
	\addplot[very thick, UNIPDRED, forget plot, dash pattern=on 5 pt off 3 pt] 
	table {%
	0 100 
	-100 100 
	};

    \addplot[
      scatter,
      only marks,
      scatter src=explicit,
      scatter/classes={1={DARKGREY, mark=square*}, 2={UNIPDRED, mark=diamond*}, 3={COMPLEMENTARY}},
      visualization depends on=\thisrow{sizedata}\as\sizedata,
      scatter/@pre marker code/.append style={
          /tikz/mark size=\sizedata
      }
    ]
    table[x=x,y=y, meta=class]{%
    x                      y                      sizedata                class
    0 0 5 1 
	100 0 6 2 
	-100 0 6 2 
	0 100 6 2 
	100 100 6 2 
	-100 100 6 2 
	28.8603 6.70813 2 3 
	22.1839 0.179283 2 3 
	-25.8695 -4.74675 2 3 
	-25.698 1.31045 2 3 
	-26.1185 5.61727 2 3 
	-18.8974 12.8042 2 3 
	-25.8278 8.77018 2 3 
	-11.9403 24.625 2 3 
	86.2329 -14.9938 2 3 
	113.411 -25.4381 2 3 
	80.796 10.6876 2 3 
	89.0457 9.21925 2 3 
	87.952 27.3012 2 3 
	89.4537 -0.761095 2 3 
	108.92 -15.5517 2 3 
	84.0229 -13.509 2 3 
	-71.3282 -2.49305 2 3 
	-114.738 26.7601 2 3 
	-115.867 28.3354 2 3 
	-117.757 24.2406 2 3 
	-103.439 -22.3846 2 3 
	-89.1148 5.80584 2 3 
	-116.494 -26.2217 2 3 
	-130.211 3.15653 2 3 
	-19.5929 85.0171 2 3 
	-22.2209 78.2847 2 3 
	13.0963 129.843 2 3 
	-15.409 112.165 2 3 
	-26.1002 118.747 2 3 
	12.3455 84.2945 2 3 
	-30.9925 96.9305 2 3 
	-33.7183 108.212 2 3 
	125.011 97.7272 2 3 
	112.293 126.281 2 3 
	113.959 76.0568 2 3 
	103.236 114.801 2 3 
	104.862 76.3655 2 3 
	131.206 109.996 2 3 
	114.825 128.336 2 3 
	90.7864 78.3714 2 3 
	-109.959 77.6437 2 3 
	-83.8184 115.384 2 3 
	-78.9347 82.8969 2 3 
	-114.208 81.4837 2 3 
	-123.238 80.8704 2 3 
	-104.259 88.5412 2 3 
	-94.1952 108.948 2 3 
	-114.781 80.0824 2 3 
    };
    \legend{\gls{iab}-donor, \gls{iab}-node, \gls{ue}}

    \end{axis}
    
\end{tikzpicture}

%% file: Figures/Sim_results/E2E_throughput_ECDF_packet_size_100.tex
\begin{tikzpicture}

\pgfplotsset{every tick label/.append style={font=\small}}

\definecolor{UNIPDRED}{RGB}{155,0,20}
\definecolor{LIGHT_GREY}{RGB}{189,195,199}
\definecolor{COMPLEMENTARY}{RGB}{0,153,153}
\definecolor{DARKGREY}{RGB}{55,65,64}
\definecolor{SAND}{RGB}{180,160,135}

\begin{axis}[
width=\fwidth,
height=\fheight,
at={(0\fwidth,0\fheight)},
scale only axis,
legend style={
    /tikz/every even column/.append style={column sep=0.2cm},
    at={(0.3,0.65)}, 
    anchor=south, 
    draw=white!80!black, 
    font=\scriptsize
    },
legend columns=2,
xlabel style={font=\footnotesize},
xlabel={UE throughut [Mbps]},
xtick={0, 1, 2, 3, 4, 5, 6, 7, 8},
xmajorgrids,
xmin=0, xmax=8.5,
xtick style={color=white!15!black},
ylabel shift = -1 pt,
ylabel style={font=\footnotesize},
ylabel={},
ymajorgrids,
ymin=0, ymax=1.05,
ytick style={color=white!15!black},
ytick={0.2,0.4,0.6,0.8,1},
]

\addplot [very thick, SAND]
table {%
0.202961564064026 0
0.252192497253418 0.000852465629577637
0.56183123588562 0.00255751609802246
0.866669297218323 0.00341010093688965
0.896506190299988 0.00426256656646729
0.964531421661377 0.00511503219604492
0.975398540496826 0.00596761703491211
0.976921319961548 0.00682008266448975
0.993403792381287 0.00852513313293457
1.05385088920593 0.0102301836013794
1.05855548381805 0.011082649230957
1.05874216556549 0.0213128328323364
1.06492495536804 0.0230178833007812
1.1454005241394 0.0238704681396484
1.15920770168304 0.0247229337692261
1.1640613079071 0.0255753993988037
1.16465044021606 0.0281330347061157
1.16465842723846 0.0289855003356934
1.16519439220428 0.029837965965271
1.16634881496429 0.0306905508041382
1.16657912731171 0.032395601272583
1.16756224632263 0.0332480669021606
1.16788184642792 0.0358055830001831
1.16833114624023 0.0366581678390503
1.16834080219269 0.0383632183074951
1.16948056221008 0.0392156839370728
1.16974532604218 0.0409207344055176
1.17054116725922 0.0417732000350952
1.17056858539581 0.0426257848739624
1.17679977416992 0.04347825050354
1.17750024795532 0.0443308353424072
1.25732815265656 0.0451833009719849
1.25757563114166 0.0537084341049194
1.25892186164856 0.0545608997344971
1.2593879699707 0.0562659502029419
1.28072679042816 0.0596760511398315
1.28277683258057 0.0605285167694092
1.2828027009964 0.0613811016082764
1.51543700695038 0.062233567237854
1.541384100914 0.0630861520767212
1.55124962329865 0.0639386177062988
1.55171287059784 0.0656436681747437
1.55198323726654 0.0699062347412109
1.55223274230957 0.0767263174057007
1.75625610351562 0.0784313678741455
1.75644981861115 0.0835464000701904
1.75716769695282 0.0852514505386353
1.75728631019592 0.0895140171051025
1.8184335231781 0.0903666019439697
1.93340647220612 0.0912190675735474
1.95776987075806 0.0920716524124146
2.01278352737427 0.0929241180419922
2.01823544502258 0.0937767028808594
2.0327730178833 0.094629168510437
2.03416466712952 0.0954816341400146
2.03462386131287 0.100596785545349
2.03826832771301 0.101449251174927
2.03864073753357 0.104006767272949
2.0396134853363 0.110826969146729
2.03963041305542 0.113384485244751
2.20001173019409 0.115089535713196
2.24309825897217 0.115942001342773
2.45997714996338 0.117647051811218
2.461092710495 0.118499517440796
2.46135830879211 0.122762203216553
2.46208715438843 0.12361466884613
2.46273350715637 0.127877235412598
2.46319723129272 0.13043475151062
2.46343350410461 0.132992267608643
2.46343636512756 0.134697318077087
2.50774335861206 0.135549902915955
2.50960659980774 0.136402368545532
2.50992274284363 0.138959884643555
2.51023030281067 0.146632552146912
2.51055574417114 0.154305219650269
2.57261538505554 0.155157685279846
2.57346606254578 0.161977767944336
2.5737726688385 0.163682818412781
2.5740122795105 0.16709291934967
2.57489943504333 0.167945384979248
2.59287309646606 0.168797969818115
2.73921680450439 0.169650435447693
2.74353766441345 0.17050302028656
2.74409031867981 0.17391300201416
2.74417591094971 0.176470637321472
2.74463605880737 0.178175568580627
2.74487280845642 0.181585669517517
2.74487280845642 0.182438135147095
2.76417517662048 0.18414318561554
2.76449728012085 0.187553286552429
2.76473951339722 0.194373369216919
2.76553177833557 0.195225954055786
2.76564264297485 0.196078419685364
2.77623224258423 0.196930885314941
2.77951049804688 0.197783470153809
2.80185532569885 0.198635935783386
2.8774573802948 0.199488520622253
2.91201424598694 0.200340986251831
3.07873964309692 0.202046036720276
3.08935785293579 0.202898502349854
3.08982825279236 0.213128685951233
3.09190583229065 0.2139812707901
3.11039876937866 0.214833736419678
3.16138291358948 0.215686321258545
3.1794867515564 0.216538786888123
3.25020360946655 0.2173912525177
3.25068521499634 0.218243837356567
3.25160574913025 0.219096302986145
3.25207996368408 0.221653938293457
3.25226044654846 0.224211454391479
3.25231909751892 0.226768970489502
3.253342628479 0.22762143611908
3.25335836410522 0.229326486587524
3.25575923919678 0.231031537055969
3.2620701789856 0.232736587524414
3.26222920417786 0.233589053153992
3.26291584968567 0.234441637992859
3.26295113563538 0.236146688461304
3.26880288124084 0.236999154090881
3.26952004432678 0.240409255027771
3.26971411705017 0.246376752853394
3.27007675170898 0.248081803321838
3.28747820854187 0.249786853790283
3.28793287277222 0.254902005195618
3.378178358078 0.255754470825195
3.39298796653748 0.256607055664062
3.39373064041138 0.25745952129364
3.39434003829956 0.26086950302124
3.4259557723999 0.261722087860107
3.45490646362305 0.263427138328552
3.45598101615906 0.26768970489502
3.45616626739502 0.271952271461487
3.45727181434631 0.272804737091064
3.45739197731018 0.276214838027954
3.4653217792511 0.277067422866821
3.46551704406738 0.277919888496399
3.46621513366699 0.278772354125977
3.46651029586792 0.281329870223999
3.47287178039551 0.282182455062866
3.47299766540527 0.287297487258911
3.47396636009216 0.288150072097778
3.47399115562439 0.289855003356934
3.47489523887634 0.290707588195801
3.47495365142822 0.29411768913269
3.55280566215515 0.294970154762268
3.55282735824585 0.295822620391846
3.55347037315369 0.296675205230713
3.55392646789551 0.302642822265625
3.55408000946045 0.305200338363647
3.55603766441345 0.306052923202515
3.55702376365662 0.312020421028137
3.56011343002319 0.312873005867004
3.56034994125366 0.317135572433472
3.56537747383118 0.317988038063049
3.56614184379578 0.321398138999939
3.56629657745361 0.323955655097961
3.57259321212769 0.324808120727539
3.58701586723328 0.325660705566406
3.58864426612854 0.326513171195984
3.58864426612854 0.327365756034851
3.59166407585144 0.329070806503296
3.59267210960388 0.329923272132874
3.59267210960388 0.331628322601318
3.6157865524292 0.332480788230896
3.68164563179016 0.333333373069763
3.70917177200317 0.334185838699341
3.76687598228455 0.335038423538208
3.77259755134583 0.335890889167786
3.8116090297699 0.336743354797363
3.95285320281982 0.33759593963623
3.95313262939453 0.34100604057312
3.95932054519653 0.341858506202698
3.96012282371521 0.34782612323761
3.97004652023315 0.348678588867188
3.97493195533752 0.349531173706055
3.97537016868591 0.352088689804077
3.97540211677551 0.355498790740967
4.07654285430908 0.356351256370544
4.0800142288208 0.358056306838989
4.08015584945679 0.361466288566589
4.13733291625977 0.362318873405457
4.14362668991089 0.363171339035034
4.17856073379517 0.364023923873901
4.17973756790161 0.364876389503479
4.18114137649536 0.369138956069946
4.18135166168213 0.370844006538391
4.19823598861694 0.371696472167969
4.43343210220337 0.374253988265991
4.43376350402832 0.375959038734436
4.4338641166687 0.377664089202881
4.43756103515625 0.378516674041748
4.43763542175293 0.383631706237793
4.56832838058472 0.38448429107666
4.57059574127197 0.387041807174683
4.5708589553833 0.39130437374115
4.60395431518555 0.392156839370728
4.6182656288147 0.393009424209595
4.6186261177063 0.39471435546875
4.61914968490601 0.39812445640564
4.61993598937988 0.398977041244507
4.61996364593506 0.399829506874084
4.62105560302734 0.400681972503662
4.62153244018555 0.404092073440552
4.73246669769287 0.404944658279419
4.73310422897339 0.406649589538574
4.73330068588257 0.410912156105042
4.8337082862854 0.411764740943909
4.99098491668701 0.412617206573486
4.99284362792969 0.415174722671509
4.99328994750977 0.417732238769531
4.99480962753296 0.418584823608398
4.99525785446167 0.420289874076843
4.99553298950195 0.421142339706421
4.99661684036255 0.421994924545288
4.99676084518433 0.423699855804443
5.17032241821289 0.424552440643311
5.24879693984985 0.425404906272888
5.41970729827881 0.426257491111755
5.49168968200684 0.427109956741333
5.4940037727356 0.4279625415802
5.49420738220215 0.428815007209778
5.49536418914795 0.429667472839355
5.49723100662231 0.430520057678223
5.49791431427002 0.4313725233078
5.50421285629272 0.432225108146667
5.50569820404053 0.433077573776245
5.62997341156006 0.433930158615112
5.63085031509399 0.439045190811157
5.6316065788269 0.439897656440735
5.66009998321533 0.440750241279602
5.660813331604 0.442455291748047
5.6609354019165 0.445865273475647
5.68189811706543 0.446717858314514
5.68688774108887 0.447570323944092
5.74201440811157 0.448422908782959
5.96551561355591 0.449275374412537
5.96557807922363 0.450127840042114
5.96841955184937 0.450980424880981
5.97161245346069 0.452685356140137
6.01341819763184 0.453537940979004
6.42634057998657 0.455242991447449
6.63571929931641 0.457800507545471
6.63647985458374 0.458652973175049
6.79553747177124 0.459505558013916
6.80161094665527 0.460358023643494
6.80394315719604 0.461210608482361
6.80474090576172 0.462915658950806
6.80669689178467 0.463768124580383
6.80806350708008 0.465473175048828
6.88397359848022 0.466325640678406
6.88428688049316 0.467178106307983
6.88578176498413 0.468030691146851
6.88593053817749 0.468883275985718
6.88760709762573 0.470588207244873
6.89998388290405 0.47144079208374
6.90590143203735 0.472293257713318
6.91876888275146 0.473145723342896
6.91991949081421 0.477408409118652
6.92025327682495 0.480818390846252
6.92038679122925 0.483376026153564
6.92038679122925 0.484228491783142
6.96480131149292 0.48508095741272
6.96578121185303 0.488491058349609
6.96609401702881 0.491048574447632
6.97554874420166 0.491901159286499
6.97604942321777 0.494458675384521
7.03075361251831 0.496163725852966
7.10637521743774 0.497016191482544
7.10752820968628 0.497868776321411
7.10972833633423 0.498721241950989
7.11386919021606 0.499573707580566
7.11498880386353 0.500426292419434
7.11578226089478 0.502983808517456
7.11615324020386 0.505541324615479
7.12094640731812 0.506393909454346
7.12096405029297 0.507246375083923
7.12387132644653 0.508098840713501
7.12517976760864 0.508951425552368
7.13388347625732 0.509804010391235
7.18222141265869 0.512361526489258
7.1822304725647 0.513213992118835
7.18277883529663 0.514066457748413
7.3571515083313 0.51491904258728
7.43635606765747 0.517476558685303
7.43645238876343 0.520034074783325
7.45985698699951 0.520886659622192
7.45991516113281 0.52173912525177
7.4609522819519 0.523444175720215
7.46114635467529 0.526854276657104
7.59792232513428 0.527706742286682
7.61994743347168 0.52855920791626
7.62002849578857 0.529411792755127
7.62103986740112 0.530264258384705
7.6213231086731 0.534526824951172
7.65454196929932 0.535379409790039
7.65667629241943 0.536231875419617
7.65774869918823 0.537084341049194
7.6579122543335 0.542199492454529
7.72764778137207 0.543051958084106
7.72875070571899 0.544757008552551
7.72908353805542 0.546462059020996
7.72914600372314 0.549872159957886
7.72991609573364 0.551577091217041
7.7323317527771 0.552429676055908
7.73250818252563 0.555839776992798
7.75897026062012 0.556692242622375
7.75935840606689 0.55839729309082
7.7607536315918 0.560102224349976
7.76098394393921 0.56180739402771
7.76100492477417 0.562659859657288
7.88925790786743 0.563512325286865
7.89457368850708 0.564364910125732
7.89519262313843 0.56521737575531
7.8969521522522 0.566069841384888
7.89737606048584 0.567775011062622
7.89823436737061 0.5686274766922
7.89847326278687 0.569479942321777
7.92821931838989 0.570332527160645
7.92979621887207 0.574594974517822
7.92980718612671 0.576300144195557
7.95089387893677 0.577152609825134
7.95202875137329 0.578005075454712
7.96519136428833 0.578857660293579
7.96846389770508 0.579710125923157
7.97418642044067 0.580562591552734
7.97646808624268 0.582267761230469
7.97681188583374 0.583120226860046
7.97761011123657 0.583972692489624
7.9778094291687 0.588235378265381
7.97958421707153 0.591645359992981
7.98170280456543 0.593350410461426
7.98264646530151 0.594202876091003
7.98287963867188 0.595055341720581
7.9837121963501 0.595907926559448
7.98383378982544 0.596760511398315
7.9848895072937 0.597612977027893
7.98517799377441 0.599318027496338
7.98613309860229 0.600170493125916
7.98673295974731 0.602728128433228
7.9877667427063 0.60528564453125
7.98811101913452 0.609548091888428
7.98812198638916 0.610400676727295
7.98888969421387 0.611253261566162
7.98906755447388 0.612958192825317
7.99013471603394 0.614663243293762
7.99044609069824 0.61551570892334
7.99170303344727 0.616368293762207
7.99174737930298 0.617220878601074
7.99272680282593 0.618073344230652
7.99467468261719 0.629155993461609
7.99509811401367 0.630861043930054
7.99566555023193 0.632566094398499
7.99620008468628 0.638533592224121
7.99736976623535 0.642796277999878
7.99797105789185 0.6453537940979
7.99840593338013 0.647058844566345
7.99900770187378 0.651321411132812
7.99938678741455 0.65558397769928
7.99994373321533 0.674339294433594
8.0001220703125 0.678601861000061
8.00043487548828 0.685421943664551
8.00069141387939 0.68883204460144
8.0010814666748 0.751065731048584
8.00146007537842 0.837169647216797
8.0027437210083 0.960784316062927
8.00304412841797 0.994032382965088
8.00409317016602 0.99658989906311
8.00410461425781 0.999147415161133
8.5 1
};
\addlegendentry{Distr}

\addplot [very thick, COMPLEMENTARY, mark=square*, mark repeat=60, mark phase=40]
table {%
0.086592435836792 0
0.189862370491028 0.000852465629577637
1.3988926410675 0.00511503219604492
1.92758643627167 0.00596761703491211
2.01948261260986 0.00682008266448975
2.31519150733948 0.00852513313293457
2.31544995307922 0.0119352340698242
2.31560254096985 0.013640284538269
2.32876396179199 0.0144927501678467
2.32946300506592 0.0153452157974243
2.32954382896423 0.018755316734314
2.56339597702026 0.0196079015731812
2.57244086265564 0.0204603672027588
2.61001586914062 0.0213128328323364
2.61001920700073 0.0221654176712036
2.62650728225708 0.0230178833007812
2.72383069992065 0.0238704681396484
2.72408556938171 0.0255753993988037
2.83915209770203 0.0264279842376709
2.83924412727356 0.029837965965271
2.84066367149353 0.0306905508041382
2.84075236320496 0.0315430164337158
2.84208655357361 0.0332480669021606
2.84252381324768 0.0366581678390503
2.84252715110779 0.0375106334686279
2.84352231025696 0.0383632183074951
2.84379839897156 0.0392156839370728
2.89895129203796 0.0400681495666504
2.90018367767334 0.0417732000350952
2.90064382553101 0.04347825050354
2.90121984481812 0.0477408170700073
2.90139818191528 0.0537084341049194
2.90536093711853 0.0545608997344971
2.90542149543762 0.0554134845733643
2.90685057640076 0.0571185350418091
2.90721011161804 0.0588235855102539
2.90721082687378 0.0596760511398315
2.94853854179382 0.0605285167694092
2.94894361495972 0.0673487186431885
2.94925427436829 0.0690536499023438
2.95113444328308 0.0699062347412109
2.95115900039673 0.0716112852096558
2.95401072502136 0.0724637508392334
2.96269488334656 0.0733163356781006
2.96521067619324 0.0750212669372559
2.96564769744873 0.0852514505386353
2.96572613716125 0.0861040353775024
2.96853137016296 0.0869565010070801
2.96902394294739 0.0912190675735474
2.97015023231506 0.0937767028808594
2.97041130065918 0.0980392694473267
2.97629880905151 0.0988917350769043
2.98126339912415 0.100596785545349
2.98271012306213 0.103154301643372
3.01194357872009 0.104859352111816
3.01945900917053 0.105711817741394
3.01967525482178 0.109121918678284
3.23590207099915 0.109974384307861
3.31172204017639 0.110826969146729
3.31184506416321 0.115942001342773
3.31228256225586 0.116794586181641
3.44960951805115 0.117647051811218
3.45018625259399 0.119352102279663
3.45119285583496 0.12702476978302
3.45123600959778 0.128729701042175
3.45227360725403 0.129582285881042
3.45250058174133 0.13043475151062
3.46077227592468 0.131287336349487
3.46423172950745 0.13384485244751
3.46475887298584 0.134697318077087
3.48296451568604 0.135549902915955
3.48536205291748 0.136402368545532
3.48624491691589 0.137254953384399
3.4865734577179 0.139812469482422
3.48713231086731 0.147485017776489
3.49969387054443 0.148337602615356
3.59618997573853 0.150042653083801
3.62584829330444 0.150895118713379
3.62839031219482 0.151747703552246
3.62898707389832 0.154305219650269
3.62980222702026 0.155157685279846
3.63049578666687 0.161977767944336
3.63050007820129 0.162830352783203
3.63140845298767 0.163682818412781
3.68788862228394 0.164535403251648
3.6882381439209 0.167945384979248
3.68843626976013 0.17050302028656
3.70117974281311 0.171355485916138
3.70218181610107 0.172208070755005
3.70249652862549 0.17732310295105
3.79234504699707 0.178175568580627
3.83652353286743 0.179028153419495
3.83732867240906 0.183290719985962
3.83733797073364 0.18414318561554
4.0497899055481 0.184995770454407
4.05855464935303 0.185848236083984
4.06005811691284 0.186700820922852
4.0601601600647 0.187553286552429
4.33814287185669 0.188405752182007
4.34171915054321 0.189258337020874
4.34918642044067 0.190110802650452
4.41729784011841 0.190963387489319
4.42240381240845 0.191815853118896
4.4225640296936 0.192668437957764
4.42333269119263 0.193520903587341
4.42361640930176 0.195225954055786
4.42408847808838 0.196078419685364
4.42409944534302 0.196930885314941
4.46884536743164 0.197783470153809
4.47032165527344 0.199488520622253
4.47187232971191 0.200340986251831
4.47289848327637 0.204603552818298
4.48064231872559 0.205456137657166
4.48233127593994 0.208013653755188
4.48614454269409 0.208866119384766
4.51637172698975 0.209718704223633
4.52773714065552 0.21057116985321
4.55808925628662 0.211423635482788
4.55847835540771 0.213128685951233
4.55887269973755 0.214833736419678
4.57647323608398 0.215686321258545
4.57751131057739 0.219948887825012
4.57756042480469 0.221653938293457
4.6439151763916 0.223358869552612
4.64689636230469 0.225063920021057
4.94020128250122 0.225916504859924
4.95096302032471 0.226768970489502
4.9925651550293 0.22762143611908
4.99554491043091 0.228474020957947
4.99593830108643 0.231884002685547
4.99978160858154 0.232736587524414
5.00021553039551 0.237851619720459
5.16222143173218 0.238704204559326
5.17207050323486 0.239556670188904
5.35001802444458 0.240409255027771
5.42053461074829 0.241261720657349
5.44690179824829 0.242114186286926
5.45254325866699 0.242966771125793
5.45396995544434 0.247229337692261
5.45398950576782 0.248081803321838
5.50202226638794 0.249786853790283
5.59459781646729 0.25063943862915
5.89242839813232 0.252344369888306
5.89318418502808 0.253196954727173
5.89476728439331 0.25404942035675
5.90379333496094 0.254902005195618
6.00147438049316 0.255754470825195
6.36559915542603 0.256607055664062
6.38446998596191 0.258311986923218
6.53258085250854 0.260017037391663
6.54047012329102 0.26086950302124
6.5412278175354 0.265984654426575
6.57111072540283 0.269394755363464
6.57166957855225 0.270247220993042
6.57168531417847 0.271099805831909
6.57509803771973 0.271952271461487
6.57516098022461 0.272804737091064
6.88826036453247 0.273657321929932
6.89411687850952 0.274509787559509
6.89417457580566 0.275362253189087
6.8947377204895 0.276214838027954
6.90114450454712 0.277067422866821
6.9041576385498 0.278772354125977
6.90453100204468 0.280477404594421
6.90724754333496 0.281329870223999
6.90814542770386 0.282182455062866
6.90814542770386 0.283034920692444
6.91855239868164 0.283887505531311
6.92195558547974 0.284739971160889
6.934157371521 0.286445021629333
6.93530178070068 0.288150072097778
6.94164991378784 0.289002537727356
6.94169187545776 0.291560173034668
6.94494199752808 0.292412638664246
6.96442937850952 0.293265104293823
6.96516466140747 0.29411768913269
6.96526575088501 0.294970154762268
6.9662127494812 0.295822620391846
6.97380304336548 0.296675205230713
7.00702571868896 0.297527670860291
7.00762462615967 0.298380255699158
7.00907039642334 0.299232721328735
7.01691770553589 0.301790237426758
7.01693487167358 0.302642822265625
7.01838445663452 0.303495287895203
7.01839303970337 0.30434775352478
7.02130317687988 0.305200338363647
7.02138900756836 0.306052923202515
7.02187871932983 0.306905388832092
7.02191305160522 0.30775785446167
7.04538726806641 0.308610439300537
7.04713439941406 0.309462904930115
7.07217168807983 0.31116795539856
7.07469034194946 0.312020421028137
7.07523107528687 0.312873005867004
7.07523965835571 0.313725471496582
7.0788516998291 0.314578056335449
7.12222719192505 0.317988038063049
7.12867450714111 0.318840622901917
7.13079071044922 0.320545673370361
7.13246536254883 0.321398138999939
7.13276672363281 0.323103189468384
7.13326263427734 0.323955655097961
7.1486120223999 0.325660705566406
7.14968013763428 0.326513171195984
7.14970684051514 0.327365756034851
7.18428897857666 0.328218221664429
7.18566465377808 0.329070806503296
7.18641138076782 0.331628322601318
7.18659114837646 0.334185838699341
7.22518539428711 0.335890889167786
7.23346996307373 0.336743354797363
7.26383781433105 0.33759593963623
7.27372217178345 0.339300870895386
7.27375030517578 0.340153455734253
7.27481937408447 0.34100604057312
7.27921962738037 0.341858506202698
7.30267381668091 0.343563556671143
7.30469036102295 0.34441602230072
7.30489444732666 0.345268487930298
7.30714511871338 0.34782612323761
7.31491041183472 0.348678588867188
7.3149847984314 0.35123610496521
7.31721305847168 0.352088689804077
7.32715940475464 0.353793621063232
7.33394575119019 0.3546462059021
7.33407688140869 0.357203722000122
7.33776998519897 0.358056306838989
7.33873653411865 0.358908772468567
7.33875513076782 0.360613822937012
7.3398904800415 0.361466288566589
7.34036922454834 0.362318873405457
7.34152364730835 0.363171339035034
7.34300756454468 0.366581439971924
7.3441162109375 0.368286371231079
7.35033130645752 0.369138956069946
7.37367916107178 0.369991540908813
7.37416219711304 0.370844006538391
7.40017938613892 0.371696472167969
7.40039873123169 0.372549057006836
7.40144824981689 0.373401522636414
7.40145778656006 0.375106573104858
7.44197607040405 0.375959038734436
7.45572948455811 0.376811623573303
7.46067142486572 0.377664089202881
7.46171855926514 0.378516674041748
7.4617280960083 0.380221605300903
7.48547792434692 0.381074190139771
7.48614168167114 0.381926655769348
7.49592638015747 0.38448429107666
7.50187206268311 0.385336756706238
7.50191116333008 0.387041807174683
7.50281286239624 0.38789427280426
7.54609870910645 0.389599323272705
7.55158805847168 0.390451788902283
7.55551338195801 0.393009424209595
7.55555295944214 0.39471435546875
7.56615829467773 0.395566940307617
7.56847238540649 0.397271990776062
7.58602428436279 0.39812445640564
7.59569025039673 0.398977041244507
7.59792232513428 0.399829506874084
7.59839487075806 0.403239488601685
7.65812683105469 0.404092073440552
7.67167663574219 0.404944658279419
7.67512273788452 0.405797123908997
7.67608737945557 0.407502174377441
7.67775058746338 0.408354640007019
7.67780160903931 0.410912156105042
7.69366836547852 0.411764740943909
7.69750595092773 0.412617206573486
7.70798587799072 0.413469791412354
7.70991134643555 0.415174722671509
7.74328231811523 0.416027307510376
7.74427461624146 0.416879773139954
7.74662637710571 0.417732238769531
7.74693965911865 0.420289874076843
7.74723243713379 0.421994924545288
7.77040672302246 0.422847390174866
7.7831015586853 0.423699855804443
7.79211235046387 0.425404906272888
7.79310703277588 0.426257491111755
7.79310703277588 0.427109956741333
7.79351949691772 0.428815007209778
7.91268491744995 0.430520057678223
7.91538381576538 0.432225108146667
7.91623973846436 0.436487674713135
7.94209432601929 0.437340140342712
7.94286346435547 0.439045190811157
7.94347906112671 0.440750241279602
7.94585371017456 0.442455291748047
7.94597482681274 0.443307757377625
7.95940446853638 0.445012807846069
7.95993375778198 0.445865273475647
7.96003341674805 0.447570323944092
7.96060752868652 0.448422908782959
7.96091651916504 0.450127840042114
7.96530151367188 0.450980424880981
7.96542263031006 0.451832890510559
7.96913814544678 0.452685356140137
7.97020053863525 0.454390525817871
7.97160577774048 0.458652973175049
7.97470664978027 0.459505558013916
7.97590303421021 0.460358023643494
7.9772777557373 0.465473175048828
7.98087072372437 0.468883275985718
7.98107004165649 0.469735741615295
7.98269081115723 0.47144079208374
7.98327922821045 0.477408409118652
7.98467779159546 0.479965925216675
7.98507785797119 0.48508095741272
7.98602199554443 0.485933542251587
7.98612213134766 0.487638473510742
7.98728895187378 0.488491058349609
7.98741054534912 0.491048574447632
7.98870038986206 0.492753624916077
7.98882293701172 0.493606090545654
7.98992347717285 0.494458675384521
7.99015712738037 0.496163725852966
7.99155855178833 0.497016191482544
7.9916467666626 0.497868776321411
7.99272680282593 0.498721241950989
7.99290466308594 0.501278758049011
7.99389553070068 0.502131223678589
7.99396181106567 0.503836393356323
7.99508619308472 0.504688858985901
7.99576568603516 0.507246375083923
7.99667882919312 0.508098840713501
7.99816083908081 0.515771508216858
7.99886274337769 0.52855920791626
7.99904108047485 0.532821893692017
7.99973249435425 0.557544708251953
8.0005464553833 0.575447559356689
8.00075817108154 0.623188376426697
8.00113677978516 0.685421943664551
8.00228595733643 0.741688013076782
8.00264263153076 0.783461213111877
8.00275421142578 0.79283881187439
8.00330066680908 0.861892580986023
8.00366973876953 0.891730546951294
8.00400447845459 0.919011116027832
8.00433921813965 0.953964233398438
8.00443935394287 0.960784316062927
8.00500869750977 0.999147415161133
8.5 1
};
\addlegendentry{\gls{msr}}

\addplot [very thick, DARKGREY, mark=*, mark repeat=60, mark phase=40]
table {%
0.898911833763123 0
1.51619553565979 0.00152206420898438
1.71427202224731 0.00304412841796875
2.73237276077271 0.00456619262695312
2.93101477622986 0.00761032104492188
3.01191163063049 0.00913238525390625
3.55738711357117 0.0106544494628906
3.73190855979919 0.012176513671875
5.38479089736938 0.0152206420898438
5.47402811050415 0.0167428255081177
5.99239015579224 0.0197869539260864
5.99287891387939 0.0213090181350708
5.99509477615356 0.0228310823440552
5.99598741531372 0.0243531465530396
5.996253490448 0.0273972749710083
5.9962682723999 0.0304414033889771
6.01512622833252 0.0319634675979614
6.01515531539917 0.0334855318069458
6.01669836044312 0.0350075960159302
6.01719570159912 0.0426179170608521
6.07254028320312 0.0441399812698364
6.28851842880249 0.0456620454788208
6.35930919647217 0.0471842288970947
6.36045360565186 0.056316614151001
6.36056756973267 0.0593607425689697
6.36144495010376 0.0608828067779541
6.36158227920532 0.0654489994049072
6.36200952529907 0.0669710636138916
6.36205577850342 0.0700151920318604
6.36418533325195 0.0715372562408447
6.36497974395752 0.0730593204498291
6.36505460739136 0.0806697607040405
6.61412954330444 0.0821918249130249
6.64430856704712 0.0837138891220093
6.64467763900757 0.086758017539978
6.64516639709473 0.0958904027938843
6.64543104171753 0.105022788047791
6.71213293075562 0.106544852256775
6.71216487884521 0.108066916465759
6.71272039413452 0.109589099884033
6.71295356750488 0.117199420928955
6.72217178344727 0.118721485137939
6.7223973274231 0.120243549346924
6.72359466552734 0.121765613555908
6.72392177581787 0.132420063018799
6.72555732727051 0.135464191436768
6.72576379776001 0.141552448272705
6.79436874389648 0.143074631690979
6.79473304748535 0.147640824317932
6.79477405548096 0.150684952735901
6.79671669006348 0.158295273780823
6.79693031311035 0.16438353061676
6.79888677597046 0.165905594825745
6.79910945892334 0.168949723243713
6.79939365386963 0.176560163497925
6.80443000793457 0.178082227706909
6.80447053909302 0.179604291915894
6.80586576461792 0.181126356124878
6.80631446838379 0.1887366771698
6.88344621658325 0.190258741378784
6.88494920730591 0.191780805587769
6.88521528244019 0.194824934005737
6.88576412200928 0.196346998214722
6.88585424423218 0.197869062423706
6.88643598556519 0.19939112663269
6.88733339309692 0.213089823722839
6.88794946670532 0.214611887931824
6.88807439804077 0.220700144767761
6.94335412979126 0.222222208976746
6.94448804855347 0.22374427318573
6.944983959198 0.225266337394714
6.94557189941406 0.229832530021667
6.97627782821655 0.231354594230652
6.98135089874268 0.232876777648926
6.98175048828125 0.235920906066895
6.98354244232178 0.237442970275879
6.98365306854248 0.242009162902832
7.05254507064819 0.243531227111816
7.05283975601196 0.249619483947754
7.05286598205566 0.251141548156738
7.06433057785034 0.252663612365723
7.10555744171143 0.254185676574707
7.12754154205322 0.255707740783691
7.14339256286621 0.257229804992676
7.15239715576172 0.25875186920166
7.15255737304688 0.264840126037598
7.1527624130249 0.269406318664551
7.1835789680481 0.270928502082825
7.18478345870972 0.272450566291809
7.18482828140259 0.273972630500793
7.18732881546021 0.275494694709778
7.18822002410889 0.280060887336731
7.18825578689575 0.2831050157547
7.18930006027222 0.284627079963684
7.19121837615967 0.286149144172668
7.19487762451172 0.287671208381653
7.19492244720459 0.289193272590637
7.19570732116699 0.292237401008606
7.19618511199951 0.298325777053833
7.19876623153687 0.299847841262817
7.19919919967651 0.307458162307739
7.2024507522583 0.308980226516724
7.2039966583252 0.312024354934692
7.20443058013916 0.313546419143677
7.21159839630127 0.315068483352661
7.21285772323608 0.316590547561646
7.21314764022827 0.325722932815552
7.21487903594971 0.327244997024536
7.21564102172852 0.328767061233521
7.21581315994263 0.333333373069763
7.27083873748779 0.334855437278748
7.27805662155151 0.336377501487732
7.28034591674805 0.339421629905701
7.28288555145264 0.340943694114685
7.28293180465698 0.342465758323669
7.28371715545654 0.345509886741638
7.31301832199097 0.347031950950623
7.31306505203247 0.348554015159607
7.31454420089722 0.350076079368591
7.31477975845337 0.35312020778656
7.31491899490356 0.356164455413818
7.32467222213745 0.359208583831787
7.32584095001221 0.360730648040771
7.32643938064575 0.362252712249756
7.32699060440063 0.368340969085693
7.32844066619873 0.369863033294678
7.33063936233521 0.371385097503662
7.33133268356323 0.385083675384521
7.33136987686157 0.389649868011475
7.34784746170044 0.391171932220459
7.3478946685791 0.392693996429443
7.34859991073608 0.394216060638428
7.34934329986572 0.400304436683655
7.34976673126221 0.407914757728577
7.35004901885986 0.410958886146545
7.35017108917236 0.414003014564514
7.35638761520386 0.415525078773499
7.35925436019897 0.418569326400757
7.35928297042847 0.420091390609741
7.36022663116455 0.421613454818726
7.36052846908569 0.424657583236694
7.36084938049316 0.430745840072632
7.36113262176514 0.436834096908569
7.36115121841431 0.438356161117554
7.36893701553345 0.439878225326538
7.36968421936035 0.456620931625366
7.37004375457764 0.45814311504364
7.37113237380981 0.459665179252625
7.37119817733765 0.461187243461609
7.37245750427246 0.462709307670593
7.3761043548584 0.464231371879578
7.37643623352051 0.474885821342468
7.38710498809814 0.476407885551453
7.41054344177246 0.477929949760437
7.42093706130981 0.479452013969421
7.42147445678711 0.48249626159668
7.42232847213745 0.502283096313477
7.42267370223999 0.505327224731445
7.42270278930664 0.509893417358398
7.42364311218262 0.511415481567383
7.42566919326782 0.512937545776367
7.42582321166992 0.515981674194336
7.42693901062012 0.51750373840332
7.4274754524231 0.519025802612305
7.42751407623291 0.522069931030273
7.42852306365967 0.523592114448547
7.49926519393921 0.525114178657532
7.50085258483887 0.537290692329407
7.50129365921021 0.538812756538391
7.50138139724731 0.543379068374634
7.50197982788086 0.546423196792603
7.5022931098938 0.55251145362854
7.50256824493408 0.555555582046509
7.50264644622803 0.558599710464478
7.50330352783203 0.560121774673462
7.50451993942261 0.561643838882446
7.50492191314697 0.563165903091431
7.50931978225708 0.564687967300415
7.51070499420166 0.567732095718384
7.51109266281128 0.57686448097229
7.51157999038696 0.581430673599243
7.51161909103394 0.584474802017212
7.52177619934082 0.585996985435486
7.54462051391602 0.58751904964447
7.55869674682617 0.589041113853455
7.55874633789062 0.590563178062439
7.57774353027344 0.593607306480408
7.57955121994019 0.595129370689392
7.58022499084473 0.596651434898376
7.64738464355469 0.598173499107361
7.66259241104126 0.599695563316345
7.66432237625122 0.602739810943604
7.66481351852417 0.604261875152588
7.66486501693726 0.605783939361572
7.66542816162109 0.607306003570557
7.66543817520142 0.608828067779541
7.66609287261963 0.610350131988525
7.66646194458008 0.613394260406494
7.66663599014282 0.617960453033447
7.66664600372314 0.619482517242432
7.6688380241394 0.621004581451416
7.66955518722534 0.627092838287354
7.66999578475952 0.631659030914307
7.67124605178833 0.633181095123291
7.70492362976074 0.634703159332275
7.70643377304077 0.637747287750244
7.70658874511719 0.645357608795166
7.70767545700073 0.64687967300415
7.73014545440674 0.648401737213135
7.74230098724365 0.649923920631409
7.76928186416626 0.651445984840393
7.81690788269043 0.652968049049377
7.81755781173706 0.654490113258362
7.8614559173584 0.657534241676331
7.8618369102478 0.659056305885315
7.92915058135986 0.662100434303284
7.93385982513428 0.665144681930542
7.93413639068604 0.669710874557495
7.93414735794067 0.671232938766479
7.93507957458496 0.672755002975464
7.93711996078491 0.674277067184448
7.93736171722412 0.678843259811401
7.93784475326538 0.68188738822937
7.93873405456543 0.692541837692261
7.93894243240356 0.694063901901245
7.94006204605103 0.695585966110229
7.9402379989624 0.697108030319214
7.94380855560303 0.698630094528198
7.94419384002686 0.701674222946167
7.9446439743042 0.704718351364136
7.94484186172485 0.707762479782104
7.94599676132202 0.709284543991089
7.94609594345093 0.712328791618347
7.94738292694092 0.713850855827332
7.94816398620605 0.721461176872253
7.94880247116089 0.722983241081238
7.94895648956299 0.730593681335449
7.95414352416992 0.732115745544434
7.95516872406006 0.735159873962402
7.9551796913147 0.74124813079834
7.95598459243774 0.742770195007324
7.95612812042236 0.748858451843262
7.95650291442871 0.75190258026123
7.95773839950562 0.753424644470215
7.95794773101807 0.754946708679199
7.96065139770508 0.756468772888184
7.96078395843506 0.757990837097168
7.96225261688232 0.759512901306152
7.96250629425049 0.762557029724121
7.96288204193115 0.76560115814209
7.96295928955078 0.767123222351074
7.9653787612915 0.768645286560059
7.96707010269165 0.770167350769043
7.97342157363892 0.771689414978027
7.97345495223999 0.773211479187012
7.97531604766846 0.774733662605286
7.97569274902344 0.777777791023254
7.97584772109985 0.785388112068176
7.97732210159302 0.788432240486145
7.98071575164795 0.789954423904419
7.98155879974365 0.792998552322388
7.98243522644043 0.794520616531372
7.98270177841187 0.797564744949341
7.98281288146973 0.80060887336731
7.98344564437866 0.803653001785278
7.98392295837402 0.817351579666138
7.98435592651367 0.820395708084106
7.98611116409302 0.831050157546997
7.9864444732666 0.83561635017395
7.98661088943481 0.837138533592224
7.98810005187988 0.838660597801208
7.98827791213989 0.843226790428162
7.98904514312744 0.844748854637146
7.98934507369995 0.847792983055115
7.99023485183716 0.849315047264099
7.99049043655396 0.856925487518311
7.99107980728149 0.858447551727295
7.99130296707153 0.863013744354248
7.99155855178833 0.866057872772217
7.99292707443237 0.867579936981201
7.99361705780029 0.87062406539917
7.99361705780029 0.872146129608154
7.99430751800537 0.875190258026123
7.9967794418335 0.878234386444092
7.99733638763428 0.881278514862061
7.99850606918335 0.885844707489014
7.99938678741455 0.893455028533936
7.99991035461426 0.896499156951904
8.00036811828613 0.902587532997131
8.00071334838867 0.907153725624084
8.00137138366699 0.910197854042053
8.0019063949585 0.936073064804077
8.00401496887207 0.980213165283203
8.00448417663574 0.981735229492188
8.00504207611084 0.995433807373047
8.00558948516846 0.998477935791016
8.5 1
};
\addlegendentry{\gls{ba}}

\addplot [very thick, UNIPDRED, mark=triangle*, mark repeat=60, mark phase=40]
table {%
0.496104836463928 0
0.792945146560669 0.00118625164031982
2.413494348526 0.00355875492095947
3.00969815254211 0.0047450065612793
3.40677261352539 0.00593113899230957
3.59254956245422 0.00711739063262939
3.63197016716003 0.00830364227294922
3.67936635017395 0.0106761455535889
4.8841814994812 0.0130486488342285
5.01563358306885 0.0142349004745483
5.01893711090088 0.0154211521148682
5.01923656463623 0.0189797878265381
5.01965761184692 0.0213522911071777
5.02117395401001 0.0225385427474976
5.05732822418213 0.0237247943878174
5.0972900390625 0.026097297668457
5.09775447845459 0.0284698009490967
5.0985312461853 0.0320284366607666
5.10007429122925 0.0332146883010864
5.10019063949585 0.0355871915817261
5.10471200942993 0.0367734432220459
5.10478544235229 0.0379596948623657
5.10802841186523 0.0391459465026855
5.10804080963135 0.0403321981430054
5.31071138381958 0.0415183305740356
5.36052179336548 0.0438908338546753
5.36055850982666 0.0450770854949951
5.36222362518311 0.0474495887756348
5.36458206176758 0.065243124961853
5.51200914382935 0.0664293766021729
5.57987928390503 0.0676156282424927
5.62029123306274 0.0688018798828125
5.62085008621216 0.0699881315231323
5.6218409538269 0.0711743831634521
5.62186765670776 0.072360634803772
5.62281179428101 0.0735468864440918
5.62752246856689 0.0747331380844116
5.62807750701904 0.0759193897247314
5.6282753944397 0.081850528717041
5.62867259979248 0.0901541709899902
5.6290168762207 0.0937129259109497
5.62963390350342 0.0960854291915894
5.62965393066406 0.103202819824219
5.63159084320068 0.104389071464539
6.19212341308594 0.105575323104858
6.27782392501831 0.106761574745178
6.2780237197876 0.109134078025818
6.27879428863525 0.110320329666138
6.27892208099365 0.113878965377808
6.27944898605347 0.115065217018127
6.80820655822754 0.116251468658447
6.82517576217651 0.117437720298767
6.82524919509888 0.119810223579407
6.82592725753784 0.122182726860046
6.82596778869629 0.123368978500366
6.82649183273315 0.125741362571716
6.8265814781189 0.129300117492676
6.82704830169678 0.131672620773315
6.82705640792847 0.132858872413635
6.9669394493103 0.134045124053955
7.03652715682983 0.135231256484985
7.06781816482544 0.136417508125305
7.08099937438965 0.137603759765625
7.12571907043457 0.138790011405945
7.23871421813965 0.139976263046265
7.24390268325806 0.141162514686584
7.24589586257935 0.142348766326904
7.24682855606079 0.143535017967224
7.24687433242798 0.144721269607544
7.24821043014526 0.145907521247864
7.24861335754395 0.147093772888184
7.27235841751099 0.148279905319214
7.28391075134277 0.149466156959534
7.28453016281128 0.150652408599854
7.28586149215698 0.151838660240173
7.28650903701782 0.153024911880493
7.28698110580444 0.155397415161133
7.2874059677124 0.156583666801453
7.29303503036499 0.157769918441772
7.29329442977905 0.158956050872803
7.2977352142334 0.160142302513123
7.29849624633789 0.161328554153442
7.29854249954224 0.162514805793762
7.29922914505005 0.164887309074402
7.29989719390869 0.167259812355042
7.31374502182007 0.168446063995361
7.31376361846924 0.169632315635681
7.31447219848633 0.170818567276001
7.314537525177 0.174377202987671
7.3249249458313 0.175563454627991
7.33001279830933 0.176749706268311
7.33910226821899 0.17793595790863
7.3440408706665 0.17912220954895
7.34673833847046 0.18030846118927
7.34716129302979 0.18149471282959
7.34924936294556 0.18268084526062
7.34978580474854 0.18386709690094
7.35611438751221 0.18623960018158
7.3588490486145 0.187425851821899
7.35982084274292 0.188612103462219
7.36206722259521 0.189798355102539
7.36209535598755 0.190984606742859
7.36450338363647 0.194543361663818
7.36464500427246 0.195729494094849
7.36526870727539 0.196915745735168
7.36555242538452 0.199288249015808
7.36668634414673 0.201660752296448
7.37866449356079 0.204033255577087
7.38023900985718 0.205219507217407
7.3803334236145 0.207591891288757
7.38111209869385 0.208778142929077
7.38196611404419 0.214709401130676
7.38420724868774 0.215895652770996
7.38438749313354 0.218268156051636
7.38674402236938 0.221826791763306
7.39285087585449 0.223013043403625
7.39502191543579 0.224199295043945
7.39514589309692 0.225385546684265
7.39735651016235 0.226571798324585
7.39740419387817 0.227758049964905
7.39855766296387 0.228944301605225
7.39873886108398 0.230130434036255
7.40102815628052 0.231316685676575
7.40654706954956 0.232502937316895
7.4083251953125 0.233689188957214
7.42284679412842 0.236061692237854
7.42917680740356 0.237247943878174
7.43052339553833 0.238434195518494
7.43088865280151 0.240806579589844
7.43092727661133 0.241992831230164
7.43201446533203 0.243179082870483
7.43451690673828 0.244365334510803
7.43535470962524 0.246737837791443
7.43771457672119 0.249110341072083
7.43787813186646 0.251482725143433
7.44568252563477 0.253855228424072
7.45406436920166 0.255041480064392
7.45498418807983 0.256227731704712
7.48908090591431 0.259786486625671
7.48917865753174 0.262158989906311
7.49041986465454 0.263345241546631
7.49044895172119 0.266903877258301
7.49122095108032 0.268090128898621
7.49456596374512 0.27046263217926
7.50485324859619 0.27164888381958
7.50506925582886 0.27758002281189
7.54712009429932 0.278766274452209
7.54716968536377 0.279952526092529
7.54843044281006 0.281138777732849
7.54847002029419 0.283511281013489
7.54893636703491 0.284697532653809
7.55761241912842 0.285883784294128
7.55840826034546 0.289442539215088
7.56083726882935 0.291814923286438
7.63076496124268 0.293001174926758
7.63105916976929 0.296559929847717
7.63320016860962 0.297746181488037
7.63329124450684 0.301304817199707
7.65491008758545 0.302491068840027
7.6550121307373 0.304863572120667
7.65585994720459 0.306049823760986
7.65853548049927 0.307236075401306
7.65875005722046 0.310794830322266
7.70025253295898 0.311980962753296
7.70176076889038 0.313167333602905
7.7021222114563 0.314353466033936
7.70291805267334 0.315539717674255
7.70306253433228 0.320284724235535
7.79486322402954 0.321470975875854
7.79998922348022 0.322657108306885
7.81506729125977 0.325029611587524
7.81689786911011 0.326215863227844
7.81756830215454 0.329774618148804
7.81845188140869 0.330960869789124
7.8205714225769 0.332147121429443
7.82686185836792 0.333333373069763
7.82757759094238 0.334519624710083
7.8276309967041 0.335705757141113
7.82912540435791 0.336892008781433
7.82929611206055 0.340450763702393
7.83950614929199 0.341637015342712
7.87191677093506 0.342823266983032
7.87425994873047 0.346381902694702
7.87448644638062 0.348754405975342
7.87698936462402 0.349940657615662
7.89731121063232 0.354685664176941
7.89991903305054 0.357058167457581
7.90107154846191 0.3582444190979
7.90148448944092 0.359430551528931
7.90252876281738 0.36061680316925
7.91486263275146 0.36180305480957
7.91521120071411 0.36536180973053
7.91685962677002 0.37010669708252
7.91749286651611 0.386714100837708
7.9198112487793 0.393831491470337
7.92504549026489 0.395017862319946
7.92541790008545 0.397390246391296
7.92547225952148 0.398576498031616
7.92625999450684 0.399762749671936
7.92644596099854 0.405694007873535
7.92736577987671 0.408066391944885
7.92771577835083 0.411625146865845
7.92888736724854 0.412811398506165
7.9329195022583 0.413997650146484
7.93967819213867 0.417556285858154
7.94007301330566 0.419928789138794
7.94173145294189 0.421115040779114
7.94198417663574 0.423487544059753
7.94462251663208 0.425860047340393
7.94711875915527 0.428232431411743
7.94770240783691 0.431791186332703
7.94843912124634 0.432977437973022
7.94898986816406 0.435349941253662
7.95119142532349 0.436536192893982
7.95254611968994 0.438908696174622
7.95748472213745 0.440094947814941
7.95804738998413 0.444839835166931
7.95990085601807 0.446026086807251
7.95994520187378 0.447212338447571
7.96057415008545 0.448398590087891
7.96241807937622 0.44958484172821
7.96251726150513 0.451957225799561
7.96326875686646 0.45314359664917
7.96447277069092 0.4543297290802
7.96494817733765 0.45670223236084
7.96636295318604 0.45788848400116
7.96637392044067 0.459074735641479
7.96830892562866 0.460260987281799
7.96920442581177 0.461447238922119
7.96965837478638 0.465005874633789
7.97139549255371 0.467378377914429
7.97172784805298 0.470937132835388
7.97227048873901 0.472123384475708
7.97230339050293 0.475682020187378
7.97314500808716 0.478054523468018
7.97328901290894 0.481613278388977
7.97665691375732 0.482799530029297
7.97672319412231 0.483985781669617
7.97750997543335 0.485172033309937
7.97752141952515 0.486358284950256
7.97923994064331 0.489916920661926
7.97952842712402 0.493475675582886
7.98053789138794 0.494661927223206
7.980881690979 0.497034430503845
7.98196935653687 0.512455463409424
7.98242425918579 0.514827966690063
7.98430061340332 0.517200469970703
7.98488903045654 0.524317979812622
7.98511123657227 0.525504112243652
7.98594427108765 0.526690363883972
7.98684453964233 0.532621622085571
7.98722219467163 0.534994125366211
7.98918962478638 0.569395065307617
7.98992347717285 0.577698707580566
7.99074697494507 0.597864747047424
7.99096870422363 0.601423501968384
7.99106931686401 0.603796005249023
7.99241542816162 0.610913395881653
7.99284887313843 0.625148296356201
7.99390649795532 0.63463819026947
7.99419593811035 0.648873090744019
7.99447393417358 0.651245594024658
7.99494171142578 0.661921739578247
7.99529790878296 0.667852878570557
7.99799346923828 0.72241997718811
8.00110340118408 0.845788836479187
8.00148296356201 0.874258518218994
8.00168323516846 0.90510082244873
8.00206279754639 0.930011868476868
8.0023078918457 0.951364159584045
8.00249767303467 0.973902702331543
8.00272083282471 0.977461457252502
8.00346851348877 0.988137602806091
8.00391483306885 0.99406886100769
8.00484085083008 0.995254993438721
8.00517559051514 0.99881374835968
8.5 1
};
\addlegendentry{\gls{mrba}}

\addplot[mark=none, black, dashed] coordinates {(0, 0.25) (8.5, 0.25)};

\draw[very thick, dashed] (8,0) -- (8,1);

\end{axis}

\end{tikzpicture}

%% file: Figures/Sim_results/E2E_first_quartile_throughput.tex
\begin{tikzpicture}
\pgfplotsset{every tick label/.append style={font=\small}}

\definecolor{UNIPDRED}{RGB}{155,0,20}
\definecolor{LIGHT_GREY}{RGB}{189,195,199}
\definecolor{COMPLEMENTARY}{RGB}{0,153,153}
\definecolor{DARKGREY}{RGB}{55,65,64}
\definecolor{SAND}{RGB}{180,160,135}

\begin{axis}[
width=\fwidth,
height=\fheight,
at={(0\fwidth,0\fheight)},
scale only axis,
legend style={
    /tikz/every even column/.append style={column sep=0.2cm},
    at={(0.5,0.05)}, 
    anchor=south, 
    draw=white!80!black, 
    font=\scriptsize
    },
legend columns=2,
xlabel style={font=\footnotesize},
xlabel={Packet size [B]},
xtick={100, 200, 300, 400, 500},
xmajorgrids,
xmin=50, xmax=500,
xtick style={color=white!15!black},
ylabel shift = -1 pt,
ylabel style={font=\footnotesize},
ylabel={E2E throughput [Mbps]},
ymajorgrids,
ymin=0, ymax=8,
ytick style={color=white!15!black},
ytick={2, 4, 6, 8}
]

\addplot [ultra thick, SAND]
table {%
50 3.46753779256683
100 3.28747817169726
200 3.94575183087198
500 4.00105152059087
};
\addlegendentry{Distr}

\addplot [ultra thick, COMPLEMENTARY, mark=square*]
table {%
50 3.99162842045843
100 5.50202239928505
200 4.93673478556716
500 3.94631392811716
};
\addlegendentry{\gls{msr}}

\addplot [ultra thick, DARKGREY, mark=*]
table {%
50 3.98958406793754
100 7.05283992388594
200 7.62966951310821
500 3.30460047092302
};
\addlegendentry{\gls{ba}}

\addplot [ultra thick, UNIPDRED, mark=triangle*]
table {%
50 3.99308883129582
100 7.43779170402193
200 7.17359363179047
500 6.47136135389916
};
\addlegendentry{\gls{mrba}}
\end{axis}

\end{tikzpicture}

%% file: Figures/Sim_results/E2E_third_quartile_throughput.tex
\begin{tikzpicture}
\pgfplotsset{every tick label/.append style={font=\small}}

\definecolor{UNIPDRED}{RGB}{155,0,20}
\definecolor{LIGHT_GREY}{RGB}{189,195,199}
\definecolor{COMPLEMENTARY}{RGB}{0,153,153}
\definecolor{DARKGREY}{RGB}{55,65,64}
\definecolor{SAND}{RGB}{180,160,135}

\begin{axis}[
width=\fwidth,
height=\fheight,
at={(0\fwidth,0\fheight)},
scale only axis,
legend style={
    /tikz/every even column/.append style={column sep=0.2cm},
    at={(0.5,0.05)}, 
    anchor=south, 
    draw=white!80!black, 
    font=\scriptsize
    },
legend columns=2,
xlabel style={font=\footnotesize},
xlabel={Packet size [B]},
xtick={0, 100, 200, 300, 400, 500},
xmajorgrids,
xmin=50, xmax=500,
xtick style={color=white!15!black},
ylabel shift = -1 pt,
ylabel style={font=\footnotesize},
ylabel={E2E throughput [Mbps]},
ymajorgrids,
ymin=0, ymax=28,
ytick style={color=white!15!black},
ytick={5, 10, 15, 20, 25}
]

\addplot [ultra thick, SAND]
table {%
50 4.00031767071162
100 8.001070377308
200 15.6118996910022
500 18.8086383406533
};
\addlegendentry{Distr}

\addplot [ultra thick, COMPLEMENTARY, mark=square*]
table {%
50 4.0010914013884
100 8.00239761467529
200 16.0017947699339
500 22.4075412739043
};
\addlegendentry{\gls{msr}}

\addplot [ultra thick, DARKGREY, mark=*]
table {%
50 4.000535188654
100 7.95612815316127
200 11.7146609875454
500 9.95776415720082
};
\addlegendentry{\gls{ba}}

\addplot [ultra thick, UNIPDRED, mark=triangle*]
table {%
50 4.00047952236148
100 7.99908013257237
200 15.5613083831987
500 25.9400756180665
};
\addlegendentry{\gls{mrba}}
\end{axis}

\end{tikzpicture}

%% file: Figures/Sim_results/E2E_third_quartile_delay.tex
\begin{tikzpicture}

\pgfplotsset{every tick label/.append style={font=\small}}

\definecolor{UNIPDRED}{RGB}{155,0,20}
\definecolor{LIGHT_GREY}{RGB}{189,195,199}
\definecolor{COMPLEMENTARY}{RGB}{0,153,153}
\definecolor{DARKGREY}{RGB}{55,65,64}
\definecolor{SAND}{RGB}{180,160,135}

\begin{axis}[
width=\fwidth,
height=\fheight,
at={(0\fwidth,0\fheight)},
scale only axis,
legend style={
    /tikz/every even column/.append style={column sep=0.2cm},
    at={(0.5,0.9)}, 
    anchor=south, 
    draw=white!80!black, 
    font=\scriptsize
    },
legend columns=2,
xlabel style={font=\footnotesize},
xlabel={Packet size [B]},
xtick={100, 200, 300, 400, 500},
xmajorgrids,
xmin=50, xmax=500,
xtick style={color=white!15!black},
ylabel shift = -1 pt,
ylabel style={font=\footnotesize},
ylabel={E2E delay [ms]},
ymajorgrids,
ymin=0, ymax=2300,
ytick style={color=white!15!black},
ytick={500, 1000, 1500, 2000}
]

\addplot [ultra thick, SAND]
table {%
50 507.177999
100 728.391519
200 773.416639
500 1500.108319
};
\addlegendentry{Distr}

\addplot [ultra thick, COMPLEMENTARY, mark=square*]
table {%
50 18.587359
100 266.508319
200 569.174879
500 1217.429119
};
\addlegendentry{\gls{msr}}

\addplot [ultra thick, DARKGREY, mark=*]
table {%
50 19.912479
100 142.562399
200 910.787359
500 2171.387359
};
\addlegendentry{\gls{ba}}

\addplot [ultra thick, UNIPDRED, mark=triangle*]
table {%
50 18.341599
100 94.579039
200 779.713519
500 1091.820799
};
\addlegendentry{\gls{mrba}}
\end{axis}

\end{tikzpicture}

%% file: Figures/Sim_results/Median_BSR_UE.tex
\begin{tikzpicture}
    
    \pgfplotsset{every tick label/.append style={font=\small}}
    
    \definecolor{UNIPDRED}{RGB}{155,0,20}
    \definecolor{LIGHT_GREY}{RGB}{189,195,199}
    \definecolor{COMPLEMENTARY}{RGB}{0,153,153}
    \definecolor{DARKGREY}{RGB}{55,65,64}
    \definecolor{SAND}{RGB}{180,160,135}
    
    \begin{axis}[
        width=\fwidth,
        height=\fheight,
        at={(0\fwidth,0\fheight)},
        scale only axis,
        legend image post style={mark indices={}},
        legend style={
            /tikz/every even column/.append style={column sep=0.2cm},
            at={(0.55,0.95)}, 
            anchor=south, 
            draw=white!80!black, 
            font=\tiny
            },
        legend columns=2,
        xlabel style={font=\footnotesize},
        xlabel={Packet size [B]},
        xtick={100, 200, 300, 400, 500},
        xmajorgrids,
        xmin=50, xmax=500,
        xtick style={color=white!15!black},
        ylabel shift = -1 pt,
        ylabel style={font=\footnotesize},
        ylabel={\gls{rlc} buffer [B]},
        ymajorgrids,
        ymin=0, ymax=20500,
        ytick style={color=white!15!black},
        ytick={0,5000,10000,15000,20000},
        ]
    
    \addplot [ultra thick, SAND]
    table {%
    50 640
    100 1040
    200 1610
    500 1600
    };
    
    \addplot [ultra thick, COMPLEMENTARY, mark=square*]
    table {%
    50 640
    100 1040
    200 1610
    500 1080
    };   
    
    \addplot [ultra thick, DARKGREY, mark=*]
    table {%
    50 640
    100 1700
    200 20218
    500 19351
    };
    
    \addplot [ultra thick, UNIPDRED, mark=triangle*]
    table {%
    50 640
    100 1127
    200 3899
    500 5100
    };
    
    \end{axis}
    
    \end{tikzpicture}
    

%% file: Figures/Sim_results/Median_BSR_node.tex
\begin{tikzpicture}

\definecolor{UNIPDRED}{RGB}{155,0,20}
\definecolor{LIGHT_GREY}{RGB}{189,195,199}
\definecolor{COMPLEMENTARY}{RGB}{0,153,153}
\definecolor{DARKGREY}{RGB}{55,65,64}
\definecolor{SAND}{RGB}{180,160,135}

\pgfplotsset{every tick label/.append style={font=\small}}

\begin{axis}[
    width=\fwidth,
    height=\fheight,
    at={(0\fwidth,0\fheight)},
    scale only axis,
    legend image post style={mark indices={}},
    legend style={
        /tikz/every even column/.append style={column sep=0.2cm},
        at={(0.5,1.12)}, 
        anchor=south, 
        draw=white!80!black, 
        font=\scriptsize
        },
    legend columns=2,
    xlabel style={font=\footnotesize},
    xlabel={Packet size [B]},
    xtick={100, 200, 300, 400, 500},
    xmajorgrids,
    xmin=50, xmax=500,
    xtick style={color=white!15!black},
    ylabel shift = -1 pt,
    ylabel style={font=\footnotesize},
    ylabel={\gls{rlc} buffer [B]},
    ymajorgrids,
    ymin=0, ymax=49538443,
    ytick style={color=white!15!black},
    ytick={0,10000000,20000000,30000000,40000000,50000000},
    ]

\addplot [ultra thick, SAND]
table {%
50 738
100 1138
200 1675000
500 27915900
};
\addlegendentry{Distr}

\addplot [ultra thick, COMPLEMENTARY, mark=square*]
table {%
50 738
100 1140
200 9926
500 26710450
};
\addlegendentry{\gls{msr}}

\addplot [ultra thick, DARKGREY, mark=*]
table {%
50 748
100 62292
200 8338710
500 47179500
};
\addlegendentry{\gls{ba}}

\addplot [ultra thick, UNIPDRED, mark=triangle*]
table {%
50 738
100 5408
200 540406
500 15162300
};
\addlegendentry{\gls{mrba}}

\end{axis}

\end{tikzpicture}

%% file: Figures/Sim_results/Third_quartile_BSR_vs_depth.tex
\begin{tikzpicture}
\tikzset{every pin edge/.style={draw=none}}
\pgfplotsset{every tick label/.append style={font=\small}}

\definecolor{UNIPDRED}{RGB}{155,0,20}
\definecolor{LIGHT_GREY}{RGB}{189,195,199}
\definecolor{COMPLEMENTARY}{RGB}{0,153,153}
\definecolor{DARKGREY}{RGB}{55,65,64}
\definecolor{SAND}{RGB}{180,160,135}

\begin{axis}[
    width=\fwidth,
    height=\fheight,
    at={(0\fwidth,0\fheight)},
    scale only axis,
    legend style={
        /tikz/every even column/.append style={column sep=0.2cm},
        at={(0.55,0.95)}, 
        anchor=south, 
        draw=white!80!black, 
        font=\scriptsize
        },
    legend columns=2,
    xlabel style={font=\footnotesize},
    xlabel={Node depth},
    xtick={0,1,2},
    xmajorgrids,
    xmin=0, xmax=2,
    xtick style={color=white!15!black},
    ylabel shift = -1 pt,
    ylabel style={font=\footnotesize},
    ylabel={\gls{rlc} buffer [B]},
    ymajorgrids,
    ymin=0, ymax=50000000,
    ytick style={color=white!15!black},
    ytick={0,10000000,20000000,30000000,40000000,50000000},
    ]

\addplot [ultra thick, SAND]
table {%
0 37581200
1 18420
2 5020
};

\addplot [ultra thick, COMPLEMENTARY, mark=square*]
table {%
0 20181500
1 465436
2 13665
};

\addplot [ultra thick, DARKGREY, mark=*]
table {%
0 42314500
1 10
2 10
};

\addplot [ultra thick, UNIPDRED, mark=triangle*]
table {%
0 14633500
1 4599
2 1534
};

\coordinate (anchor) at (axis cs:0.55, -0.04);
\end{axis}

\node[pin=10 :{%
\begin{tikzpicture}[trim axis left,trim axis right]
\begin{axis}[
    xmajorgrids,
    ymajorgrids,
    axis background/.style={fill=gray!0},
    xtick style={font=\tiny},
    ytick style={font=\tiny},
    xtick={1},
    ytick={0, 200000, 400000},
    xmin=0.9,xmax=1.1,
    ymin=0,ymax=465436,
    line join=round,
    enlargelimits,width = 3.4cm
]

\addplot [ultra thick, forget plot, DARKGREY, mark=*, table/row sep=\\]
table {%
0 42314500\\
1 10\\
2 10\\
};

\addplot [ultra thick, forget plot, UNIPDRED, mark=triangle*, table/row sep=\\]
table {%
0 14633500\\
1 4599\\
2 1534\\
};

\addplot [ultra thick, forget plot, COMPLEMENTARY, mark=square*, table/row sep=\\]
table {%
0 20181500\\
1 465436\\
2 13665\\
};

\addplot [ultra thick, forget plot, SAND, table/row sep=\\]
table {%
0 37581200\\
1 18420\\
2 5020\\
};

\end{axis}
\end{tikzpicture}%
}] at (anchor) {};

\end{tikzpicture}

%% file: Figures/Sim_results/TCP_E2E_delay_ECDF_packet_size_256.tex
\begin{tikzpicture}

    \pgfplotsset{every tick label/.append style={font=\small}}

    \definecolor{UNIPDRED}{RGB}{155,0,20}
    \definecolor{LIGHT_GREY}{RGB}{189,195,199}
    \definecolor{COMPLEMENTARY}{RGB}{0,153,153}
    \definecolor{DARKGREY}{RGB}{55,65,64}
    \definecolor{SAND}{RGB}{180,160,135}

    \begin{axis}[
        width=\fwidth,
        height=\fheight,
        at={(0\fwidth,0\fheight)},
        scale only axis,
        legend image post style={mark indices={}},
        legend style={
            /tikz/every even column/.append style={column sep=0.2cm},
            at={(0.7,0.05)}, 
            anchor=south, 
            draw=white!80!black, 
            font=\scriptsize
            },
        legend columns=2,
        xlabel style={font=\footnotesize},
        ylabel={},
        xtick={500, 1000, 1500, 2000},
        xmajorgrids,
        ymin=0, ymax=1.05,
        xtick style={color=white!15!black},
        ylabel shift = -1 pt,
        ylabel style={font=\footnotesize},
        xlabel={Per UE delay [ms]},
        ymajorgrids,
        xmin=350, xmax=2250,
        ytick style={color=white!15!black},
        ytick={0, 0.2, 0.4, 0.6, 0.8, 1},
        ]

\addplot [very thick, SAND]
table {%
468.243682861328 0
525.4853515625 0.00438916683197021
530.960388183594 0.0068051815032959
533.381225585938 0.00764846801757812
541.114562988281 0.0102576017379761
554.193664550781 0.0181885957717896
563.935363769531 0.0262007713317871
566.356262207031 0.0281230211257935
575.095764160156 0.0360682010650635
578.739501953125 0.0393539667129517
597.245788574219 0.0559717416763306
599.387451171875 0.0579164028167725
618.843688964844 0.075462818145752
620.881225585938 0.0773200988769531
622.2978515625 0.0785778760910034
624.612487792969 0.0806667804718018
630.66455078125 0.0861411094665527
632.089538574219 0.0874069929122925
635.708312988281 0.0907069444656372
637.093688964844 0.0919160842895508
639.129089355469 0.09377121925354
662.5478515625 0.114867687225342
664.831176757812 0.116948366165161
667.474975585938 0.119329929351807
668.214538574219 0.119980216026306
671.75830078125 0.12321925163269
692.387451171875 0.141913652420044
694.756225585938 0.144018888473511
713.945739746094 0.16130518913269
716.085388183594 0.163245797157288
719.760375976562 0.166580319404602
739.541625976562 0.184370517730713
743.143676757812 0.187656402587891
763.247863769531 0.205761551856995
765.585388183594 0.207854509353638
785.5478515625 0.225825667381287
787.856262207031 0.227916598320007
807.447814941406 0.245526075363159
809.706237792969 0.247596621513367
813.447814941406 0.250949501991272
818.691589355469 0.255740880966187
820.710388183594 0.257547378540039
822.147827148438 0.258829593658447
824.439514160156 0.260908365249634
841.6416015625 0.276437044143677
843.65625 0.27824342250824
846.381225585938 0.280708312988281
850.029113769531 0.284016370773315
870.560424804688 0.302542209625244
873.327026367188 0.304982662200928
874.745788574219 0.306248664855957
876.841613769531 0.308183073997498
893.193664550781 0.322866439819336
895.279113769531 0.324729800224304
898.974975585938 0.328060269355774
900.435363769531 0.329387187957764
902.822875976562 0.331496357917786
903.566650390625 0.332164883613586
906.295776367188 0.334593176841736
907.687438964844 0.335826635360718
909.645751953125 0.337563991546631
911.095764160156 0.338876724243164
913.2978515625 0.340829372406006
919.783264160156 0.346537351608276
922.443664550781 0.348906636238098
924.622863769531 0.35079038143158
939.147827148438 0.363451838493347
941.556213378906 0.365567207336426
945.46875 0.368976831436157
946.872863769531 0.370196104049683
950.91455078125 0.373697280883789
962.947814941406 0.383926391601562
965.531188964844 0.386084318161011
968.441589355469 0.388559341430664
970.893676757812 0.390623927116394
975.110412597656 0.394210338592529
985.889526367188 0.403212189674377
989.006225585938 0.405754208564758
990.364562988281 0.406910419464111
992.574951171875 0.408745408058167
993.929138183594 0.409875154495239
996.491577148438 0.41202712059021
1006.38122558594 0.42030143737793
1009.46661376953 0.422833323478699
1010.89575195312 0.424017906188965
1013.28533935547 0.426013469696045
1017.73327636719 0.429711699485779
1026.94787597656 0.437254548072815
1029.52917480469 0.439402341842651
1030.88537597656 0.440528035163879
1033.92700195312 0.443059921264648
1043.18530273438 0.450578451156616
1045.84362792969 0.452744483947754
1054.43115234375 0.459608674049377
1056.87707519531 0.461608171463013
1065.31457519531 0.468321919441223
1070.44165039062 0.472351431846619
1071.76037597656 0.473399877548218
1076.97912597656 0.477474093437195
1078.24780273438 0.478445410728455
1081.58947753906 0.481038331985474
1082.87707519531 0.482044100761414
1090.54577636719 0.488012075424194
1093.12915039062 0.49003803730011
1094.43957519531 0.49107027053833
1099.54162597656 0.495087623596191
1107.97705078125 0.501801371574402
1111.28125 0.504404306411743
1161.58325195312 0.543134570121765
1167.79577636719 0.547680139541626
1170.52490234375 0.549647092819214
1177.34790039062 0.554599165916443
1179.98950195312 0.556555986404419
1186.21667480469 0.561123967170715
1187.46044921875 0.562026143074036
1193.78125 0.566675424575806
1196.93322753906 0.568971633911133
1212.875 0.58067798614502
1219.21875 0.585347652435303
1221.98327636719 0.587414145469666
1223.23950195312 0.588344812393188
1229.68322753906 0.593018412590027
1234.74157714844 0.596440315246582
1237.46252441406 0.598344326019287
1240.46044921875 0.600457668304443
1246.19580078125 0.604535818099976
1249.28540039062 0.606659293174744
1250.43530273438 0.607382774353027
1258.01037597656 0.612503290176392
1262.49572753906 0.615500569343567
1265.28955078125 0.617307066917419
1269.74365234375 0.620131492614746
1272.83532714844 0.6220862865448
1277.39782714844 0.625014543533325
1281.14782714844 0.627442717552185
1289.94165039062 0.633201479911804
1294.79577636719 0.636422157287598
1298.00830078125 0.638570070266724
1302.87292480469 0.641796827316284
1307.23120117188 0.644743204116821
1308.36877441406 0.645476818084717
1312.02490234375 0.647897005081177
1317.44787597656 0.651373744010925
1321.26037597656 0.653818249702454
1327.00622558594 0.657589673995972
1331.79577636719 0.660741329193115
1335.40832519531 0.663137078285217
1339.89575195312 0.665996074676514
1343.21875 0.668182611465454
1343.89367675781 0.668641805648804
1346.66870117188 0.67046856880188
1355.32702636719 0.676223278045654
1359.94580078125 0.679208278656006
1363.78540039062 0.681699514389038
1368.58740234375 0.684865355491638
1372.29577636719 0.687297701835632
1377.08532714844 0.690457463264465
1380.17492675781 0.692481279373169
1381.33325195312 0.693239212036133
1384.2666015625 0.695171713829041
1389.19372558594 0.698459506034851
1392.31665039062 0.700503706932068
1396.73742675781 0.703423738479614
1400.32287597656 0.705825567245483
1451.69165039062 0.737748503684998
1455.0478515625 0.739682912826538
1459.69580078125 0.742328643798828
1469.8291015625 0.747768402099609
1473.25622558594 0.749593019485474
1478.37084960938 0.752368807792664
1481.62084960938 0.754020929336548
1487.59362792969 0.757005929946899
1696.59362792969 0.854613542556763
1701.97082519531 0.857007265090942
1704.6728515625 0.858196020126343
1718.89367675781 0.864351034164429
1721.58532714844 0.865454435348511
1726.6416015625 0.867527008056641
1729.33532714844 0.868634462356567
1735.28332519531 0.871119618415833
1755.03747558594 0.879243612289429
1760.06665039062 0.881312131881714
1761.0625 0.881730794906616
1763.8125 0.882893085479736
1769.96667480469 0.885420918464661
1772.69787597656 0.886560916900635
1789.5166015625 0.893563270568848
1805.02490234375 0.899612426757812
1811.12707519531 0.901983857154846
1823.02282714844 0.906271457672119
1829.12084960938 0.908455848693848
1842.7353515625 0.91330623626709
1845.12707519531 0.914125204086304
1852.46252441406 0.916675329208374
1861.36669921875 0.91949987411499
1869.77709960938 0.921848773956299
1871.93115234375 0.922431945800781
1884.67077636719 0.926012396812439
1891.94165039062 0.928038358688354
1901.07080078125 0.93054986000061
1934.59790039062 0.939937829971313
1945.53125 0.942989826202393
1953.06665039062 0.945141792297363
2001.30627441406 0.958445310592651
2003.48120117188 0.959044814109802
2014.37915039062 0.962092876434326
2016.56872558594 0.962712526321411
2027.42700195312 0.965754508972168
2034.68530273438 0.967772245407104
2044.77490234375 0.970273733139038
2199.99780273438 0.999997973442078
2200 1
};
\addlegendentry{Distr}

\addplot [very thick, COMPLEMENTARY, mark=square*, mark repeat=35, mark phase=20]
table {%
468.243682861328 0
519.5478515625 0.00676190853118896
520.295776367188 0.00704514980316162
527.470825195312 0.00911414623260498
529.229125976562 0.0109275579452515
531.908325195312 0.0131776332855225
533.097839355469 0.0139131546020508
539.99365234375 0.0172921419143677
540.193664550781 0.0176383256912231
544.347839355469 0.0240107774734497
546.668701171875 0.026020884513855
548.281188964844 0.0275667905807495
551.808349609375 0.0299268960952759
553.270812988281 0.0328378677368164
554.060424804688 0.0343129634857178
555.974975585938 0.0376526117324829
559.747863769531 0.0432776212692261
562.939514160156 0.0463064908981323
565.497863769531 0.0502558946609497
566.874938964844 0.0528205633163452
579.5478515625 0.075320839881897
580.814575195312 0.0774410963058472
582.885375976562 0.0815832614898682
584.674987792969 0.0851470232009888
586.277038574219 0.0881916284561157
658.943664550781 0.224432468414307
659.997863769531 0.226241946220398
661.989501953125 0.229640603065491
781.295776367188 0.454293370246887
782.40625 0.456236600875854
782.82080078125 0.457074403762817
783.991577148438 0.459155321121216
784.381225585938 0.459843635559082
785.995788574219 0.462900161743164
788.308349609375 0.467042207717896
797.597839355469 0.483130693435669
799.095764160156 0.48558521270752
813.643676757812 0.511637687683105
815.47705078125 0.514902591705322
816.239501953125 0.516271352767944
817.583251953125 0.518639445304871
820.816650390625 0.524213314056396
822.897827148438 0.52795422077179
824.181213378906 0.530078411102295
824.560424804688 0.530711650848389
826.410400390625 0.533992290496826
837.897827148438 0.554246544837952
839.166625976562 0.556366682052612
840.770812988281 0.558907866477966
845.241577148438 0.565874338150024
846.377014160156 0.567667961120605
859.143676757812 0.59035325050354
860.356262207031 0.592257022857666
868.170776367188 0.605548739433289
869.539489746094 0.607932567596436
880.241577148438 0.626129388809204
881.368713378906 0.627958536148071
887.293701171875 0.637674570083618
888.524963378906 0.639586329460144
893.187438964844 0.646910667419434
894.260375976562 0.648661136627197
900.795776367188 0.659533739089966
902.1416015625 0.661567449569702
906.360412597656 0.668038129806519
908.918701171875 0.671987533569336
913.747863769531 0.679614782333374
915.170776367188 0.681809782981873
925.037414550781 0.698991775512695
926.41455078125 0.701167106628418
931.197814941406 0.708711862564087
932.662475585938 0.710922479629517
933.441589355469 0.712358236312866
934.50830078125 0.714171648025513
939.847839355469 0.72268009185791
941.24365234375 0.724729537963867
943.058349609375 0.727392435073853
945.866638183594 0.731581807136536
949.339538574219 0.736640453338623
951.247863769531 0.739370346069336
953.710388183594 0.742753267288208
956.6416015625 0.746785163879395
958.691589355469 0.749605655670166
959.324951171875 0.750490665435791
962.72705078125 0.755242586135864
965.593688964844 0.759160399436951
966.997863769531 0.761064291000366
968.606262207031 0.763345718383789
968.972900390625 0.763908267021179
971.7978515625 0.767747521400452
973.358337402344 0.770009279251099
973.97705078125 0.770835399627686
976.256225585938 0.773801326751709
978.691589355469 0.776901006698608
979.879089355469 0.778745889663696
988.760375976562 0.791388511657715
991.833251953125 0.795691967010498
993.762451171875 0.798126816749573
996.495788574219 0.801639556884766
998.289489746094 0.803842306137085
1000.82495117188 0.806985378265381
1002.65832519531 0.809203863143921
1005.24365234375 0.812594652175903
1006.88122558594 0.814632177352905
1009.14782714844 0.81740939617157
1010.94366455078 0.819773435592651
1013.10626220703 0.822353959083557
1015.53948974609 0.825032711029053
1017.94573974609 0.827951431274414
1019.86041259766 0.830193638801575
1028.49572753906 0.839595079421997
1030.53332519531 0.841601133346558
1034.30627441406 0.845511198043823
1036.07495117188 0.84734034538269
1038.73120117188 0.850125312805176
1083.85827636719 0.891377091407776
1088.11450195312 0.893572092056274
1125.95629882812 0.909719586372375
1141.72705078125 0.914235353469849
1143.97912597656 0.914856910705566
1152.82287597656 0.917425513267517
1225.52075195312 0.933061599731445
1285.80627441406 0.941869020462036
1309.46044921875 0.94518506526947
1333.44787597656 0.948721408843994
1369.43322753906 0.951529979705811
1474.66662597656 0.959908604621887
1675.51867675781 0.976012825965881
1701.68322753906 0.978333711624146
1730.72497558594 0.980040788650513
1810.69787597656 0.985331535339355
1839.0791015625 0.98687744140625
1892.33740234375 0.98877739906311
1932.70837402344 0.99033510684967
1945.68322753906 0.990877866744995
2094.2041015625 0.995385885238647
2141.89794921875 0.997458934783936
2198.92504882812 0.999956727027893
2199.94360351562 0.999996066093445
2200 1
};
\addlegendentry{\gls{msr}}

\addplot [very thick, DARKGREY, mark=*, mark repeat=35, mark phase=20]
table {%
468.243682861328 0
522.324951171875 0.00576484203338623
529.097839355469 0.00818371772766113
530.370788574219 0.00908052921295166
535.558349609375 0.0113303661346436
540.410400390625 0.0132479667663574
542.595764160156 0.0158329010009766
543.512451171875 0.0169955492019653
545.441589355469 0.0190792083740234
547.412475585938 0.0204260349273682
553.22705078125 0.0240678787231445
562.247863769531 0.035676121711731
564.512451171875 0.0378564596176147
565.677062988281 0.0396230220794678
661.2978515625 0.177139401435852
662.724975585938 0.17918074131012
663.747863769531 0.180642366409302
665.179138183594 0.182683706283569
668.5791015625 0.187515497207642
724.795776367188 0.267537593841553
726.139526367188 0.269452095031738
728.308349609375 0.272550463676453
728.639526367188 0.273030638694763
729.870788574219 0.274779081344604
731.495788574219 0.277092218399048
732.49365234375 0.278505563735962
733.762451171875 0.280335545539856
811.839538574219 0.391875505447388
812.356262207031 0.39261531829834
814.489501953125 0.395689487457275
815.497863769531 0.397114872932434
816.918701171875 0.399153232574463
857.691589355469 0.457366228103638
858.964538574219 0.45915699005127
861.120788574219 0.462225079536438
864.524963378906 0.467059850692749
866.541625976562 0.469934701919556
893.491577148438 0.507884740829468
895.206237792969 0.510240197181702
910.843688964844 0.531635522842407
912.179138183594 0.533438444137573
930.697814941406 0.558789730072021
932.022888183594 0.56058657169342
934.4228515625 0.563869118690491
935.481201171875 0.565333724021912
937.237426757812 0.567716360092163
955.941589355469 0.59333348274231
957.256225585938 0.59512722492218
971.697814941406 0.614758968353271
973.314575195312 0.616869926452637
985.324951171875 0.632591009140015
987.918701171875 0.63601553440094
992.185363769531 0.641626358032227
992.681213378906 0.642266511917114
1003.89575195312 0.656858325004578
1005.33953857422 0.658643007278442
1007.32287597656 0.661137342453003
1008.24578857422 0.662290930747986
1009.75622558594 0.664184331893921
1010.68121337891 0.66534698009491
1013.52911376953 0.66895866394043
1021.44366455078 0.678945302963257
1022.94573974609 0.680835723876953
1023.43951416016 0.681469798088074
1026.33740234375 0.685117721557617
1027.24365234375 0.686253190040588
1028.73120117188 0.688140630722046
1036.29577636719 0.697704315185547
1037.99365234375 0.699721574783325
1043.13330078125 0.705724954605103
1045.50622558594 0.708409547805786
1050.52709960938 0.71409285068512
1052.8916015625 0.716774463653564
1057.36669921875 0.721805572509766
1059.67907714844 0.724438786506653
1064.34790039062 0.729690313339233
1066.08325195312 0.731631994247437
1070.79370117188 0.736883401870728
1072.63330078125 0.738855361938477
1073.09155273438 0.739338517189026
1077.54370117188 0.744124889373779
1080.13952636719 0.746894121170044
1084.0791015625 0.75114905834198
1086.01037597656 0.75322961807251
1086.83325195312 0.75412654876709
1088.61669921875 0.756016969680786
1092.59790039062 0.760326266288757
1094.58117675781 0.762434005737305
1097.6416015625 0.765405654907227
1099.68322753906 0.767356395721436
1102.6416015625 0.770167827606201
1104.67077636719 0.772115707397461
1107.27282714844 0.774576783180237
1108.0478515625 0.77531361579895
1110.37915039062 0.777515053749084
1111.17907714844 0.778312206268311
1113.24365234375 0.780266046524048
1116.1103515625 0.782971858978271
1123.00830078125 0.789548993110657
1125.99780273438 0.79240870475769
1128.71252441406 0.794915199279785
1129.0166015625 0.795184016227722
1131.88122558594 0.797738790512085
1134.62915039062 0.800163745880127
1137.13330078125 0.802410364151001
1139.72705078125 0.804699420928955
1141.86242675781 0.80657172203064
1142.61242675781 0.807224035263062
1151.13330078125 0.814812779426575
1153.48742675781 0.816742420196533
1156.32287597656 0.819082736968994
1157.44787597656 0.820019006729126
1160.81665039062 0.822760939598083
1161.53125 0.823364973068237
1163.89367675781 0.825324773788452
1189.45629882812 0.842519640922546
1192.97912597656 0.844609260559082
1193.63537597656 0.845004916191101
1197.98327636719 0.847592830657959
1208.76037597656 0.853964805603027
1213.53540039062 0.856471180915833
1238.12084960938 0.86672043800354
1244.68530273438 0.869103074073792
1246.23327636719 0.869679927825928
1259.45837402344 0.874484300613403
1269.38330078125 0.878087043762207
1271.86669921875 0.878986835479736
1283.99365234375 0.883386850357056
1290.63330078125 0.885793566703796
1294.77709960938 0.887303471565247
1300.35827636719 0.88930869102478
1304.41247558594 0.890764236450195
1312.05627441406 0.893146753311157
1313.32287597656 0.893478989601135
1320.01452636719 0.895221471786499
1339.40625 0.900204181671143
1349.41455078125 0.902342081069946
1385.37915039062 0.909598827362061
1402.32080078125 0.912579298019409
1407.03955078125 0.913316249847412
1443.34155273438 0.918914914131165
1448.03125 0.919633626937866
1467.64782714844 0.92266857624054
1499.32080078125 0.927584886550903
1563.39575195312 0.937477707862854
1591.70629882812 0.941826343536377
1653.74572753906 0.951453447341919
1685.47705078125 0.956387758255005
1702.21875 0.958978891372681
1788.875 0.97244119644165
1919.58947753906 0.986833572387695
2154.66259765625 0.998493194580078
2200 1
};
\addlegendentry{\gls{ba}}

\addplot [very thick, UNIPDRED, mark=triangle*, mark repeat=35, mark phase=20]
table {%
468.243682861328 0
519.147827148438 0.00662744045257568
519.877014160156 0.00692105293273926
528.431213378906 0.0100339651107788
532.147827148438 0.0137029886245728
540.022888183594 0.016846776008606
545.697814941406 0.0258920192718506
551.520812988281 0.0295610427856445
555.087463378906 0.0344080924987793
555.4853515625 0.0351147651672363
557.181213378906 0.0376946926116943
561.897827148438 0.0449633598327637
563.316650390625 0.0468635559082031
565.683288574219 0.0497485399246216
567.174987792969 0.0519692897796631
567.933288574219 0.0532670021057129
570.556213378906 0.057233452796936
575.322875976562 0.0644479990005493
577.3291015625 0.0672171115875244
582.593688964844 0.0752079486846924
586.972900390625 0.0818315744400024
588.8291015625 0.0845080614089966
590.341613769531 0.0869683027267456
594.697814941406 0.0935571193695068
596.147827148438 0.0957469940185547
597.345764160156 0.0975466966629028
598.464538574219 0.0993040800094604
599.627014160156 0.101262211799622
600.295776367188 0.102235436439514
601.529113769531 0.104046821594238
602.231201171875 0.105147480964661
603.345764160156 0.106966614723206
603.608337402344 0.107414603233337
604.916625976562 0.109426736831665
609.695739746094 0.116884589195251
614.824951171875 0.124921798706055
616.145751953125 0.12726616859436
616.539489746094 0.127945899963379
623.847839355469 0.139636635780334
625.168701171875 0.141467332839966
627.239501953125 0.144920110702515
633.324951171875 0.154714584350586
634.747863769531 0.157120704650879
641.416625976562 0.167892217636108
642.841613769531 0.170302271842957
643.524963378906 0.171341180801392
644.65625 0.173287749290466
646.99365234375 0.176987648010254
647.772888183594 0.178277611732483
654.087463378906 0.188496947288513
656.097839355469 0.19189178943634
658.47705078125 0.195603370666504
658.870788574219 0.196290731430054
660.791625976562 0.199604511260986
667.983276367188 0.211453676223755
670.127014160156 0.214972019195557
670.893676757812 0.216296792030334
677.268737792969 0.226631999015808
678.643676757812 0.229011058807373
679.0478515625 0.229752540588379
680.291625976562 0.231853604316711
682.697814941406 0.235615372657776
688.7978515625 0.245297789573669
689.895751953125 0.247112989425659
697.795776367188 0.260290741920471
698.997863769531 0.262187004089355
701.131225585938 0.265593528747559
702.3916015625 0.267706155776978
703.120788574219 0.268895626068115
704.437438964844 0.271236181259155
705.677062988281 0.273190379142761
706.874938964844 0.275090575218201
709.043701171875 0.278624415397644
709.410400390625 0.27919602394104
710.647827148438 0.281107783317566
712.747863769531 0.28456449508667
714.443664550781 0.287136673927307
720.724975585938 0.29725170135498
722.339538574219 0.299781441688538
722.597839355469 0.300217866897583
724.022888183594 0.302608489990234
724.747863769531 0.303809642791748
725.818725585938 0.305621027946472
732.697814941406 0.316956520080566
733.7978515625 0.318686723709106
735.118713378906 0.320826292037964
736.447814941406 0.32317841053009
737.037414550781 0.324163198471069
743.212463378906 0.33417010307312
744.197814941406 0.336062669754028
747.845764160156 0.342025756835938
755.616638183594 0.354894518852234
757.74365234375 0.358405232429504
764.697814941406 0.369651794433594
765.824951171875 0.371409177780151
771.75830078125 0.380967974662781
773.145751953125 0.383138537406921
773.910400390625 0.384478688240051
775.345764160156 0.386896371841431
776.795776367188 0.3893141746521
783.8291015625 0.400935411453247
784.943664550781 0.402804732322693
785.206237792969 0.403268098831177
786.629089355469 0.405457973480225
791.795776367188 0.414147853851318
791.910400390625 0.414340972900391
801.32080078125 0.430253148078918
802.49365234375 0.432145595550537
803.285339355469 0.433485746383667
805.277038574219 0.436869025230408
805.99365234375 0.4380122423172
808.07080078125 0.441449522972107
808.827026367188 0.44273567199707
810.24365234375 0.445141792297363
810.408325195312 0.445481657981873
811.597839355469 0.447559475898743
812.90625 0.449776411056519
813.6416015625 0.450989127159119
815.7978515625 0.454526901245117
823.522888183594 0.467268109321594
824.524963378906 0.469025373458862
826.143676757812 0.471755981445312
826.5478515625 0.472485899925232
827.843688964844 0.474606275558472
828.037414550781 0.474911332130432
829.99365234375 0.478236675262451
830.772888183594 0.479538202285767
831.947814941406 0.481504082679749
832.737426757812 0.482875108718872
834.697814941406 0.486231327056885
836.574951171875 0.489475607872009
837.310424804688 0.490730762481689
838.435363769531 0.492557644844055
849.364562988281 0.511281371116638
850.833251953125 0.51372230052948
850.889526367188 0.513857483863831
861.627014160156 0.532079219818115
862.933288574219 0.534211158752441
870.260375976562 0.546160697937012
871.347839355469 0.547937154769897
872.895751953125 0.550416707992554
877.927062988281 0.558284044265747
879.345764160156 0.560462236404419
880.056213378906 0.561570644378662
881.595764160156 0.564046382904053
882.285339355469 0.565077543258667
884.695739746094 0.568819999694824
884.947814941406 0.569179058074951
886.291625976562 0.571322679519653
887.056213378906 0.572655081748962
888.568725585938 0.575111389160156
889.281188964844 0.576239109039307
890.7978515625 0.578707098960876
892.793701171875 0.581704139709473
898.012451171875 0.589930534362793
899.197814941406 0.591896414756775
992.956237792969 0.734379529953003
994.941589355469 0.737140893936157
997.645751953125 0.740674734115601
998.974975585938 0.7424476146698
1001.77081298828 0.746070265769958
1003.59161376953 0.748433828353882
1004.17706298828 0.749109745025635
1005.8916015625 0.75127649307251
1008.49786376953 0.754624962806702
1010.49786376953 0.757061958312988
1014.10418701172 0.761264085769653
1064.63537597656 0.822151064872742
1066.37707519531 0.824217319488525
1066.71667480469 0.82460355758667
1069.58325195312 0.827913403511047
1069.86877441406 0.828257083892822
1076.94787597656 0.836390852928162
1078.80627441406 0.838291049003601
1080.99572753906 0.840840101242065
1081.05627441406 0.840944290161133
1083.19372558594 0.8430415391922
1115.62084960938 0.871795415878296
1118.8603515625 0.873992919921875
1126.09790039062 0.879222393035889
1130.44787597656 0.882080316543579
1166.14367675781 0.902553677558899
1170.96667480469 0.905195474624634
1174.08740234375 0.906570315361023
1178.33325195312 0.90848982334137
1179.84790039062 0.90910005569458
1188.09362792969 0.911996603012085
1189.16247558594 0.912471771240234
1235.77282714844 0.925448656082153
1285.07287597656 0.934092044830322
1298.74572753906 0.936208605766296
1325.58325195312 0.939595699310303
1341.70837402344 0.941550016403198
1343.15625 0.941781759262085
1367.43530273438 0.94446587562561
1373.14575195312 0.945049047470093
1403.0791015625 0.947520852088928
1734.14782714844 0.967978715896606
1779.19580078125 0.971056938171387
1807.67907714844 0.973161816596985
1847.99365234375 0.976050615310669
1851.63745117188 0.976351976394653
1887.41247558594 0.979163646697998
1893.77490234375 0.979650259017944
1923.78332519531 0.981782197952271
2191.61254882812 0.999378204345703
2199.99780273438 0.999996185302734
2200 1
};
\addlegendentry{\gls{mrba}}
\end{axis}

\end{tikzpicture}

%% file: Figures/Sim_results/TCP_E2E_throughput_ECDF_packet_size_256.tex
\begin{tikzpicture}

\pgfplotsset{every tick label/.append style={font=\small}}

    \definecolor{UNIPDRED}{RGB}{155,0,20}
    \definecolor{LIGHT_GREY}{RGB}{189,195,199}
    \definecolor{COMPLEMENTARY}{RGB}{0,153,153}
    \definecolor{DARKGREY}{RGB}{55,65,64}
    \definecolor{SAND}{RGB}{180,160,135}

\begin{axis}[
    width=\fwidth,
    height=\fheight,
    at={(0\fwidth,0\fheight)},
    scale only axis,
    legend image post style={mark indices={}},
    legend style={
        /tikz/every even column/.append style={column sep=0.2cm},
        at={(0.7,0.05)}, 
        anchor=south, 
        draw=white!80!black, 
        font=\scriptsize
        },
    legend columns=2,
    xlabel style={font=\footnotesize},
    ylabel={},
    xtick={50, 100, 150},
    xmajorgrids,
    ymin=0, ymax=1.05,
    xtick style={color=white!15!black},
    ylabel shift = -1 pt,
    ylabel style={font=\footnotesize},
    xlabel={Per UE throughput [Mbps]},
    ymajorgrids,
    xmin=0, xmax=150,
    ytick style={color=white!15!black},
    ytick={0, 0.2, 0.4, 0.6, 0.8, 1},
    ]

\addplot [very thick, SAND]
table {%
0.0287342071533203 0
0.0371237993240356 0.00171089172363281
0.0878909826278687 0.00342178344726562
0.346015810966492 0.00769889354705811
0.458461880683899 0.0102652311325073
0.606310606002808 0.0153977870941162
0.63634729385376 0.0205303430557251
0.638597965240479 0.0248075723648071
0.640201807022095 0.0256630182266235
0.680343508720398 0.0273737907409668
0.685259580612183 0.0316510200500488
0.695359230041504 0.0367835760116577
0.696218252182007 0.0384944677352905
0.725967526435852 0.0410606861114502
0.739611864089966 0.0436270236968994
0.74266242980957 0.0444824695587158
0.81693172454834 0.0513259172439575
0.828238248825073 0.0547477006912231
0.837233304977417 0.05560302734375
0.902937650680542 0.0573139190673828
0.911258459091187 0.0615911483764648
0.924090623855591 0.0650128126144409
0.986952781677246 0.0718562602996826
1.06532347202301 0.0744225978851318
1.07107734680176 0.0752780437469482
1.11456513404846 0.0778442621231079
1.15134358406067 0.0821214914321899
1.15668904781342 0.0838323831558228
1.18633329868317 0.0863986015319824
1.22763669490814 0.0915312767028809
1.25698006153107 0.0940974950790405
1.26538777351379 0.0966638326644897
1.28493738174438 0.0983747243881226
1.31196665763855 0.11120617389679
1.33330678939819 0.113772392272949
1.339968085289 0.118905067443848
1.36357545852661 0.122326731681824
1.38029265403748 0.132591962814331
1.40238654613495 0.136869072914124
1.41183066368103 0.142001748085022
1.41406440734863 0.14970064163208
1.47023451328278 0.157399535179138
1.48385643959045 0.161676645278931
1.49459838867188 0.165098428726196
1.50480329990387 0.167664647102356
1.54704535007477 0.17878532409668
1.57047092914581 0.181351542472839
1.59738326072693 0.184773325920105
1.61019337177277 0.187339544296265
1.61209285259247 0.188194990158081
1.63452970981598 0.191616773605347
1.63764750957489 0.193327665328979
1.63855862617493 0.195038557052612
1.64258754253387 0.197604775428772
1.64376473426819 0.200171113014221
1.67859470844269 0.206159114837646
1.73291981220245 0.211291670799255
1.75324082374573 0.215568900108337
1.76444911956787 0.218990564346313
1.77288520336151 0.221556901931763
1.78008162975311 0.223267793655396
1.79736995697021 0.225834012031555
1.79992830753326 0.226689457893372
1.89328503608704 0.232677459716797
1.90057456493378 0.23438835144043
1.94071424007416 0.239520907402039
1.95721864700317 0.244653582572937
2.01270151138306 0.247219800949097
2.08526825904846 0.25834047794342
2.10953712463379 0.261762142181396
2.16797113418579 0.27544903755188
2.19129419326782 0.288280606269836
2.20371103286743 0.290846824645996
2.20943331718445 0.297690391540527
2.22924160957336 0.302822947502136
2.24127173423767 0.305389165878296
2.24942779541016 0.307955503463745
2.25076746940613 0.308810949325562
2.2893078327179 0.317365288734436
2.31514883041382 0.319931507110596
2.45897102355957 0.329341292381287
2.46524667739868 0.330196738243103
2.51809358596802 0.333618521690369
2.54055237770081 0.336184740066528
2.55224299430847 0.337895631790161
2.55329179763794 0.339606523513794
2.57448101043701 0.343883633613586
2.62732362747192 0.348160862922668
2.64115452766418 0.350727081298828
2.7036726474762 0.355859756469727
2.71103739738464 0.358425974845886
2.73175072669983 0.363558530807495
2.76669001579285 0.370402097702026
2.79180455207825 0.372968316078186
2.82877373695374 0.375534653663635
2.85294270515442 0.379811763763428
2.88162803649902 0.383233547210693
2.88338303565979 0.384944438934326
2.88890051841736 0.385799884796143
2.92370080947876 0.388366103172302
2.93459367752075 0.391787886619568
3.01517820358276 0.395209550857544
3.02518486976624 0.397775888442993
3.06865835189819 0.400342226028442
3.07408118247986 0.401197671890259
3.09739351272583 0.403763890266418
3.11805748939514 0.408041000366211
3.1366491317749 0.411462783813477
3.17191219329834 0.414029121398926
3.20227575302124 0.415740013122559
3.29009771347046 0.422583341598511
3.29140448570251 0.423438787460327
3.30214762687683 0.426005125045776
3.31021308898926 0.427716016769409
3.38134956359863 0.437981128692627
3.40473246574402 0.440547466278076
3.43035578727722 0.443969249725342
3.43400287628174 0.444824695587158
3.46373605728149 0.447390913963318
3.46420311927795 0.449101805686951
3.46662187576294 0.449957251548767
3.48528385162354 0.452523469924927
3.5211923122406 0.457656145095825
3.56206941604614 0.460222482681274
3.58200645446777 0.46364414691925
3.71693634986877 0.475620150566101
3.72956109046936 0.47818648815155
3.75654077529907 0.48075270652771
3.77893567085266 0.483319044113159
3.82819890975952 0.485885381698608
3.8352382183075 0.488451719284058
3.85153150558472 0.491017937660217
3.85821294784546 0.49272882938385
3.87212491035461 0.495295166969299
3.91328120231628 0.498716831207275
3.96807312965393 0.502994060516357
3.96858763694763 0.50470495223999
4.00316190719604 0.50727117061615
4.02444267272949 0.50983738899231
4.05338954925537 0.512403726577759
4.05980014801025 0.514114618301392
4.29127979278564 0.535500407218933
4.30277299880981 0.538066744804382
4.3059229850769 0.538922071456909
4.38899421691895 0.541488409042358
4.49381875991821 0.544910192489624
4.53920841217041 0.54833197593689
4.60790014266968 0.553464412689209
4.67319250106812 0.555175304412842
4.68145990371704 0.556886196136475
4.70034217834473 0.559452533721924
4.70824003219604 0.56030797958374
4.74091148376465 0.562874317169189
4.74702548980713 0.563729763031006
4.83432865142822 0.568006873130798
4.93627452850342 0.573994874954224
4.93891525268555 0.57485032081604
5.12614059448242 0.580838322639465
5.14018583297729 0.585970878601074
5.20229721069336 0.588537216186523
5.2465295791626 0.592814445495605
5.28982353210449 0.594525218009949
5.30438852310181 0.597946882247925
5.33540105819702 0.600513219833374
5.34512376785278 0.602224111557007
5.389564037323 0.605645895004272
5.41673994064331 0.609067559242249
5.45519065856934 0.611633896827698
5.4810791015625 0.614200115203857
5.55653953552246 0.616766452789307
5.59104919433594 0.620188236236572
5.71786689758301 0.627887010574341
5.82896041870117 0.63045334815979
5.84658050537109 0.631308794021606
6.14880037307739 0.635586023330688
6.27813482284546 0.641573905944824
6.29823350906372 0.643284797668457
6.34835433959961 0.64499568939209
6.35257911682129 0.645851135253906
6.43140411376953 0.649272918701172
6.57165861129761 0.650983810424805
6.57402276992798 0.651839256286621
6.89169645309448 0.65953803062439
6.89881801605225 0.660393476486206
7.08784437179565 0.664670705795288
7.14395332336426 0.668092370033264
7.14762544631958 0.668947815895081
7.27264261245728 0.677502155303955
7.28915309906006 0.679213047027588
7.45855331420898 0.68349015712738
7.49694776535034 0.68605637550354
7.5797290802002 0.688622713088989
7.74595022201538 0.695466279983521
7.82546472549438 0.697177052497864
7.8604040145874 0.699743390083313
7.96884489059448 0.702309608459473
7.97087907791138 0.703165054321289
8.01082706451416 0.705731391906738
8.03297233581543 0.708297729492188
8.11168670654297 0.710864067077637
8.18515205383301 0.713430285453796
8.20366096496582 0.715996503829956
8.21247386932373 0.717707395553589
8.28819561004639 0.720273733139038
8.44493579864502 0.72711718082428
8.45096015930176 0.728828072547913
8.59292030334473 0.731394290924072
8.60062789916992 0.733105182647705
8.64025783538818 0.736526966094971
8.64641189575195 0.737382411956787
8.75086879730225 0.739948749542236
8.78127002716064 0.74165952205658
9.07768058776855 0.749358415603638
9.46374607086182 0.75363564491272
9.58797836303711 0.757057309150696
9.59193801879883 0.757912755012512
9.60428333282471 0.758768200874329
9.73576545715332 0.761334419250488
9.7401237487793 0.762189865112305
9.8993968963623 0.76817798614502
10.8964338302612 0.781864881515503
10.9009275436401 0.782720327377319
11.0518398284912 0.786997437477112
12.075475692749 0.79469633102417
12.1972923278809 0.797262668609619
12.3709831237793 0.799828886985779
12.8203964233398 0.801539778709412
12.9326686859131 0.805816888809204
13.2163190841675 0.808383226394653
13.4739799499512 0.810094118118286
14.563196182251 0.815226674079895
15.7830591201782 0.818648338317871
17.2529735565186 0.828913688659668
18.4278106689453 0.832335352897644
19.0228977203369 0.83575701713562
21.0073757171631 0.838323354721069
23.0308666229248 0.841745138168335
23.9310722351074 0.843456029891968
28.4373836517334 0.846022248268127
30.119026184082 0.849443912506104
34.7817916870117 0.852010250091553
34.8945808410645 0.852865695953369
35.3691482543945 0.853721141815186
37.7247734069824 0.856287479400635
38.9159622192383 0.859709143638611
39.2842330932617 0.860564589500427
43.5693359375 0.86484169960022
47.188304901123 0.869974374771118
53.8663673400879 0.872540712356567
54.7224464416504 0.874251484870911
62.5528221130371 0.878528594970703
64.2262878417969 0.882805824279785
64.3802185058594 0.883661270141602
65.6859893798828 0.886227607727051
65.7560424804688 0.887083053588867
68.3482666015625 0.889649271965027
70.4195327758789 0.894781827926636
70.8212432861328 0.896492719650269
75.9479217529297 0.901625394821167
77.5449142456055 0.907613277435303
77.5718383789062 0.908468723297119
78.5177230834961 0.911035060882568
78.7301177978516 0.911890506744385
79.5807342529297 0.914456844329834
79.7080993652344 0.91531229019165
83.2162017822266 0.918733954429626
83.3014602661133 0.920444846153259
84.6517333984375 0.923011064529419
84.8869323730469 0.924721956253052
84.923698425293 0.926432847976685
85.0861968994141 0.928143739700317
85.0998687744141 0.928999185562134
85.6565017700195 0.937553405761719
85.9748840332031 0.940119743347168
86.0080184936523 0.940975189208984
86.3420181274414 0.945252418518066
86.5098114013672 0.952095746994019
86.6039199829102 0.953806638717651
87.2917709350586 0.957228422164917
87.371208190918 0.959794759750366
87.3864593505859 0.960650205612183
88.9585342407227 0.970059871673584
90.6937408447266 0.975192546844482
90.7225570678711 0.976047992706299
91.6213302612305 0.981180429458618
92.257926940918 0.983746767044067
92.4870300292969 0.9854576587677
95.2200622558594 0.991445660591125
95.5282592773438 0.993156552314758
96.1391220092773 0.994011998176575
100.658157348633 0.995722770690918
101.063446044922 0.998289108276367
101.78197479248 0.999144554138184
150 1
};
\addlegendentry{Distr}

\addplot [very thick, COMPLEMENTARY, mark=square*, mark repeat=55, mark phase=40]
table {%
0.00999653339385986 0
0.18618106842041 0.00342178344726562
0.23947811126709 0.00598800182342529
0.333441495895386 0.00855433940887451
0.352973461151123 0.0119760036468506
0.353565216064453 0.0136868953704834
0.3748779296875 0.0179641246795654
0.377195358276367 0.0188194513320923
0.414880871772766 0.0213857889175415
0.417248725891113 0.0222412347793579
0.492755651473999 0.0265183448791504
0.49326479434967 0.0282292366027832
0.505341529846191 0.0307955741882324
0.524819016456604 0.0350726842880249
0.536711454391479 0.0367835760116577
0.53900146484375 0.0393499135971069
0.54201865196228 0.0410606861114502
0.557332038879395 0.0436270236968994
0.56901478767395 0.0479042530059814
0.608031272888184 0.0504704713821411
0.630350112915039 0.0573139190673828
0.653141856193542 0.0607357025146484
0.654970765113831 0.0633019208908081
0.663637280464172 0.0658682584762573
0.669018745422363 0.0675791501998901
0.677386999130249 0.0692900419235229
0.740627527236938 0.0752780437469482
0.748180866241455 0.0786997079849243
0.764818906784058 0.0804105997085571
0.801711320877075 0.0889649391174316
0.914310216903687 0.0966638326644897
0.924806118011475 0.100085496902466
0.951181888580322 0.102651834487915
0.972347259521484 0.106928944587708
0.992884039878845 0.109495282173157
0.994224309921265 0.110350728034973
1.01191318035126 0.112917065620422
1.01392924785614 0.113772392272949
1.04868018627167 0.118049621582031
1.04951167106628 0.12061595916748
1.07780027389526 0.124037623405457
1.0795419216156 0.126603960990906
1.08158051967621 0.127459406852722
1.09834039211273 0.130881071090698
1.09953451156616 0.132591962814331
1.1333817243576 0.136013746261597
1.1355607509613 0.136869072914124
1.16420412063599 0.139435410499573
1.16506230831146 0.142001748085022
1.17537152767181 0.144567966461182
1.18738555908203 0.147134304046631
1.1935453414917 0.151411414146423
1.21464061737061 0.154833197593689
1.2455917596817 0.160821199417114
1.27421951293945 0.16424298286438
1.27777481079102 0.165953755378723
1.2993575334549 0.169375538825989
1.31242048740387 0.171941876411438
1.34737312793732 0.177074432373047
1.36374473571777 0.183062434196472
1.37591075897217 0.189905881881714
1.39020097255707 0.192472219467163
1.3945449590683 0.195038557052612
1.41226994991302 0.198460221290588
1.42124319076538 0.20530366897583
1.42166769504547 0.207014560699463
1.42530107498169 0.209580898284912
1.43128776550293 0.213858008384705
1.4334329366684 0.216424345970154
1.44793796539307 0.221556901931763
1.45022130012512 0.224123239517212
1.45887184143066 0.226689457893372
1.47568380832672 0.230966687202454
1.48030436038971 0.233532905578613
1.4849009513855 0.236099243164062
1.49957513809204 0.237810134887695
1.51197147369385 0.242942690849304
1.53943014144897 0.245509028434753
1.55896532535553 0.254063367843628
1.57504296302795 0.25834047794342
1.58209371566772 0.26090669631958
1.58523571491241 0.263473033905029
1.59612202644348 0.266039371490479
1.61339962482452 0.271171927452087
1.64078629016876 0.274593710899353
1.68211603164673 0.282292604446411
1.69648885726929 0.284003496170044
1.69832026958466 0.284858822822571
1.71271181106567 0.28742516040802
1.71503400802612 0.288280606269836
1.74169671535492 0.291702270507812
1.75224173069 0.295124053955078
1.77508783340454 0.298545837402344
1.79124236106873 0.30196750164032
1.82570469379425 0.309666395187378
1.83333110809326 0.313088178634644
1.84223878383636 0.315654397010803
1.86225974559784 0.318220734596252
1.8699277639389 0.320786952972412
1.87197005748749 0.321642398834229
1.90125238895416 0.326775074005127
1.90247106552124 0.32848584651947
1.92903995513916 0.335329294204712
1.96044969558716 0.337040185928345
1.96845149993896 0.339606523513794
1.97000324726105 0.34046196937561
1.98619890213013 0.342172861099243
2.00732254981995 0.344739079475403
2.01108598709106 0.346449971199036
2.02600526809692 0.349016189575195
2.03227829933167 0.351582527160645
2.03924655914307 0.354148864746094
2.05326581001282 0.356715202331543
2.06524014472961 0.358425974845886
2.08829712867737 0.360992312431335
2.09221315383911 0.362703204154968
2.11047506332397 0.365269422531128
2.17317175865173 0.381522655487061
2.1764178276062 0.383233547210693
2.22221517562866 0.385799884796143
2.24425578117371 0.387510657310486
2.24502372741699 0.389221549034119
2.25452923774719 0.391787886619568
2.26138210296631 0.394354104995728
2.33922433853149 0.402908444404602
2.34462738037109 0.404619336128235
2.36183786392212 0.408041000366211
2.3919415473938 0.41060733795166
2.39443349838257 0.412318229675293
2.50585627555847 0.420017123222351
2.56709408760071 0.426860570907593
2.59053492546082 0.428571462631226
2.60071778297424 0.430282354354858
2.61894178390503 0.432848572731018
2.64813160896301 0.437981128692627
2.66045069694519 0.43969202041626
2.69474935531616 0.441402912139893
2.70122909545898 0.443969249725342
2.76047611236572 0.447390913963318
2.77058959007263 0.449101805686951
2.79223203659058 0.453378915786743
2.79541373252869 0.455945253372192
2.81599116325378 0.458511590957642
2.85624408721924 0.46621036529541
2.87566137313843 0.468776702880859
2.88347268104553 0.471343040466309
2.89856624603271 0.473909378051758
2.90361595153809 0.475620150566101
2.91160178184509 0.479041934013367
2.92605042457581 0.485885381698608
2.93691158294678 0.489307165145874
2.93951654434204 0.490162491798401
2.9579393863678 0.493584275245667
2.96014165878296 0.494439721107483
2.97429776191711 0.497005939483643
2.98164582252502 0.499572277069092
3.01345372200012 0.511548280715942
3.02704834938049 0.514114618301392
3.03350949287415 0.515825510025024
3.04821515083313 0.518391847610474
3.08305382728577 0.523524403572083
3.11295318603516 0.526090621948242
3.13176465034485 0.528656959533691
3.16039896011353 0.531223297119141
3.26941204071045 0.544054746627808
3.32011437416077 0.54576563835144
3.32367873191833 0.547476530075073
3.34370732307434 0.550042748451233
3.3781156539917 0.551753640174866
3.38152265548706 0.554319858551025
3.42304062843323 0.558597087860107
3.43638968467712 0.561163425445557
3.43995952606201 0.562874317169189
3.50880289077759 0.567151427268982
3.51529049873352 0.569717645645142
3.60191416740417 0.578272104263306
3.63293266296387 0.585115432739258
3.63512921333313 0.586826324462891
3.64581871032715 0.58939266204834
3.65057849884033 0.591103553771973
3.67266201972961 0.593669891357422
3.7233350276947 0.605645895004272
3.76039958000183 0.609067559242249
3.88232445716858 0.61591100692749
3.88508534431458 0.616766452789307
3.92045140266418 0.618477344512939
3.92416906356812 0.620188236236572
3.96395230293274 0.622754573822021
4.00456380844116 0.627031564712524
4.06628370285034 0.63045334815979
4.10646963119507 0.632164239883423
4.10910558700562 0.633875131607056
4.15332555770874 0.636441469192505
4.17224311828613 0.639863133430481
4.20528984069824 0.641573905944824
4.2072286605835 0.642429351806641
4.24568700790405 0.645851135253906
4.27427434921265 0.650983810424805
4.27686262130737 0.651839256286621
4.37341785430908 0.655260920524597
4.38978004455566 0.658682584762573
4.39742803573608 0.65953803062439
4.44929456710815 0.662104368209839
4.45268392562866 0.663815259933472
4.46167325973511 0.665526151657104
4.51201343536377 0.668092370033264
4.63683319091797 0.674935817718506
4.69321870803833 0.677502155303955
4.73041486740112 0.680068492889404
4.73605680465698 0.681779384613037
4.74980735778809 0.684345602989197
4.79808378219604 0.690333604812622
4.85303020477295 0.696321725845337
4.98732042312622 0.705731391906738
4.99089765548706 0.707442283630371
5.03670310974121 0.711719393730164
5.11339139938354 0.713430285453796
5.11800956726074 0.714285731315613
5.18045091629028 0.716851949691772
5.26939105987549 0.721984624862671
5.28773212432861 0.726261734962463
5.30210399627686 0.729683518409729
5.32029581069946 0.732249736785889
5.32697200775146 0.735671520233154
5.33921098709106 0.738237857818604
5.34131526947021 0.73909330368042
5.37160110473633 0.74165952205658
5.56476449966431 0.746792078018188
5.57259225845337 0.747647523880005
5.95217037200928 0.760478973388672
5.96735286712646 0.762189865112305
6.07881021499634 0.764756202697754
7.5417308807373 0.783575773239136
7.54809951782227 0.785286545753479
7.81147241592407 0.790419101715088
7.88670825958252 0.792129993438721
7.98633766174316 0.795551776885986
7.98801898956299 0.796407222747803
8.29807281494141 0.803250670433044
8.8161563873291 0.807527780532837
8.96269035339355 0.812660455703735
10.2218618392944 0.82121467590332
10.4444980621338 0.822925567626953
11.1452760696411 0.825491905212402
13.229211807251 0.829769015312195
13.7269802093506 0.834046125411987
15.7746105194092 0.838323354721069
18.2060222625732 0.840889692306519
19.6877937316895 0.842600584030151
23.9410285949707 0.845166802406311
36.4297790527344 0.851154804229736
36.7490730285645 0.852010250091553
44.260124206543 0.857998371124268
56.4426498413086 0.863130807876587
57.0303115844727 0.863986253738403
61.5657272338867 0.867408037185669
62.0811042785645 0.868263483047485
65.5272064208984 0.870829820632935
67.1641540527344 0.872540712356567
70.7332534790039 0.875962376594543
70.8874664306641 0.877673149108887
80.2640075683594 0.885372161865234
81.2116851806641 0.887938380241394
83.1876602172852 0.89136016368866
83.3083953857422 0.893070936203003
84.0433502197266 0.895637273788452
84.2809600830078 0.896492719650269
85.406494140625 0.899059057235718
88.3571243286133 0.907613277435303
88.6335983276367 0.909324169158936
90.2629470825195 0.912745952606201
90.4132537841797 0.913601398468018
92.6400299072266 0.91787850856781
95.169189453125 0.920444846153259
95.2564849853516 0.922155618667603
95.7640380859375 0.923866510391235
98.1523513793945 0.926432847976685
100.48136138916 0.92985463142395
100.53630065918 0.930710077285767
101.270843505859 0.933276295661926
101.687141418457 0.934987187385559
102.825752258301 0.937553405761719
102.868507385254 0.938408851623535
103.011054992676 0.939264297485352
104.325057983398 0.942686080932617
104.41088104248 0.943541526794434
105.435089111328 0.946107864379883
105.576637268066 0.947818636894226
106.645004272461 0.953806638717651
106.791854858398 0.955517530441284
107.874740600586 0.95893931388855
107.896598815918 0.959794759750366
108.254486083984 0.961505532264709
109.135124206543 0.964071869850159
112.857063293457 0.976047992706299
113.185020446777 0.978614211082458
113.92440032959 0.980324983596802
113.953605651855 0.982035875320435
114.083236694336 0.982891321182251
115.803565979004 0.9854576587677
119.775741577148 0.991445660591125
121.263778686523 0.994011998176575
121.631225585938 0.994867324829102
125.164413452148 0.996578216552734
129.905914306641 0.998289108276367
135.386734008789 0.999144554138184
150 1
};
\addlegendentry{\gls{msr}}

\addplot [very thick, DARKGREY, mark=*, mark repeat=55, mark phase=40]
table {%
0.0696563720703125 0
0.311533331871033 0.00513696670532227
0.316915273666382 0.00599312782287598
0.387442946434021 0.0102739334106445
0.454594612121582 0.0162671804428101
0.464829206466675 0.0188356637954712
0.636717796325684 0.0291095972061157
0.711424827575684 0.040239691734314
0.746590971946716 0.0428081750869751
0.749353647232056 0.0453766584396362
0.771653532981873 0.0488014221191406
0.830410718917847 0.0539383888244629
0.849928379058838 0.056506872177124
0.850812196731567 0.0582191944122314
0.917634248733521 0.0642123222351074
0.970981597900391 0.068493127822876
1.05107533931732 0.0779109001159668
1.0563600063324 0.0804795026779175
1.06739413738251 0.0839041471481323
1.11290383338928 0.0873287916183472
1.1177659034729 0.0890411138534546
1.15769827365875 0.0933219194412231
1.1699732542038 0.0976027250289917
1.18856453895569 0.101027369499207
1.20295202732086 0.106164336204529
1.22331547737122 0.107876658439636
1.22370254993439 0.109589099884033
1.23226273059845 0.113013744354248
1.24515020847321 0.115582227706909
1.25626420974731 0.11815071105957
1.26355528831482 0.120719194412231
1.268763422966 0.125
1.27019810676575 0.125856161117554
1.28707468509674 0.128424644470215
1.30726647377014 0.130993127822876
1.3276914358139 0.136986255645752
1.34555721282959 0.138698577880859
1.34784412384033 0.140410900115967
1.3780996799469 0.145547986030579
1.39534199237823 0.14811646938324
1.44044077396393 0.154965758323669
1.46707224845886 0.157534241676331
1.46896088123322 0.158390402793884
1.49399757385254 0.161815047264099
1.55029273033142 0.172089099884033
1.56810462474823 0.174657583236694
1.57465612888336 0.176369905471802
1.59235429763794 0.178938388824463
1.60818552970886 0.181506872177124
1.63409328460693 0.184931516647339
1.64376735687256 0.1875
1.68809986114502 0.190068483352661
1.69023942947388 0.190924644470215
1.71822988986969 0.193493127822876
1.72113943099976 0.195205450057983
1.7548576593399 0.200342416763306
1.75666356086731 0.201198577880859
1.77856075763702 0.205479502677917
1.78652250766754 0.208047986030579
1.78929591178894 0.21061646938324
1.82196140289307 0.215753436088562
1.85106873512268 0.218321919441223
1.86017620563507 0.221746563911438
1.87282288074493 0.22688353061676
1.90926504135132 0.231164336204529
1.91361975669861 0.232876777648926
1.93579065799713 0.235445261001587
1.94444441795349 0.238013744354248
2.01315426826477 0.251712322235107
2.08014035224915 0.254280805587769
2.12184834480286 0.263698577880859
2.12950205802917 0.267123222351074
2.17874670028687 0.273972630500793
2.18379330635071 0.274828791618347
2.20839905738831 0.277397274971008
2.21835994720459 0.280821919441223
2.24642205238342 0.284246563911438
2.25558423995972 0.287671208381653
2.30991864204407 0.296232938766479
2.39442944526672 0.303938388824463
2.41246604919434 0.309075355529785
2.48355984687805 0.313356161117554
2.50174379348755 0.31934928894043
2.53034210205078 0.321917772293091
2.53328824043274 0.322773933410645
2.55698418617249 0.325342416763306
2.55930352210999 0.326198577880859
2.58669900894165 0.329623222351074
2.60790967941284 0.333904147148132
2.60941076278687 0.334760308265686
2.62648916244507 0.336472630500793
2.66448068618774 0.340753436088562
2.68899440765381 0.344178080558777
2.69984149932861 0.347602725028992
2.73224377632141 0.351027369499207
2.7366406917572 0.354452013969421
2.7414116859436 0.356164455413818
2.75254225730896 0.357876777648926
2.75938892364502 0.360445261001587
2.76616358757019 0.363013744354248
2.78969502449036 0.370719194412231
2.81162285804749 0.373287677764893
2.81454372406006 0.375
2.84025549888611 0.378424644470215
2.86007690429688 0.385273933410645
2.8755624294281 0.386986255645752
2.88379383087158 0.393835544586182
2.90058469772339 0.39811646938324
2.90908885002136 0.399828791618347
2.92084145545959 0.402397274971008
2.93625354766846 0.404965758323669
2.95961689949036 0.408390402793884
2.99329161643982 0.412671208381653
3.00736045837402 0.416095852851868
3.01116633415222 0.417808175086975
3.01127505302429 0.419520616531372
3.03540658950806 0.427226066589355
3.04127717018127 0.428938388824463
3.04611349105835 0.43065071105957
3.08279800415039 0.434931516647339
3.08793950080872 0.436643838882446
3.10698628425598 0.440068483352661
3.12016296386719 0.442636966705322
3.14275145530701 0.446061611175537
3.17038035392761 0.448630094528198
3.20692777633667 0.451198577880859
3.20924258232117 0.453767061233521
3.30894899368286 0.462328791618347
3.31716418266296 0.464041113853455
3.37465858459473 0.467465758323669
3.4335765838623 0.474315047264099
3.4706666469574 0.483732938766479
3.50122404098511 0.488013744354248
3.56874370574951 0.499143838882446
3.61984086036682 0.503424644470215
3.70660018920898 0.511986255645752
3.73805284500122 0.513698577880859
3.74458122253418 0.516267061233521
3.77490520477295 0.520547866821289
3.80513882637024 0.52311635017395
3.81010627746582 0.524828791618347
3.82820391654968 0.527397274971008
3.83641028404236 0.530821919441223
3.85226511955261 0.533390402793884
3.867182970047 0.535958886146545
3.90454483032227 0.546232938766479
3.91837000846863 0.553938388824463
3.94217419624329 0.558219194412231
3.95549535751343 0.564212322235107
3.96389222145081 0.571061611175537
3.97590851783752 0.573630094528198
3.97642278671265 0.575342416763306
3.9848313331604 0.577910900115967
3.98681545257568 0.579623222351074
3.99314141273499 0.582191705703735
3.99547004699707 0.583904027938843
4.0025520324707 0.586472630500793
4.00800657272339 0.589041113853455
4.01993751525879 0.592465758323669
4.02826166152954 0.595034241676331
4.05071020126343 0.597602725028992
4.05400943756104 0.599315047264099
4.07351160049438 0.601027369499207
4.08330059051514 0.606164455413818
4.09033632278442 0.608732938766479
4.09617567062378 0.610445261001587
4.1198935508728 0.615582227706909
4.12802314758301 0.619863033294678
4.2008638381958 0.632705450057983
4.22314691543579 0.635273933410645
4.22483396530151 0.636130094528198
4.24548721313477 0.639554738998413
4.24699258804321 0.640410900115967
4.29077625274658 0.646404027938843
4.32253360748291 0.652397274971008
4.3362774848938 0.657534241676331
4.34541845321655 0.660102725028992
4.35670185089111 0.662671208381653
4.36532545089722 0.666952133178711
4.3923864364624 0.672945261001587
4.3969087600708 0.675513744354248
4.4653525352478 0.686643838882446
4.49653577804565 0.692636966705322
4.56634759902954 0.702910900115967
4.6351375579834 0.71061635017395
4.64681577682495 0.714041113853455
4.64809417724609 0.716609597206116
4.66063022613525 0.719178080558777
4.68862342834473 0.721746563911438
4.71120119094849 0.724315047264099
4.71311950683594 0.726027369499207
4.74286365509033 0.731164455413818
4.76998090744019 0.733732938766479
4.77347612380981 0.734589099884033
4.863685131073 0.742294549942017
4.86648607254028 0.74315071105957
4.97535610198975 0.751712322235107
4.97999715805054 0.752568483352661
5.01765489578247 0.755136966705322
5.03837108612061 0.75684928894043
5.19705724716187 0.760273933410645
5.21294784545898 0.762842416763306
5.34609889984131 0.766267061233521
5.355393409729 0.767123222351074
5.47445392608643 0.770547866821289
5.49709510803223 0.772260189056396
5.61100339889526 0.775684952735901
5.61903810501099 0.777397274971008
5.74880504608154 0.783390402793884
5.76765632629395 0.785102725028992
5.77962303161621 0.785958886146545
6.22670650482178 0.796232938766479
6.52709245681763 0.801369905471802
6.73258876800537 0.80565071105957
6.75512981414795 0.806506872177124
6.84916734695435 0.809075355529785
6.89095830917358 0.810787677764893
6.94398069381714 0.813356161117554
6.95037508010864 0.814212322235107
7.25365877151489 0.817636966705322
7.26115560531616 0.818493127822876
7.51085662841797 0.821061611175537
7.61523962020874 0.826198577880859
7.70913743972778 0.827910900115967
7.71693325042725 0.828767061233521
8.92582225799561 0.833904027938843
14.3271532058716 0.836472630500793
14.6339664459229 0.838184952735901
19.0534343719482 0.841609597206116
19.3417930603027 0.842465758323669
25.0876770019531 0.845034241676331
31.5812377929688 0.847602725028992
39.6345062255859 0.851027369499207
39.8307342529297 0.85188364982605
69.0308227539062 0.871575355529785
69.1410522460938 0.872431516647339
73.0238876342773 0.874143838882446
85.3019180297852 0.879280805587769
88.3098297119141 0.884417772293091
93.5864028930664 0.887842416763306
94.561767578125 0.890410900115967
94.6445541381836 0.891267061233521
95.9405670166016 0.892979383468628
95.9667510986328 0.894691705703735
96.9508056640625 0.897260189056396
98.451530456543 0.900684952735901
102.807052612305 0.904109597206116
105.302589416504 0.910102725028992
105.469093322754 0.910958886146545
107.453186035156 0.915239810943604
107.663475036621 0.917808294296265
109.336303710938 0.929794549942017
109.417366027832 0.93065071105957
111.857528686523 0.938356161117554
111.910758972168 0.939212322235107
112.350234985352 0.941780805587769
112.555335998535 0.943493127822876
113.313323974609 0.946061611175537
113.317153930664 0.946917772293091
113.361633300781 0.947773933410645
114.869255065918 0.954623222351074
115.671020507812 0.957191705703735
116.544242858887 0.959760189056396
118.423759460449 0.966609597206116
118.933265686035 0.970890402793884
118.938987731934 0.971746563911438
119.710784912109 0.974315047264099
125.102378845215 0.990582227706909
126.936195373535 0.992294549942017
127.476036071777 0.994006872177124
132.674407958984 0.997431516647339
133.01042175293 0.998287677764893
137.007217407227 0.999143838882446
150 1
};
\addlegendentry{\gls{ba}}

\addplot [very thick, UNIPDRED, mark=triangle*, mark repeat=55, mark phase=40]
table {%
0.0289767980575562 0
0.0303722620010376 0.000854015350341797
0.202624678611755 0.00597774982452393
1.18308234214783 0.0153714418411255
1.32052397727966 0.0179333686828613
1.33299481868744 0.0196412801742554
1.51515924930573 0.0247651338577271
1.53620231151581 0.0264730453491211
1.56433546543121 0.0290349721908569
1.6222243309021 0.0315968990325928
1.67487752437592 0.0358667373657227
1.77332091331482 0.0392826795578003
1.77959299087524 0.0401365756988525
1.84140121936798 0.0418446063995361
1.92658805847168 0.0435525178909302
1.97898852825165 0.0478223562240601
2.26137137413025 0.0631937980651855
2.36267256736755 0.0666097402572632
2.37676692008972 0.069171667098999
2.43156504631042 0.0717335939407349
2.43582963943481 0.0725874900817871
2.68806314468384 0.0853971242904663
2.71901226043701 0.0871050357818604
2.72243714332581 0.0879590511322021
2.7951352596283 0.090520977973938
2.84231758117676 0.0939368009567261
2.90548300743103 0.0964987277984619
2.94233179092407 0.09991455078125
2.95507764816284 0.102476477622986
3.03104996681213 0.111016273498535
3.06229972839355 0.112724184989929
3.08932113647461 0.116994023323059
3.27855062484741 0.119555950164795
3.29119801521301 0.120409965515137
3.51719903945923 0.123825788497925
3.62557029724121 0.129803538322449
3.68084859848022 0.134073495864868
3.79968762397766 0.140051245689392
3.82330679893494 0.141759157180786
3.928382396698 0.146883010864258
3.93496823310852 0.148590922355652
3.98987174034119 0.152006864547729
3.99186944961548 0.152860760688782
4.04614448547363 0.157130718231201
4.21592712402344 0.159692525863647
4.2297797203064 0.160546541213989
4.40879821777344 0.165670394897461
4.4314341545105 0.168232321739197
4.83414125442505 0.184457778930664
4.9084734916687 0.194705367088318
4.95826148986816 0.197267293930054
4.97684049606323 0.19982922077179
4.99368476867676 0.202391147613525
4.99542570114136 0.203245043754578
5.03466701507568 0.206660985946655
5.06493330001831 0.208368897438049
5.07297563552856 0.210930824279785
5.14538717269897 0.216908574104309
5.18001174926758 0.218616604804993
5.18820858001709 0.221178531646729
5.22044849395752 0.223740339279175
5.23146677017212 0.227156281471252
5.36516904830933 0.232280135154724
5.39812088012695 0.236549973487854
5.42824411392212 0.23911190032959
5.45597887039185 0.241673827171326
5.47463989257812 0.245943665504456
5.49918460845947 0.248505592346191
5.50106954574585 0.249359488487244
5.52552509307861 0.251921415328979
5.58635759353638 0.257045269012451
5.61265420913696 0.259607195854187
5.61499357223511 0.260461091995239
5.70484972000122 0.266438961029053
5.78890895843506 0.277540564537048
5.84387302398682 0.280956506729126
5.85040473937988 0.283518314361572
5.88862133026123 0.285226345062256
5.89476203918457 0.28693425655365
5.92391109466553 0.290350198745728
5.92849159240723 0.293766021728516
5.97092485427856 0.297181844711304
5.9902663230896 0.301451683044434
6.01384544372559 0.303159713745117
6.08385372161865 0.305721640586853
6.08783626556396 0.306575536727905
6.13667869567871 0.309137463569641
6.14739608764648 0.311699390411377
6.20123815536499 0.315969228744507
6.22123527526855 0.31767725944519
6.22696113586426 0.318531155586243
6.37836503982544 0.32621693611145
6.43457221984863 0.333902597427368
6.45638656616211 0.337318539619446
6.48637676239014 0.340734481811523
6.48954010009766 0.342442393302917
6.52597761154175 0.345858216285706
6.52997827529907 0.346712231636047
6.54628705978394 0.348420143127441
6.54918909072876 0.350128054618835
6.61159133911133 0.352689981460571
6.67447996139526 0.355251908302307
6.69886064529419 0.361229658126831
6.72348690032959 0.363791704177856
6.75147819519043 0.368061542510986
6.83377456665039 0.371477365493774
6.86317300796509 0.37403929233551
6.92678356170654 0.37830913066864
6.94471836090088 0.380871057510376
7.00943040847778 0.383432984352112
7.01608896255493 0.384286880493164
7.12304210662842 0.390264749526978
7.12679147720337 0.391972661018372
7.20395994186401 0.397096514701843
7.30439853668213 0.400512456893921
7.3853702545166 0.404782295227051
7.41659498214722 0.410760045051575
7.45997714996338 0.413321971893311
7.46936893463135 0.414175987243652
7.76869821548462 0.425277471542358
7.77313804626465 0.4261314868927
7.83106088638306 0.43040132522583
7.84770393371582 0.432963252067566
7.93827629089355 0.436379194259644
7.94990110397339 0.437233090400696
8.05379772186279 0.439795017242432
8.05981731414795 0.440649032592773
8.09958076477051 0.443210959434509
8.10349559783936 0.444064855575562
8.18374443054199 0.447480797767639
8.30713844299316 0.453458547592163
8.39748573303223 0.456874489784241
8.39969253540039 0.457728385925293
8.45501518249512 0.460290312767029
8.5699634552002 0.464560270309448
8.58772087097168 0.467976093292236
8.6650333404541 0.470538020133972
8.6731595993042 0.473099946975708
8.7882719039917 0.475661754608154
8.8262939453125 0.47822380065918
8.83227825164795 0.479931712150574
8.92188930511475 0.483347535133362
8.95973682403564 0.485909461975098
8.98604106903076 0.486763477325439
9.08950901031494 0.489325404167175
9.11531257629395 0.491887331008911
9.12015151977539 0.493595242500305
9.16714382171631 0.497011065483093
9.23460006713867 0.499572992324829
9.36495971679688 0.506404757499695
9.50808525085449 0.509820699691772
9.51611423492432 0.511528611183167
9.64209651947021 0.519214391708374
9.79027462005615 0.52177619934082
9.8191614151001 0.524338245391846
10.066014289856 0.53031599521637
10.2450361251831 0.542271614074707
10.3088493347168 0.544833421707153
10.4731740951538 0.549103260040283
10.6227836608887 0.551665306091309
10.6430931091309 0.552519202232361
10.7966222763062 0.555081129074097
10.8345251083374 0.556789040565491
10.8842706680298 0.559350967407227
11.0377817153931 0.564474821090698
11.2224426269531 0.567036747932434
11.3645582199097 0.572160482406616
11.4389629364014 0.576430320739746
11.4535903930664 0.57813835144043
11.5039510726929 0.580700278282166
11.5102796554565 0.581554174423218
12.0305042266846 0.590947866439819
12.0670118331909 0.592655897140503
12.4528026580811 0.601195573806763
12.6929979324341 0.604611396789551
12.8239183425903 0.608027338981628
12.8267736434937 0.608881235122681
13.2901153564453 0.614859104156494
13.4663591384888 0.61742103099823
13.5174312591553 0.619128942489624
13.8038330078125 0.622544765472412
13.8139915466309 0.624252796173096
13.8239774703979 0.625106811523438
16.1122589111328 0.648164033889771
16.1256885528564 0.649017930030823
16.2921371459961 0.651579856872559
18.2468070983887 0.674637079238892
18.2654819488525 0.676344990730286
18.6654186248779 0.678906917572021
18.8549823760986 0.681468844413757
19.0438899993896 0.687446594238281
19.6508903503418 0.693424463272095
19.9932956695557 0.696840286254883
21.0874576568604 0.70452606678009
21.0904846191406 0.705379962921143
21.3305225372314 0.709649801254272
21.3897647857666 0.711357831954956
21.4990520477295 0.71306574344635
21.8399963378906 0.715627670288086
21.8872127532959 0.716481685638428
22.2527885437012 0.719897508621216
22.2762203216553 0.720751523971558
22.5051898956299 0.723313331604004
22.522029876709 0.724167346954346
23.0651779174805 0.730145215988159
23.0733013153076 0.730999231338501
23.4162578582764 0.736122965812683
24.2881107330322 0.739538908004761
24.3065128326416 0.741246819496155
25.3439922332764 0.753202438354492
26.0461235046387 0.75661826133728
26.0489463806152 0.757472276687622
26.4485015869141 0.76088809967041
26.4760704040527 0.761742115020752
27.0505523681641 0.766011953353882
27.0820560455322 0.766865968704224
27.361457824707 0.76942777633667
27.7471179962158 0.774551630020142
27.7794189453125 0.775405645370483
28.2970504760742 0.781383514404297
28.2990036010742 0.782237410545349
28.8982772827148 0.789069175720215
28.9202098846436 0.789923191070557
29.0689373016357 0.792484998703003
29.4219284057617 0.802732706069946
29.7742919921875 0.807002544403076
29.8001537322998 0.807856559753418
30.2231960296631 0.81298041343689
30.2261695861816 0.813834309577942
30.2766094207764 0.816396236419678
30.4112644195557 0.818104267120361
30.8912830352783 0.827497839927673
31.1179485321045 0.830059766769409
31.6326141357422 0.836891531944275
31.6500930786133 0.837745428085327
32.2227058410645 0.843723297119141
33.3905563354492 0.850555062294006
33.5965690612793 0.854825019836426
33.6460113525391 0.85653281211853
34.9136848449707 0.863364696502686
35.0228233337402 0.865926504135132
35.3288040161133 0.869342446327209
35.3761825561523 0.871050357818604
36.076847076416 0.873612284660339
36.0905342102051 0.874466180801392
36.7064933776855 0.877028226852417
36.7268295288086 0.877882122993469
37.3939590454102 0.883005976676941
37.4003524780273 0.883859872817993
37.5398597717285 0.884713888168335
37.9956665039062 0.886421918869019
38.0220985412598 0.887275815010071
39.5665512084961 0.90008544921875
39.5688209533691 0.900939345359802
40.0488739013672 0.90435528755188
40.0507926940918 0.905209302902222
40.1097526550293 0.906063199043274
40.6536102294922 0.909479141235352
40.6662864685059 0.910333037376404
40.8944702148438 0.912040948867798
40.9101715087891 0.91289496421814
41.19091796875 0.915456891059875
41.5601081848145 0.919726729393005
42.4286880493164 0.925704479217529
42.9274063110352 0.927412509918213
43.0010261535645 0.928266525268555
43.6782646179199 0.931682348251343
43.7923851013184 0.934244155883789
44.629508972168 0.938514113426208
45.2001724243164 0.940222024917603
45.2591209411621 0.941076040267944
46.7017097473145 0.947053790092468
47.0148849487305 0.950469732284546
47.9066162109375 0.953031539916992
48.0280609130859 0.95644748210907
48.1250762939453 0.957301378250122
51.0027160644531 0.965841174125671
51.4002342224121 0.967549085617065
52.0696144104004 0.970111012458801
54.4871444702148 0.979504704475403
54.5195388793945 0.980358600616455
56.4846000671387 0.986336469650269
60.1672096252441 0.990606307983398
60.2209014892578 0.99146032333374
61.9503746032715 0.994876146316528
62.4876708984375 0.99573016166687
66.9180297851562 0.998291969299316
77.0769271850586 0.999145984649658
150 1
};
\addlegendentry{\gls{mrba}}
\end{axis}

\end{tikzpicture}

%% file: Figures/Sim_results/Perf_vs_alloc_period.tex
\begin{tikzpicture}
\pgfplotsset{every tick label/.append style={font=\small}}

\definecolor{UNIPDRED}{RGB}{155,0,20}
\definecolor{LIGHT_GREY}{RGB}{189,195,199}
\definecolor{COMPLEMENTARY}{RGB}{0,153,153}
\definecolor{DARKGREY}{RGB}{55,65,64}
\definecolor{SAND}{RGB}{180,160,135}

\begin{axis}[
width=\fwidth,
height=\fheight,
at={(0\fwidth,0\fheight)},
scale only axis,
legend style={
  at={(0.5,0.98)}, 
  anchor=south, 
  draw=white!80!black, 
  font=\footnotesize},
  /tikz/every even column/.append style={column sep=0.3cm},
  legend image post style={scale=0.7},
  legend columns=4,
xlabel style={font=\footnotesize},
xlabel={Per \gls{ue} throughput [Mbps]},
xmajorgrids,
xmin=2.5, xmax=8.5,
xtick={3, 4, 5, 6, 7, 8},
xtick style={color=white!15!black},
ylabel style={font=\footnotesize},
ylabel={Delay [ms]},
ymajorgrids,
ymin=0, ymax=800,
ytick={100, 300, 500, 700},
ytick style={color=white!15!black}
]

\addplot[
  scatter,
  only marks,
  scatter src=explicit,
  scatter/classes={1={SAND}, 2={COMPLEMENTARY}, 3={DARKGREY}, 4={UNIPDRED}},
  mark=diamond*,
  mark size=6,
  forget plot
]
table[x=x,y=y, meta=class]{%
x                      y              class
7.43779 94.579039 4
7.05284 142.562399 3
5.50202 266.508319 2
3.28748 728.391519 1
};

\addplot[
  scatter,
  only marks,
  scatter src=explicit,
  scatter/classes={1={SAND}, 2={COMPLEMENTARY}, 3={DARKGREY}, 4={UNIPDRED}},
  forget plot,
  mark=10-pointed star,
  mark size=6,
]
table[x=x,y=y, meta=class]{%
x                      y              class
6.256849 279.149919 4
6.021586 456.087359 3
4.18902 477.641599 2
3.28748 728.391519 1
};

\addplot[
  scatter,
  only marks,
  scatter src=explicit,
  scatter/classes={1={SAND}, 2={COMPLEMENTARY}, 3={DARKGREY}, 4={UNIPDRED}},
  mark size=6,
]
table[x=x,y=y, meta=class]{%
x                      y              class
4.002868 493.541599 4
4.557969 602.633279 3
3.46291 614.112479 2
3.28748 728.391519 1
};
\legend{Distr,\gls{msr},\gls{ba},\gls{mrba}}

\end{axis}

\begin{axis}[
width=\fwidth,
height=\fheight,
at={(0\fwidth,0\fheight)},
scale only axis,
legend style={
  at={(0.97,0.91)}, 
  anchor=north east, 
  draw=white!80!black,
  font=\footnotesize},
xlabel style={font=\footnotesize},
xmajorgrids,
xmin=2.5, xmax=8.5,
xtick style={color=white!15!black},
ylabel style={font=\footnotesize},
ymajorgrids,
ymin=1, ymax=5.4,
ytick style={color=white!15!black},
hide y axis,
hide x axis,
]

\addplot[
  scatter,
  only marks,
  scatter src=explicit,
  mark size=5,
  color=black,
]
table[x=x,y=y]{%
x                      y
-20 -20 
};
\addlegendentry{4 Subframes}

\addplot[
  scatter,
  only marks,
  scatter src=explicit,
  mark size=5,
  mark=10-pointed star,
  color=black,
]
table[x=x,y=y]{%
x                      y
-20 -21 
};
\addlegendentry{2 Subframes}

\addplot[
  scatter,
  only marks,
  scatter src=explicit,
  mark size=5,
  mark=diamond*,
  color=black,
]
table[x=x,y=y]{%
x                      y
-20 -22
};
\addlegendentry{1 Subframe}

\end{axis}

\end{tikzpicture}

%% file: Figures/Sim_results/MWM_runtime_vs_nodes.tex
\begin{tikzpicture}
\pgfplotsset{every tick label/.append style={font=\small}}

    \definecolor{UNIPDRED}{RGB}{155,0,20}
    \definecolor{LIGHT_GREY}{RGB}{189,195,199}
    \definecolor{COMPLEMENTARY}{RGB}{0,153,153}
    \definecolor{DARKGREY}{RGB}{55,65,64}
    \definecolor{SAND}{RGB}{180,160,135}

    \begin{axis}[
        width=\fwidth,
        height=\fheight,
        at={(0\fwidth,0\fheight)},
        scale only axis,
        legend style={
            /tikz/every even column/.append style={column sep=0.2cm},
            at={(0.25,0.7)}, 
            anchor=south, 
            draw=white!80!black, 
            font=\scriptsize
            },
        legend columns=2,
        xlabel style={font=\footnotesize},
        xlabel={Number of \gls{iab}-nodes $\in \, \mathcal{V}$},
        xtick={1, 2, 3, 4, 5, 6, 7, 8},
        xmajorgrids,
        xmin=0.5, xmax=8.5,
        xtick style={color=white!15!black},
        ylabel shift = -1 pt,
        ylabel style={font=\footnotesize},
        ylabel={Runtime [$\mu$s]},
        ymajorgrids,
        ymin=1.45, ymax=6.25,
        ytick style={color=white!15!black},
        ytick={2, 3, 4, 5, 6},
        ]

\addplot [very thick,
          UNIPDRED,
          error bars/.cd, 
          y dir=both, 
          y explicit,
          error bar style={line width=1.2pt},
          error mark options={
            rotate=90,
            UNIPDRED,
            mark size=4pt,
            line width=1.2pt
          }]
plot coordinates {%
    (1, 1.82969) +- (1, 0.1748)
    (2, 2.46836) +- (2, 0.1749) 
    (4, 3.88117) +- (4, 0.1748) 
    (6, 4.51493) +- (6, 0.1748) 
    (8, 5.80029) +- (8, 0.1749) 
};
\end{axis}

\end{tikzpicture}